\documentclass{aa}
\usepackage{graphicx}
\usepackage{natbib} 
\usepackage[utf8]{inputenc}
\usepackage{xcolor}

\newcommand{\ie}{{\em i.e.}}
\newcommand{\eg}{{\em e.g.}}

\newcommand{\emm}[1]{\ensuremath{#1}}
\newcommand{\emr}[1]{\emm{\mathrm{#1}}}
\newcommand{\chem}[1]{\emr{#1}}
\newcommand{\unit}[1]{\emr{\,#1}}
\newcommand{\e}[1]{\emm{\times 10^{#1}}}
\newcommand{\radec}[6]{\emr{#1^{h}#2^{m}#3^{s},#4^{\circ}#5^{'}#6^{''}}}

\newcommand{\paren}[1]  {\emm{\left(  #1 \right) }}
\newcommand{\cbrace}[1] {\emm{\left\{ #1 \right\}}}

\newcommand{\Msol}{\unit{M_\odot}}
\newcommand{\Lsol}{\unit{L_\odot}}
\renewcommand{\deg}{\emm{^{\circ}}}
\newcommand{\au}{\unit{AU}}
\newcommand{\mpc}{\unit{mpc}}
\newcommand{\pc}{\unit{pc}}
\newcommand{\kpc}{\unit{kpc}}
\newcommand{\Mpc}{\unit{Mpc}}

\newcommand{\cm}{\unit{cm}}
\newcommand{\mm}{\unit{mm}}
\newcommand{\mum}{\unit{\mu m}}
\newcommand{\pccm}{\unit{cm^{-3}}}
\newcommand{\pscm}{\unit{cm^{-2}}}
\newcommand{\pspc}{\unit{pc^{-2}}}
\newcommand{\pcpc}{\unit{pc^{-3}}}

\newcommand{\kHz}{\unit{kHz}}
\newcommand{\MHz}{\unit{MHz}}
\newcommand{\GHz}{\unit{GHz}}
\newcommand{\ps}{\unit{s^{-1}}}
\newcommand{\kms}{\unit{km\,s^{-1}}}

\newcommand{\Kkms}{\unit{K\,km\,s^{-1}}}
\newcommand{\mKkms}{\unit{mK\,km\,s^{-1}}}
\newcommand{\Kkmspcpc}{\unit{K\,km\,s^{-1}\,pc^{2}}}
\newcommand{\pscmpKkms}{\unit{cm^{-2}/(K\,km\,s^{-1})}}
\newcommand{\K}{\unit{K}}
\newcommand{\mK}{\unit{mK}}

\newcommand{\magn}{\unit{mag}}

\newcommand{\Hi}{\ion{H}{i}}
\newcommand{\Hii}{\ion{H}{ii}}
\renewcommand{\H}{\chem{H}}
\newcommand{\HH}{\chem{H_2}}
\newcommand{\NN}{\chem{N_2}}
\newcommand{\twCp}{\chem{^{12}C^+}}
\newcommand{\thCp}{\chem{^{13}C^+}}
\newcommand{\twCO}{\chem{^{12}CO}}
\newcommand{\thCO}{\chem{^{13}CO}}
\newcommand{\CeiO}{\chem{C^{18}O}}
\newcommand{\CseO}{\chem{C^{17}O}}
\newcommand{\twCS}{\chem{^{12}CS}}
\newcommand{\ttSO}{\chem{^{32}SO}}
\newcommand{\tfSO}{\chem{^{34}SO}}
\newcommand{\NNHp}{\chem{N_{2}H^{+}}}
\newcommand{\HCOp}{\chem{HCO^{+}}}
\newcommand{\HthCOp}{\chem{H^{13}CO^{+}}}
\newcommand{\HCN}{\chem{HCN}}
\newcommand{\HthCN}{\chem{H^{13}CN}}
\newcommand{\HNC}{\chem{HNC}}
\newcommand{\HNthC}{\chem{HN^{13}C}}
\newcommand{\CN}{\chem{CN}}
\newcommand{\twCN}{\chem{^{12}CN}}
\newcommand{\methanol}{\chem{CH_{3}OH}}
\newcommand{\CCH}{\chem{C_{2}H}}
\newcommand{\CCCHH}{\chem{C_{3}H_{2}}}
\newcommand{\cCCCHH}{\chem{c-C_{3}H_{2}}}
\newcommand{\SiO}{\chem{SiO}}

\newcommand{\Jone}{\chem{(1-0)}}
\newcommand{\Jtwo}{\chem{(2-1)}}
\newcommand{\T}[4]{\emm{#1-#3}}
\newcommand{\J}[2]{\T{#1}{}{#2}{}}
\newcommand{\Aij}{\emm{A_{ij}}}
\newcommand{\Eu}{\emm{E_\emr{u}}}

\newcommand{\Mvir}[1][]{\emm{M_\emr{vir}^\emr{#1}}}
\newcommand{\dv}{\emm{\sigma}}
\newcommand{\R}{\emm{R}}

\newcommand{\Gz}[1][]{\emm{G_{0}^\emr{#1}}}
\newcommand{\Td}[1][]{\emm{T_\emr{d}^\emr{#1}}}

\newcommand{\W}[1][]{\emm{W^\emr{#1}}}
\newcommand{\Wco}[1][]{\emm{W^\emr{#1}_\emr{CO}}}
\newcommand{\Xco}{\emm{X_\emr{CO}}}
\newcommand{\NH}{\emm{N_\emr{H}}}
\newcommand{\Ebv}{\emm{E_\emr{B-V}}}
\newcommand{\Rv}[1][]{\emm{R_\emr{V}^\emr{#1}}}
\newcommand{\Av}[1][]{\emm{A_\emr{V}^\emr{#1}}}
\newcommand{\Ak}{\emm{A_\emr{K}}}
\newcommand{\tauk}{\emm{\tau_\emr{850}}}

\newcommand{\Tsys}{\emm{T_\emr{sys}}}
\newcommand{\Tas}{\emm{T_\emr{A}^{*}}}
\newcommand{\Tmb}{\emm{T_\emr{mb}}}
\newcommand{\Feff}{\emm{F_\emr{eff}}}
\newcommand{\Beff}{\emm{B_\emr{eff}}}

\newcommand{\ncrit}{\emm{n_\emr{crit}}}

\newcommand{\TabStars}{%
  \begin{table*}
    \caption{Properties of the stars exciting the \Hii{} regions in the
      observed field of view.}  
    \centering{} %
    \tiny{%
      \begin{tabular}{lccccccc}
        \hline \hline
        \Hii{} region & Star & Type & $\alpha,\delta$ (J2000) & $(\delta x,\delta y)$ & Parallax & Distance & $V_\emr{LSR}$ \\
                      &      &      &                         & $('','')$             & mas      & pc       & \kms{}        \\   
        \hline
        IC\,434   & $\sigma$Ori & O9.5V B        & \radec{05}{38}{44.779}{-02}{36}{00.12} & $(-33.35,00.07)$ & $2.5806\pm0.0088^{(1)}$ & $387.5\pm1.3$$^{(1)}$ & $15.0 \pm 1.6$ \\
        IC\,435   & HD\,38087   & B5V D          & \radec{05}{43}{00.573}{-02}{18}{45.38} & $(+32.87,01.33)$ &     $5.90\pm1.29^{(2)}$ & $169\pm37$$^{(2)}$    & $18.1 \pm 4.5$ \\
        NGC\,2023 & HD\,37903   & B1.5V C        & \radec{05}{41}{38.388}{-02}{15}{32.48} & $(+13.72,09.42)$ &   $2.776\pm0.271^{(3)}$ & $362\pm35$$^{(3)}$    & $-7.7 \pm 2$   \\
        NGC\,2024 & IRS2b       & O8V-B2V        & \radec{05}{41}{45.50}{-01}{54}{28.7}   & $(+20.54,29.43)$ &                     --- & $415$$^{(4)}$         & ---            \\
                  & Alnitak     & O9.7Ib+B0III C & \radec{05}{40}{45.527}{-01}{56}{33.26} & $(+05.49,31.04)$ &       $3.4\pm0.2^{(5)}$ & $294\pm21$$^{(5)}$    & $3.7 \pm 1.3$  \\
        \hline
      \end{tabular}
      \tablefoot{%
        \tablefoottext{1}{\citet{schaefer16}}
        \tablefoottext{2}{\citet{vanleeuwen07}}
        \tablefoottext{3}{Gaia DR1~\citet{gaia16,lindegren16,fabricius16,brown16}}
        \tablefoottext{4}{\citet{anthony-twarog82}}
        \tablefoottext{5}{\citet{hummel13}}}
    }
    \label{tab:stars}
  \end{table*}}

\newcommand{\TabOrionB}{%
  \begin{table*}
    \caption{Typical properties of the South-Western edge of Orion\,B.}
    \centering{} %
    \begin{tabular}{lcc}
      \hline \hline
      Parameter & Value & Notes\\
      \hline
      Distance                             & $400\pc$                                           & $1''=2\mpc$             \\
      Systemic velocity                    & $10.5\kms$                                         & LSR, radio convention   \\
      Projection center                    & \radec{05}{40}{54.270}{-02}{28}{00.00}             & $\alpha,\delta$(J2000), mane of the Horsehead \\
      Offset range \& Field of view        & $[-5.2',+43.3'] \times [-19.5',45.3']$             & $49' \times 65'$ or $5.6 \times 7.5\pc$       \\
      $\Wco[min] - \Wco[mean] - \Wco[max]$ & $0-61-288\Kkms$                                    & in $[-2,+18]\kms{}$     \\ 
      $\Av[min]  - \Av[mean]  - \Av[max]$  & $0.7-4.7-222\magn$                                 & $\Ak/\Av = 0.13$        \\ 
      $\Td[min]  - \Td[mean]  - \Td[max]$  & $16-26-99\K$                                       & \\
      $\Gz[min]  - \Gz[mean]  - \Gz[max]$  & $4-45-36000$                                     & Inter-Stellar Radiation Field~\citep{habing68} \\
      ~~~CO-traced mass                    & $11000\Msol$                                       & Standard \Xco{} \& Helium dealt with    \\ 
      ~Dust-traced mass                    & $3900\Msol$                                        & Standard \NH/\Av{} \& \Hi{} gas negligible \\ 
      Virial traced mass                   & Between $6200$ and $9500\Msol$                     & Depending on the assumed density radial profile \\
      Imaged surface                       & $43\pc^2$                                          & $= S$\\ 
      Typical volume                       & $280\pc^3$                                         & $= S^{3/2}$\\ 
      ~~CO-traced mean column density      & $260\Msol\pc^{-2}$ |  $12\times 10^{21}\HH{}\pscm$ & Standard \Xco{} \& Helium dealt with   \\ 
      Dust-traced mean column density      &  $92\Msol\pc^{-2}$ |   $4\times 10^{21}\HH{}\pscm$ & Standard \NH/\Av{} \& \Hi{} gas negligible \\ 
      ~~CO-traced mean volume density      &  $40\Msol\pc^{-3}$ | $590\HH{}\pccm$               & Standard \Xco{} \& Helium dealt with    \\ 
      Dust-traced mean volume density      &  $14\Msol\pc^{-3}$ | $210\HH{}\pccm$               & Standard \NH/\Av{} \& \Hi{} gas negligible \\
      \hline
    \end{tabular}
    \label{tab:l1630}
  \end{table*}}

\newcommand{\TabLuminosity}{%
  \begin{table*}
    \caption{Line intensities and luminosities in the [-2,+18\kms] velocity
      range and including all pixels.}
    \centering{} %
    \tiny %
    \begin{tabular}{c|cc|cc|rr|c}
      \hline
      \hline
      Species     & Simplified$^b$  & Complete$^c$        & \Aij{} & $\Eu/k$ & Intensity & Relative   & Luminosity\\
                  & quantum numbers & quantum numbers     & \ps{}  & \K{}    & \mKkms{}  & to \twCO{} & \Lsol{}   \\ 
      \hline
      \twCO{}     & \J{1}{0}           & $J=1-0$                       &  $7.2\times10^{-8}$  & 5.5  & 60\,430 & 100.00 & 1.0\e{-2} \\
      \thCO{}     & \J{1}{0}           & $J=1-0$                       &  $3.2\times10^{-8}$  & 5.3  &  9\,198 &  15.22 & 1.4\e{-3} \\
      \HCOp{}     & \J{1}{0}           & $J=1-0$                       &  $4.2\times10^{-5}$  & 4.3  &  1\,630 &   2.70 & 1.3\e{-4} \\
      \HCN{}      & \J{1}{0}           & $J=1-0, F=2-1$                &  $2.4\times10^{-5}$  & 4.3  &  1\,540 &   2.55 & 1.2\e{-4} \\
      \CN{}       & \J{1}{0}           & $N=1-0, J=3/2-1/2, F=5/2-3/2$ &  $1.2\times10^{-5}$  & 5.4  &     776 &   1.28 & 1.3\e{-4} \\
      \CeiO{}     & \J{1}{0}           & $J=1-0$                       &  $6.3\times10^{-8}$  & 5.3  &     556 &   0.92 & 8.0\e{-5} \\
      \twCS{}     & \J{2}{1}           & $J=2-1$                       &  $1.7\times10^{-5}$  & 7.0  &     513 &   0.85 & 5.3\e{-5} \\
      \CCH{}      & \J{1}{0}           & $N=1-0, J=3/2-1/2, F= 2- 1$   &  $1.5\times10^{-6}$  & 4.2  &     457 &   0.76 & 3.2\e{-5} \\
      \HNC{}      & \J{1}{0}           & $J=1-0, F=2-1$                &  $2.7\times10^{-5}$  & 4.4  &     445 &   0.74 & 3.6\e{-5} \\
      \ttSO{}     & \T{2}{3}{1}{2}     & $J=3-2, K=2-1$                &  $1.1\times10^{-5}$  & 9.2  &     283 &   0.47 & 3.0\e{-5} \\
      \CseO{}     & \J{1}{0}           & $J=1-0, F=7/2-5/2$            &  $6.7\times10^{-8}$  & 5.4  &     215 &   0.36 & 3.3\e{-5} \\
      \cCCCHH{}   & \T{2}{1,2}{1}{0,1} & $J=2-1, K_{+}=1-0, K_{-}=2-1$ &  $2.3\times10^{-5}$  & 6.4  &     149 &   0.25 & 1.1\e{-5} \\
      \NNHp{}     & \J{1}{0}           & $J=1-0, F1=2-1, F=3-2$        &  $3.6\times10^{-5}$  & 4.5  &      67 &   0.11 & 6.0\e{-6} \\
      \methanol{} & \T{2}{0}{1}{0}     & $J=2-1, K =0-0$               &  $3.4\times10^{-6}$  & 7.0  &      65 &   0.11 & 6.4\e{-6} \\ 
      \HthCN{}    & \J{1}{0}           & $J=1-0, F=2-1$                &  $2.2\times10^{-5}$  & 4.1  &      48 &   0.08 & 3.3\e{-6} \\
      \HthCOp{}   & \J{1}{0}           & $J=1-0$                       &  $3.9\times10^{-5}$  & 4.2  &      25 &   0.04 & 1.8\e{-6} \\
      \HNthC{}    & \J{1}{0}           & $J=1-0, F=2-1$                &  $1.9\times10^{-5}$  & 4.2  &     --- &    --- &       --- \\
      \SiO{}      & \J{2}{1}           & $J=2-1$                       &  $2.9\times10^{-5}$  & 6.3  &     --- &    --- &       --- \\
      \hline
    \end{tabular}
    \tablefoot{%
      \tablefoottext{a}{The lines are sorted by decreasing value of their intensity.}
      \tablefoottext{b}{Simplified transition used everywhere else in the paper.}
      \tablefoottext{c}{Complete list of quantum numbers associated to the
        transition whose frequency is listed in Table~\ref{tab:obs}. This
        frequency is the one used to fix the velocity scale.}}
    \label{tab:luminosity}
  \end{table*}}

\newcommand{\TabAvMaskProp}{%
  \begin{table*}
    \caption{Properties of the \Av{} masks sorted by increasing range of visual extinction.}
    \centering{} %
    \tiny{%
      \begin{tabular}{llcccc} 
      \hline 
      \hline 
      Parameter                            & Unit                              & $1\le\Av<2$         & $2\le\Av<6$         & $6\le\Av<15$        & $15\le\Av<222$      \\ 
      \hline 
      $\Wco[min] - \Wco[mean] - \Wco[max]$ & $\Kkms$                           & $0-6.8-46.9$        & $1.3-62.6-211$      & $24.8-122-261$      & $41.5-137-288$      \\ 
      $\Av[min]  - \Av[mean]  - \Av[max]$  & $\magn$                           & $1-1.4-2$           & $2-3.6-6$           & $6-8.5-15$          & $15-28.9-222$       \\ 
      $\Td[min]  - \Td[mean]  - \Td[max]$  & $\K$                              & $19-24-45$          & $19-26-67$          & $18-29-99$          & $16-26-95$          \\ 
      $\Gz[min]  - \Gz[mean]  - \Gz[max]$  & ISRF~\citep{habing68}             & $8.9-30-680$        & $8.4-46-5100$       & $6-72-36000$        & $4-47-28000$        \\ 
      ~~CO-traced mass                     & $\Msol$                           & $320$ $(3\%)$       & $5600$ $(51\%)$     & $4400$ $(40\%)$     & $830$ $(8\%)$       \\ 
      Dust-traced mass                     & $\Msol$                           & $300$ $(8\%)$       & $1500$ $(38\%)$     & $1400$ $(36\%)$     & $790$ $(20\%)$      \\ 
      Emitting surface                     & $\pc^2$                           & $11$ $(25\%)$       & $21$ $(48\%)$       & $8.4$ $(20\%)$      & $1.4$ $(3.3\%)$     \\ 
      Typical volume                       & $\pc^3$                           & $36$ $(13\%)$       & $93$ $(33\%)$       & $24$ $(8.6\%)$      & $1.7$ $(0.6\%)$     \\ 
      ~~CO-traced mean column density      & $\Msol\pc^{-2}|10^{21}\HH{}\pscm$ & $30$ | $1.4$        & $270$ | $12$        & $530$ | $24$        & $590$ | $27$        \\ 
      Dust-traced mean column density      & $\Msol\pc^{-2}|10^{21}\HH{}\pscm$ & $28$ | $1.3$        & $71$ | $3.3$        & $160$ | $7.6$       & $560$ | $26$        \\ 
      ~~CO-traced mean volume density      & $\Msol\pc^{-3}|\HH{}\pccm$        & $8.9$ | $130$       & $60$ | $890$        & $180$ | $2700$      & $500$ | $7500$      \\ 
      Dust-traced mean volume density      & $\Msol\pc^{-3}|\HH{}\pccm$        & $8.4$ | $130$       & $16$ | $230$        & $57$ | $850$        & $470$ | $7100$      \\ 
      \hline 
      \end{tabular}}
    \label{tab:av:mask:prop}
  \end{table*}}

\newcommand{\TabTdustMaskProp}{%
  \begin{table*}
    \caption{Properties of the \Td{} masks sorted by increasing range of dust temperature.}
    \centering{} %
    \tiny{%
      \begin{tabular}{llcccc} 
      \hline 
      \hline 
      Parameter                            & Unit                              & $16\le\Td<19.5$     & $19.5\le\Td<23.5$   & $23.5\le\Td<32$     & $32\le\Td<100$      \\ 
      \hline 
      $\Wco[min] - \Wco[mean] - \Wco[max]$ & $\Kkms$                           & $0.2-63.8-149$      & $0-46.5-197$        & $0-57.4-273$        & $0.2-114-288$       \\ 
      $\Av[min]  - \Av[mean]  - \Av[max]$  & $\magn$                           & $1.1-17.2-127$      & $0.7-4-69$          & $0.8-4.1-222$       & $0.8-7.2-186$       \\ 
      $\Td[min]  - \Td[mean]  - \Td[max]$  & $\K$                              & $16-19-20$          & $20-22-24$          & $24-27-32$          & $32-40-99$          \\ 
      $\Gz[min]  - \Gz[mean]  - \Gz[max]$  & ISRF~\citep{habing68}             & $4-8.6-10$          & $10-18-26$          & $26-50-120$         & $120-400-36000$     \\ 
      ~~CO-traced mass                     & $\Msol$                           & $220$ $(2\%)$       & $3900$ $(35\%)$     & $4300$ $(39\%)$     & $2700$ $(25\%)$     \\ 
      Dust-traced mass                     & $\Msol$                           & $260$ $(7\%)$       & $1500$ $(38\%)$     & $1400$ $(36\%)$     & $770$ $(20\%)$      \\ 
      Emitting surface                     & $\pc^2$                           & $0.8$ $(1.8\%)$     & $19$ $(45\%)$       & $18$ $(41\%)$       & $5.5$ $(13\%)$      \\ 
      Typical volume                       & $\pc^3$                           & $0.7$ $(0.3\%)$     & $84$ $(30\%)$       & $73$ $(26\%)$       & $13$ $(4.6\%)$      \\ 
      ~~CO-traced mean column density      & $\Msol\pc^{-2}|10^{21}\HH{}\pscm$ & $280$ | $13$        & $200$ | $9.3$       & $250$ | $12$        & $490$ | $23$        \\ 
      Dust-traced mean column density      & $\Msol\pc^{-2}|10^{21}\HH{}\pscm$ & $330$ | $16$        & $78$ | $3.6$        & $81$ | $3.7$        & $140$ | $6.4$       \\ 
      ~~CO-traced mean volume density      & $\Msol\pc^{-3}|\HH{}\pccm$        & $310$ | $4600$      & $46$ | $690$        & $59$ | $890$        & $210$ | $3100$      \\ 
      Dust-traced mean volume density      & $\Msol\pc^{-3}|\HH{}\pccm$        & $380$ | $5600$      & $18$ | $270$        & $19$ | $290$        & $59$ | $890$        \\ 
      \hline 
      \end{tabular}}
    \label{tab:tdust:mask:prop}
  \end{table*}}

\newcommand{\TabAvMaskedFlux}{%
  \begin{table*}
    \caption{Percentage of the total line fluxes inside the four \Av{} 
      mask regions, integrated over $[9,12\kms]$.}
    \centering{} %
    \tiny{%
      \begin{tabular}{llccccc} 
      \hline 
      \hline 
      Species     & Transition         & $0\le\Av<222$     & $1\le\Av<2$       & $2\le\Av<6$       & $6\le\Av<15$      & $15\le\Av<222$    \\ 
      \hline 
      \twCO{}     & \J{1}{0}           & $100\%$           & $2.5\%$           & $52\%$            & $38\%$            & $7.6\%$           \\ 
      \CCH{}      & \J{1}{0}           & $100\%$           & $4.4\%$           & $41\%$            & $37\%$            & $17\%$            \\ 
      \cCCCHH{}   & \T{2}{1,2}{1}{0,1} & $100\%$           & $4.5\%$           & $40\%$            & $38\%$            & $17\%$            \\ 
      \HCOp{}     & \J{1}{0}           & $100\%$           & $2.8\%$           & $41\%$            & $40\%$            & $16\%$            \\ 
      \hline 
      \thCO{}     & \J{1}{0}           & $100\%$           & $1.4\%$           & $38\%$            & $45\%$            & $15\%$            \\ 
      \HCN{}      & \J{1}{0}           & $100\%$           & $1.7\%$           & $36\%$            & $44\%$            & $18\%$            \\ 
      \twCN{}     & \J{1}{0}           & $100\%$           & $2.6\%$           & $33\%$            & $45\%$            & $19\%$            \\ 
      \hline 
      \HNC{}      & \J{1}{0}           & $100\%$           & $2.1\%$           & $29\%$            & $41\%$            & $27\%$            \\ 
      \CseO{}     & \J{1}{0}           & $100\%$           & $5.6\%$           & $25\%$            & $43\%$            & $26\%$            \\ 
      \twCS{}     & \J{2}{1}           & $100\%$           & $0.68\%$          & $25\%$            & $42\%$            & $32\%$            \\ 
      \ttSO{}     & \T{2}{3}{1}{2}     & $100\%$           & $0.86\%$          & $24\%$            & $44\%$            & $31\%$            \\ 
      \CeiO{}     & \J{1}{0}           & $100\%$           & $0.49\%$          & $23\%$            & $48\%$            & $29\%$            \\ 
      \hline 
      \methanol{} & \J{2}{1}           & $99\%$            & $4.4\%$           & $5.8\%$           & $41\%$            & $48\%$            \\ 
      \HthCOp{}   & \J{1}{0}           & $98\%$            & $0.67\%$          & $7.1\%$           & $34\%$            & $56\%$            \\ 
      \NNHp{}     & \J{1}{0}           & $100\%$           & $-11\%$           & $8.2\%$           & $17\%$            & $88\%$            \\ 
      \hline 
      \end{tabular}
      \tablefoot{%
        \tablefoottext{a}{The lines are sorted by decreasing value of the
          flux coming from the diffuse and translucent lines of sight.}}}
    \label{tab:av:mask:flux}
  \end{table*}}

\newcommand{\TabTdustMaskedFlux}{%
  \begin{table*}
    \caption{Percentage of the total line fluxes inside the four \Td{} 
      mask regions, integrated over $[9,12\kms]$.}
    \centering{} %
    \tiny{%
      \begin{tabular}{llccccc} 
      \hline 
      \hline 
      Species     & Transition         & $16\le\Td<100$    & $16\le\Td<19.5$   & $19.5\le\Td<23.5$ & $23.5\le\Td<32$   & $32\le\Td<100$    \\ 
      \hline 
      \CCH{}      & \J{1}{0}           & $100\%$           & $3.2\%$           & $25\%$            & $39\%$            & $32\%$            \\ 
      \cCCCHH{}   & \T{2}{1,2}{1}{0,1} & $100\%$           & $4.2\%$           & $25\%$            & $39\%$            & $31\%$            \\ 
      \twCN{}     & \J{1}{0}           & $100\%$           & $4\%$             & $27\%$            & $33\%$            & $36\%$            \\ 
      \HCN{}      & \J{1}{0}           & $100\%$           & $4.2\%$           & $29\%$            & $34\%$            & $32\%$            \\ 
      \hline 
      \HCOp{}     & \J{1}{0}           & $100\%$           & $4.6\%$           & $30\%$            & $36\%$            & $29\%$            \\ 
      \HNC{}      & \J{1}{0}           & $100\%$           & $7.9\%$           & $31\%$            & $32\%$            & $29\%$            \\ 
      \twCO{}     & \J{1}{0}           & $100\%$           & $2.4\%$           & $38\%$            & $38\%$            & $22\%$            \\ 
      \hline 
      \twCS{}     & \J{2}{1}           & $100\%$           & $8.4\%$           & $35\%$            & $28\%$            & $29\%$            \\ 
      \thCO{}     & \J{1}{0}           & $100\%$           & $4.6\%$           & $41\%$            & $33\%$            & $21\%$            \\ 
      \ttSO{}     & \T{2}{3}{1}{2}     & $100\%$           & $11\%$            & $38\%$            & $28\%$            & $23\%$            \\ 
      \HthCOp{}   & \J{1}{0}           & $100\%$           & $16\%$            & $33\%$            & $24\%$            & $26\%$            \\ 
      \hline 
      \CseO{}     & \J{1}{0}           & $100\%$           & $7.9\%$           & $44\%$            & $29\%$            & $19\%$            \\ 
      \methanol{} & \J{2}{1}           & $100\%$           & $18\%$            & $35\%$            & $29\%$            & $18\%$            \\ 
      \CeiO{}     & \J{1}{0}           & $100\%$           & $9.1\%$           & $44\%$            & $27\%$            & $20\%$            \\ 
      \NNHp{}     & \J{1}{0}           & $100\%$           & $27\%$            & $39\%$            & $15\%$            & $20\%$            \\ 
      \hline 
      \end{tabular}
      \tablefoot{%
        \tablefoottext{a}{The lines are sorted by decreasing value of the
          flux coming from the warm and hot dust lines of sight.}}}
    \label{tab:tdust:mask:flux}
  \end{table*}}

\newcommand{\TabCorrAreaAv}{%
  \begin{table*}[p]
    \caption{Properties of the joint distributions of visual extinction,
      and line integrated intensities.}
    \centering{} %
    \tiny{%
      \begin{tabular}{llccc} 
      \hline 
      \hline 
      Species     & Transition         & Filling factor & $\W[2.5\%]-\W[med]-\W[97.5\%]$ & $\Av[2.5\%]-\Av[med]-\Av[97.5\%]$ \\ 
                  &                    & \%             & $\Kkms{}$                      & $\magn{}$                         \\ 
      \hline 
      \twCO{}     & \J{1}{0}           & $99$    & $1.4-53-180$      & $0.95-3.3-17$     \\ 
      \thCO{}     & \J{1}{0}           & $84$    & $0.31-7.5-38$     & $1.2-3.7-19$      \\ 
      \HCOp{}     & \J{1}{0}           & $68$    & $0.26-1.7-8.3$    & $1.8-4.4-22$      \\ 
      \HCN{}      & \J{1}{0}           & $57$    & $0.23-1.5-8.5$    & $2.2-5.1-24$      \\ 
      \twCS{}     & \J{2}{1}           & $41$    & $0.15-0.75-4.5$   & $2.9-6.3-29$      \\ 
      \CeiO{}     & \J{1}{0}           & $38$    & $0.2-1-5.5$       & $3.4-6.6-31$      \\ 
      \twCN{}     & \J{1}{0}           & $29$    & $0.21-1.2-6.1$    & $3.2-7.1-35$      \\ 
      \CCH{}      & \J{1}{0}           & $23$    & $0.19-1-3.4$      & $2.4-6.9-38$      \\ 
      \ttSO{}     & \T{2}{3}{1}{2}     & $27$    & $0.12-0.59-3.4$   & $3.4-7.5-36$      \\ 
      \HNC{}      & \J{1}{0}           & $33$    & $0.16-0.86-4.4$   & $2.9-6.9-33$      \\ 
      \CseO{}     & \J{1}{0}           & $9.9$   & $0.16-0.77-2.6$   & $6.1-12-54$       \\ 
      \cCCCHH{}   & \T{2}{1,2}{1}{0,1} & $5.8$   & $0.13-0.59-2$     & $3.2-10-62$       \\ 
      \methanol{} & \J{2}{1}           & $4.6$   & $0.089-0.39-2.4$  & $3.4-15-69$       \\ 
      \HthCOp{}   & \J{1}{0}           & $2.7$   & $0.14-0.76-2.6$   & $9.4-25-91$       \\ 
      \NNHp{}     & \J{1}{0}           & $2.4$   & $0.15-0.97-6.1$   & $12-26-93$        \\ 
      \tfSO{}     & \T{2}{3}{1}{2}     & $1$     & $0.07-0.29-1$     & $5.6-31-130$      \\ 
      \HthCN{}    & \J{1}{0}           & $0.32$  & $0.23-1.2-5.9$    & $24-46-180$       \\ 
      \HNthC{}    & \J{1}{0}           & $0.18$  & $0.16-0.61-1.9$   & $35-60-190$       \\ 
      \hline 
      \end{tabular}
      \tablefoot{%
        \tablefoottext{a}{The lines are sorted by decreasing value of their
          surface filling factor.}}}
    \label{tab:corr:area:av}
  \end{table*}}

\newcommand{\TabPseudoAbundance}{%
  \begin{table}
    \caption{Minimum, median, and maximum values of the abundances
      derived for each species.}
    \centering{} %
    \tiny{%
      \begin{tabular}{llc} 
      \hline 
      \hline 
      Species     & Transition         & $(\W/\NH)\times(N_\emr{species}/1\Kkms)$  \\ 
                  &                    & [Pseudo-Abundance]                        \\ 
                  &                    & min$-$med$-$max                           \\ 
      \hline 
      \twCO{}     & \J{1}{0}           & $1\e{-7} - 1\e{-5} - 3\e{-5}$             \\ 
      \thCO{}     & \J{1}{0}           & $1\e{-7} - 1\e{-6} - 4\e{-6}$             \\ 
      \CeiO{}     & \J{1}{0}           & $2\e{-8} - 1\e{-7} - 1\e{-6}$             \\ 
      \CCH{}      & \J{1}{0}           & $1\e{-9} - 1\e{-8} - 1\e{-7}$             \\ 
      \CseO{}     & \J{1}{0}           & $1\e{-8} - 5\e{-8} - 4\e{-7}$             \\ 
      \methanol{} & \J{2}{1}           & $1\e{-10} - 1\e{-9} - 3\e{-8}$            \\ 
      \ttSO{}     & \T{2}{3}{1}{2}     & $1\e{-10} - 1\e{-9} - 5\e{-9}$            \\ 
      \twCN{}     & \J{1}{0}           & $2\e{-10} - 2\e{-9} - 2\e{-8}$            \\ 
      \HCN{}      & \J{1}{0}           & $1\e{-10} - 1\e{-9} - 4\e{-9}$            \\ 
      \twCS{}     & \J{2}{1}           & $1\e{-10} - 4\e{-10} - 5\e{-9}$           \\ 
      \cCCCHH{}   & \T{2}{1,2}{1}{0,1} & $2\e{-11} - 3\e{-10} - 3\e{-9}$           \\ 
      \HCOp{}     & \J{1}{0}           & $4\e{-11} - 3\e{-10} - 2\e{-9}$           \\ 
      \NNHp{}     & \J{1}{0}           & $2\e{-11} - 2\e{-10} - 3\e{-9}$           \\ 
      \HNC{}      & \J{1}{0}           & $3\e{-11} - 2\e{-10} - 2\e{-9}$           \\ 
      \HthCOp{}   & \J{1}{0}           & $1\e{-11} - 3\e{-11} - 4\e{-10}$          \\ 
      \hline 
      \end{tabular} 
      \tablefoot{%
        \tablefoottext{a}{The lines are sorted by decreasing value of the
          median pseudo-abundance.}}}
    \label{tab:abundance}
  \end{table}}

\newcommand{\TabWyOverWx}{%
  \begin{table}
    \caption{Minimum, median, and maximum values of the ratios of diverse
      line integrated intensities.}
    \centering 
    \tiny{%
      \begin{tabular}{llllc} 
      \hline 
      \hline 
      Species \#1  & Trans.            & Species \#2  & Trans.            & $\W(\#1)/\W(\#2)$    \\ 
                  &                    &             &                    & min$-$med$-$max      \\ 
      \hline 
      \twCO{}     & \J{1}{0}           & \thCO{}     & \J{1}{0}           & $1.2-6.9-46$         \\ 
      \twCO{}     & \J{1}{0}           & \HCOp{}     & \J{1}{0}           & $1.6-37-140$         \\ 
      \twCO{}     & \J{1}{0}           & \CeiO{}     & \J{1}{0}           & $2.9-65-410$         \\ 
      \twCO{}     & \J{1}{0}           & \twCN{}     & \J{1}{0}           & $1.9-68-210$         \\ 
      \twCO{}     & \J{1}{0}           & \CCH{}      & \J{1}{0}           & $1.8-66-320$         \\ 
      \twCO{}     & \J{1}{0}           & \HNC{}      & \J{1}{0}           & $1.8-66-320$         \\ 
      \twCO{}     & \J{1}{0}           & \twCS{}     & \J{2}{1}           & $2.4-84-350$         \\ 
      \twCO{}     & \J{1}{0}           & \ttSO{}     & \T{2}{3}{1}{2}     & $2.7-110-400$        \\ 
      \hline 
      \thCO{}     & \J{1}{0}           & \HCOp{}     & \J{1}{0}           & $0.16-6-30$          \\ 
      \thCO{}     & \J{1}{0}           & \CCH{}      & \J{1}{0}           & $0.4-13-91$          \\ 
      \thCO{}     & \J{1}{0}           & \twCN{}     & \J{1}{0}           & $0.47-13-57$         \\ 
      \thCO{}     & \J{1}{0}           & \CeiO{}     & \J{1}{0}           & $0.57-15-50$         \\ 
      \thCO{}     & \J{1}{0}           & \HNC{}      & \J{1}{0}           & $0.4-13-91$          \\ 
      \thCO{}     & \J{1}{0}           & \twCS{}     & \J{2}{1}           & $0.71-17-70$         \\ 
      \thCO{}     & \J{1}{0}           & \ttSO{}     & \T{2}{3}{1}{2}     & $0.67-24-98$         \\ 
      \thCO{}     & \J{1}{0}           & \CseO{}     & \J{1}{0}           & $0.65-34-85$         \\ 
      \hline 
      \HNC{}      & \J{1}{0}           & \twCN{}     & \J{1}{0}           & $0.33-1.7-6.4$       \\ 
      \twCS{}     & \J{2}{1}           & \HNC{}      & \J{1}{0}           & $0.22-1.1-7.3$       \\ 
      \twCN{}     & \J{1}{0}           & \HNC{}      & \J{1}{0}           & $0.22-1.1-3.9$       \\ 
      \CeiO{}     & \J{1}{0}           & \twCS{}     & \J{2}{1}           & $0.14-1.2-6.2$       \\ 
      \CeiO{}     & \J{1}{0}           & \HNC{}      & \J{1}{0}           & $0.16-1.3-8.5$       \\ 
      \HCN{}      & \J{1}{0}           & \twCN{}     & \J{1}{0}           & $0.33-1.7-6.4$       \\ 
      \HCOp{}     & \J{1}{0}           & \HCN{}      & \J{1}{0}           & $0.33-1.7-6.4$       \\ 
      \HCN{}      & \J{1}{0}           & \HNC{}      & \J{1}{0}           & $0.33-1.7-6.4$       \\ 
      \HCOp{}     & \J{1}{0}           & \HNC{}      & \J{1}{0}           & $0.33-1.7-6.4$       \\ 
      \CeiO{}     & \J{1}{0}           & \CseO{}     & \J{1}{0}           & $0.43-4.3-8.6$       \\ 
      \HCOp{}     & \J{1}{0}           & \NNHp{}     & \J{1}{0}           & $0.33-1.7-6.4$       \\ 
      \hline 
      \end{tabular}
      \tablefoot{%
        \tablefoottext{a}{In each group, the lines are sorted by increasing 
          value of the $\W(\#1)/\W(\#2)$ median.}}}
    \label{tab:ratio:wy:over:wx}
  \end{table}}

\newcommand{\TabFillingFactor}{%
  \begin{table}
    \caption{Critical density, and percentage of total flux originating
      from gas in filaments and dense cores for each line.}
    \centering{} %
    \tiny{%
      \begin{tabular}{llccc} 
      \hline 
      \hline 
      Species     & Transition         & $\ncrit{}$  & $F_{6\le\Av<15}$    & $F_{15\le\Av}$      \\ 
                  &                    & \pccm{}     & \% of $F_\emr{tot}$ & \% of $F_\emr{tot}$ \\ 
      \hline 
      \twCO{}     & \J{1}{0}           & $2\e{3}$    & $38$                & $7.6$               \\ 
      \thCO{}     & \J{1}{0}           & $2\e{3}$    & $45$                & $15$                \\ 
      \HCOp{}     & \J{1}{0}           & $2\e{5}$    & $40$                & $16$                \\ 
      \CCH{}      & \J{1}{0}           & $1\e{5}$    & $37$                & $17$                \\ 
      \cCCCHH{}   & \T{2}{1,2}{1}{0,1} & $1\e{6}$    & $38$                & $17$                \\ 
      \HCN{}      & \J{1}{0}           & $3\e{6}$    & $44$                & $18$                \\ 
      \twCN{}     & \J{1}{0}           & $3\e{5}$    & $45$                & $19$                \\ 
      \CseO{}     & \J{1}{0}           & $2\e{3}$    & $43$                & $26$                \\ 
      \HNC{}      & \J{1}{0}           & $4\e{5}$    & $41$                & $27$                \\ 
      \CeiO{}     & \J{1}{0}           & $2\e{3}$    & $48$                & $29$                \\ 
      \ttSO{}     & \T{2}{3}{1}{2}     & $2\e{5}$    & $44$                & $31$                \\ 
      \twCS{}     & \J{2}{1}           & $2\e{5}$    & $42$                & $32$                \\ 
      \methanol{} & \J{2}{1}           & $3\e{4}$    & $41$                & $48$                \\ 
      \HthCOp{}   & \J{1}{0}           & $2\e{5}$    & $34$                & $56$                \\ 
      \NNHp{}     & \J{1}{0}           & $2\e{5}$    & $17$                & $88$                \\ 
      \hline 
      \end{tabular}
      \tablefoot{%
        \tablefoottext{a}{The lines are sorted by increasing value of
          $F_{15\le\Av}/F_\emr{tot}$ (last column).}
        \tablefoottext{b}{The typical volume density of the regions 
          with $6\le\Av<15$, and $15\le\Av$ are 1\,500, and 
          7\,300\,\HH\pccm, respectively.}}}
    \label{tab:filling:factor}
  \end{table}}

\newcommand{\TabEnvironment}{%
  \begin{table*}
    \caption{Visual extinction, dust temperature, and far UV illumination
      for three different regions of 1 square-degree area in Orion\,B.}
    \centering{} %
    \tiny{%
      \begin{tabular}{lccc} 
      \hline 
      \hline 
      Environment    & $\Av[min]-\Av[5\%]-\Av[med]-\Av[95\%]-\Av[max]$  & $\Td[min]-\Td[5\%]-\Td[med]-\Td[95\%]-\Td[max]$  & $\Gz[min]-\Gz[5\%]-\Gz[med]-\Gz[95\%]-\Gz[max]$  \\ 
                     & \magn{}                                          & \K{}                                             & ISRF~\citep{habing68}                            \\ 
      \hline 
      UV illuminated & $0.76-1-3.3-12-230$           & $16.4-20.1-24.1-40.5-101$     & $4.4-12-30-400-38000$         \\ 
      Translucent    & $1.1-1.5-2.4-3.5-7.7$         & $16-17.2-17.9-18.6-20.3$      & $3.9-5.6-6.9-8.1-13$          \\ 
      UV shieded     & $1.5-2.2-3.4-7.9-34$          & $12.9-15.2-17.1-18.5-20.3$    & $1.3-3-5.3-8-13$              \\ 
      \hline 
      \end{tabular}}
    \label{tab:dust:env}
  \end{table*}}

\newcommand{\TabRatioComparison}{%
  \begin{table*}
    \caption{Flux ratios in nearby galaxies and in Orion B.}
    \tiny{
      \begin{tabular}{lcccccccll}
        \hline
        \hline
        Source     & HCN/\HCOp{} & \twCO/HCN  & \twCO/\HCOp{} & \twCO/\NNHp{} & \twCN/HCN & \CCH/\thCO{} & HCN/HNC   & Observ.   & Ref. \\
        \hline
        ULIRGs     & $1.5-2.4$   & $6.3-8.6$  & $9.7-21$     & $40$          & $0.4-0.5$ & $0.8-1.4$    & $1.4-3.0$ & MOPRA-22m & 1 \\
        M51 P2     & $1.2$       & $32$       & $39$         & $225$         & $0.5$     & $0.04$       & $2.7$     & IRAM-30m  & 2 \\
        AGNs       & $0.8-2.0$   & $3.4-25$   & $5.7-20$     & $41-68$       & $0.9-1.6$ & $0.2-0.7$    & $2.2-3.0$ & IRAM-30m  & 3 \\
        Starbursts & $0.7-1.2$   & $16-23$    & $16-24$      & $100-325$     & $1.2-1.4$ & $0.2-0.7$    & $1.9-2.5$ & IRAM-30m  & 3 \\
        M82        & $0.5$       & -          & -            & -             & -         & -            & -         & CARMA     & 4 \\
        NGC253 P7  & $1.2$       & $6.0$      & $7.0$        & -             & $0.6$     & -            & -         & ALMA      & 5 \\
        Maffei2B   & $0.6$       & -          &  -           & -             & -         & -            & $4.3$     & BIMA,OVRO & 6 \\
        LMC        & $0.4-0.7$   & $11-120$   & $6.0-48$     & $59-167$      & $0.2-0.3$ & $0.03-0.27$  & $1.1-3.4$ & IRAM-30m  & 7 \\
        OrionB     & $0.9$       & $39$       & $37$         & $900$         & $0.5$     & $0.05$       & $3.5$     & IRAM-30m  & 8\\
        \hline
      \end{tabular}
      \tablefoot{%
        \tablefoottext{1}{\citet{nishimura16}.}
        \tablefoottext{2}{\citet{watanabe14}.}
        \tablefoottext{3}{\citet{aladro15}.}
        \tablefoottext{4}{\citet{salas14}.}
        \tablefoottext{5}{\citet{meier15}.}
        \tablefoottext{6}{\citet{meier12}.}
        \tablefoottext{7}{\citet{nishimura16}.}
        \tablefoottext{8}{This work.}}}
    \label{tab:ratio:comp}
  \end{table*}}

\newcommand{\FigRGB}{%
  \begin{figure*}
    \centering %
    \includegraphics[width=\linewidth]{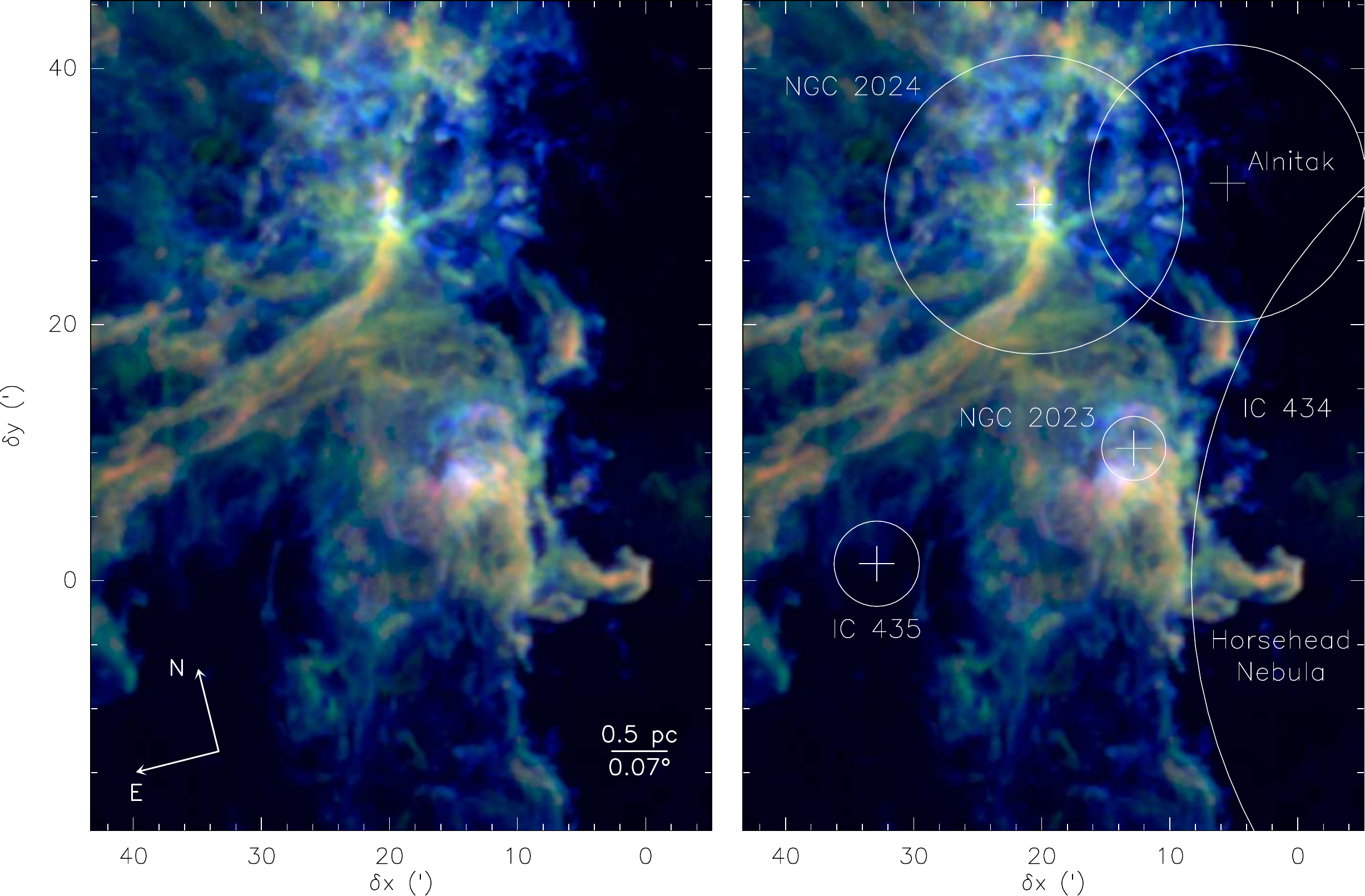}
    \caption{Composite image of the \twCO{} (blue), \thCO{} (green),
      \CeiO{} (red) \Jone{} peak-intensity main-beam temperature. The
      circles show the typical extensions of the \Hii{} regions and the
      crosses show the position of the associated exciting stars (see
      Table~\ref{tab:stars}). The $\sigma$-Ori star that excites the
      IC\,434 \Hii{} region is located 0.5\deg{} East from the Horsehead
      nebula.}
    \label{fig:rgb}
  \end{figure*}}

\newcommand{\FigAreaIma}{%
  \begin{figure*}
    \centering %
    \includegraphics[width=\linewidth]{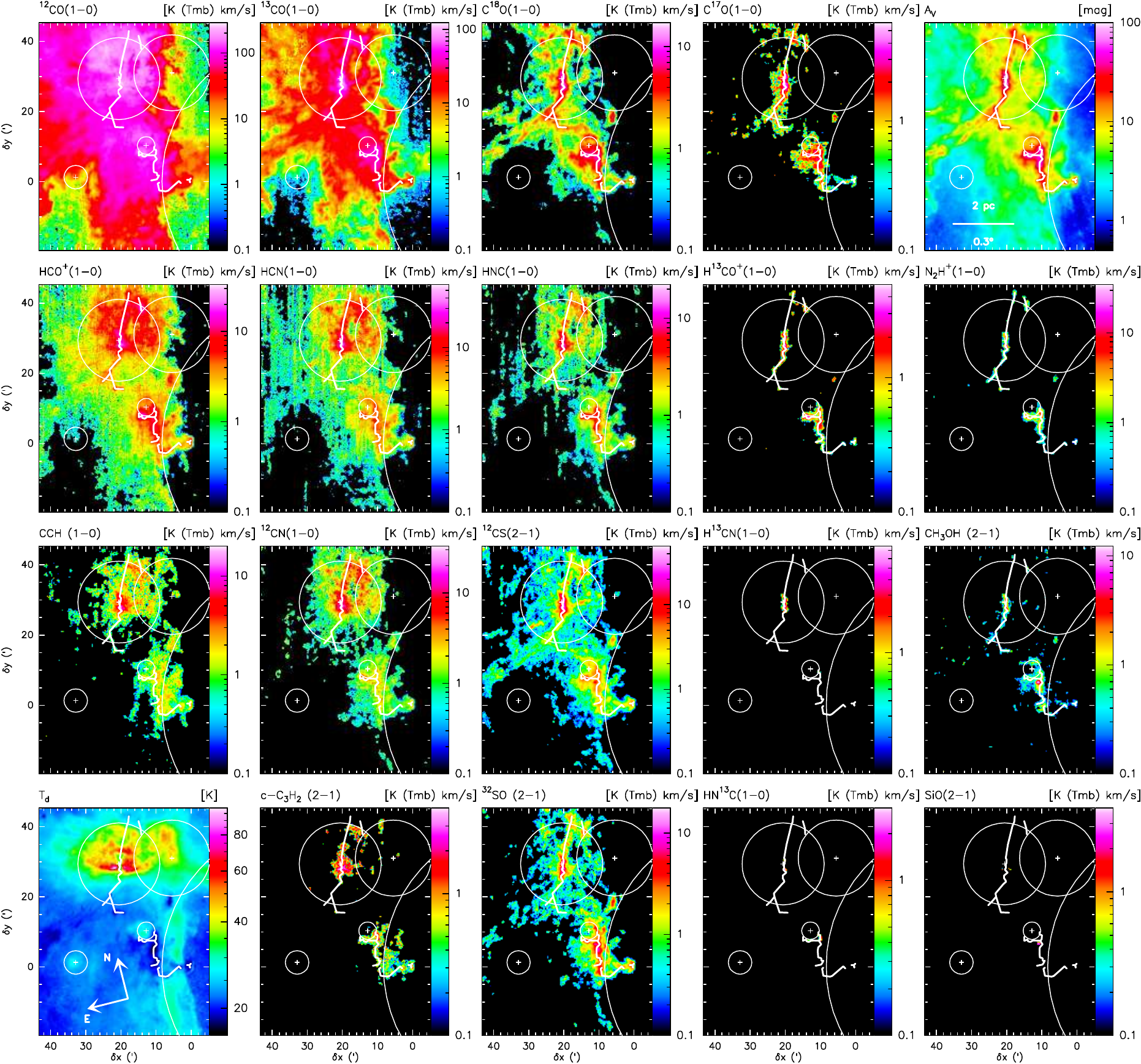}
    \caption{Spatial distribution of the line integrated intensity for some
      of the detected lines in the 3mm band, plus the dust temperature
      (bottom left panel) and the visual extinction (top right corner).
      Continuum data comes from the publicly available SED fit done
      by~\cite{lombardi14a} on the Herschel Gould Belt Survey data (PI:
      P. André). The color scales are logarithmic to reveal the
      distribution of faint signal. Pixels with a signal-to-noise lower
      than 4 were blanked out.  In addition to the circles and crosses that
      show the approximate boundaries of the \Hii{} regions and the
      associated exciting stars, we overlaid broken lines that were
      somewhat arbitrarily drawn by connecting the \NNHp{} \Jone{}
      emission.}
    \label{fig:area}
  \end{figure*}}

\newcommand{\FigMeanSpectra}{%
  \begin{figure}
    \centering %
    \includegraphics[width=\linewidth]{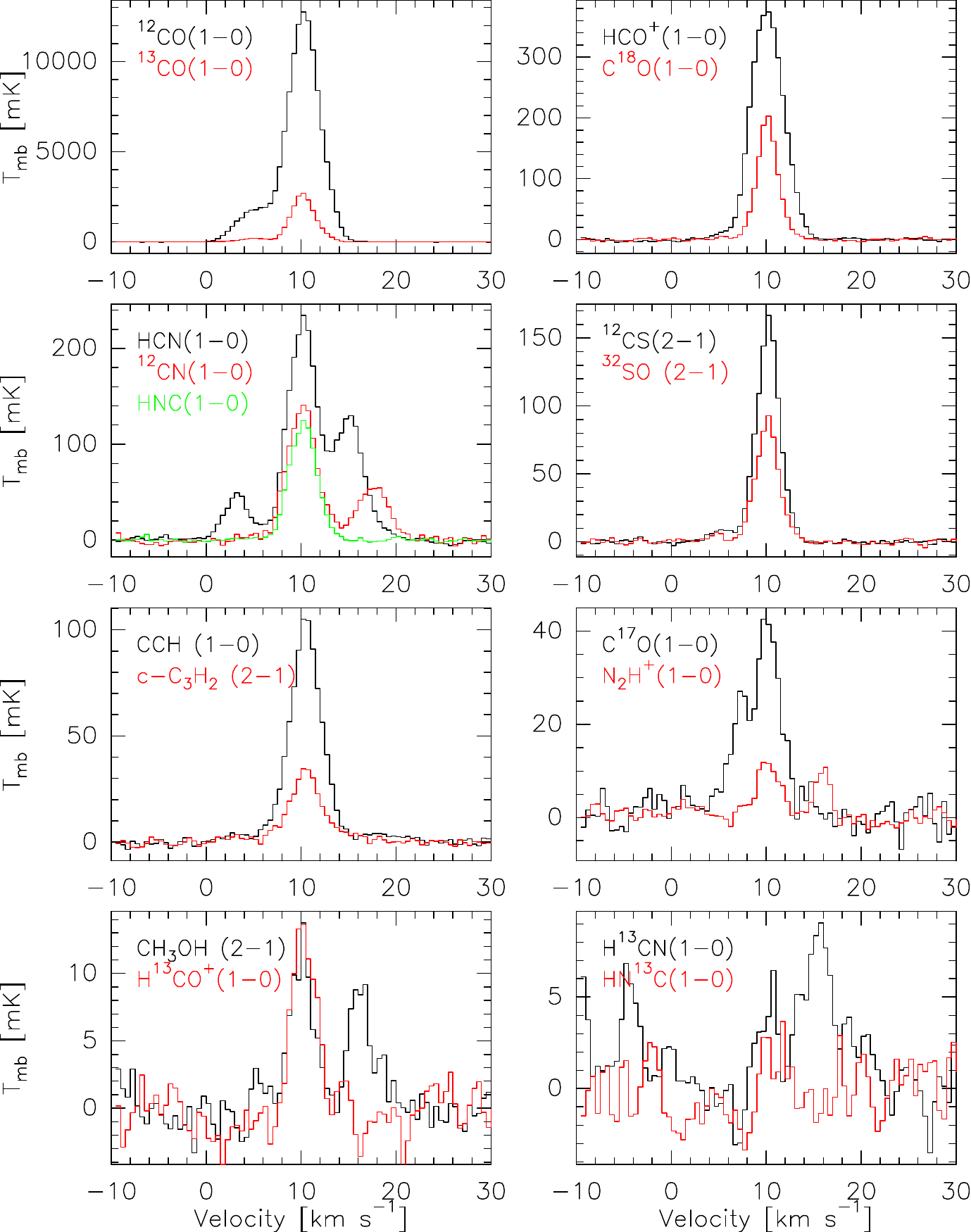}
    \caption{Spectra averaged over the mapped field of view.}
    \label{fig:mean:spec}
  \end{figure}}

\newcommand{\FigTwoComponents}{%
  \begin{figure}
    \centering %
    \includegraphics[width=\linewidth]{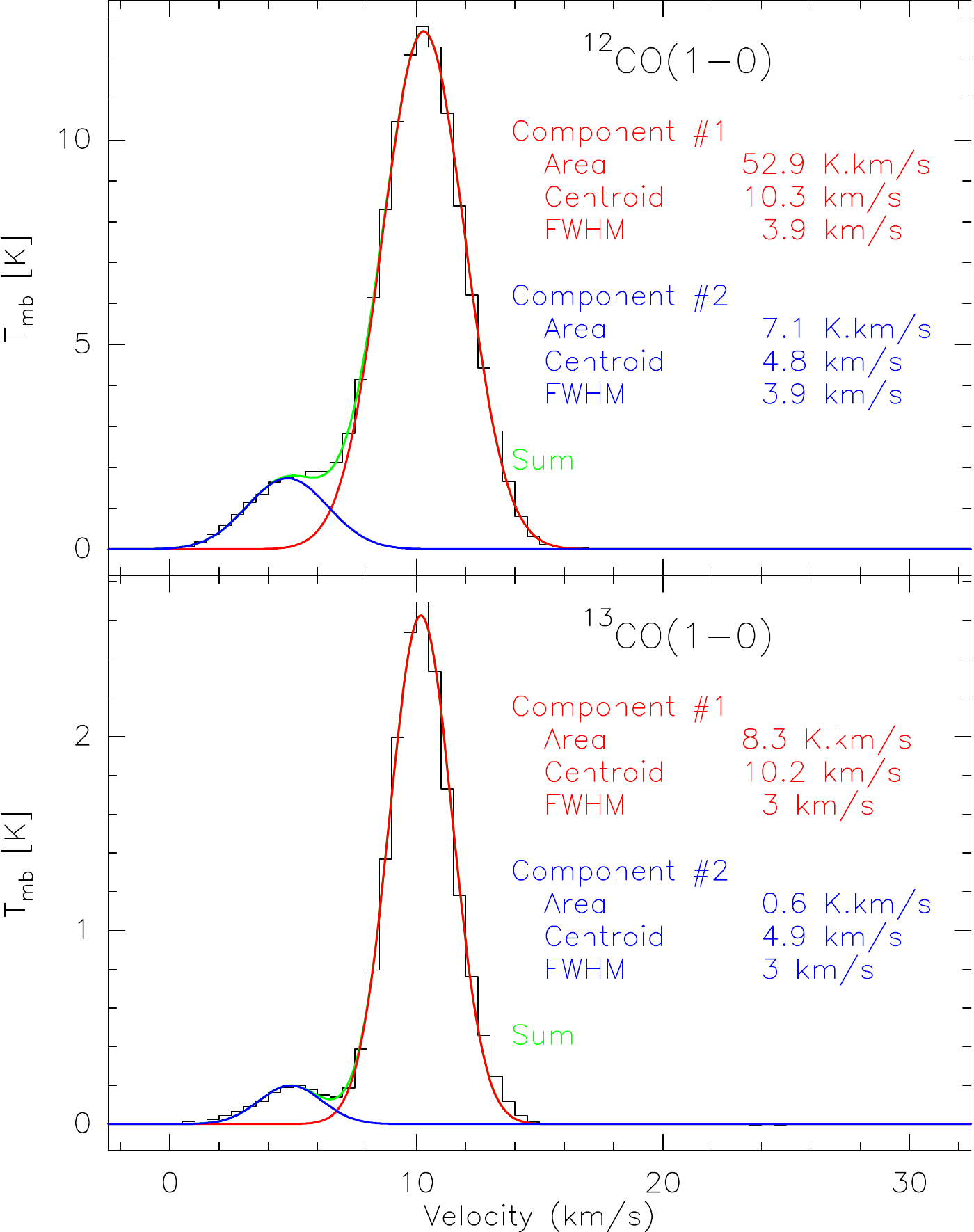}
    \caption{Gaussian fits of the two main velocity components that appears
      in the western edge of Orion\,B.}
    \label{fig:two:components}
  \end{figure}}

\newcommand{\FigHalo}{%
  \begin{figure}
    \centering %
    \includegraphics[width=\linewidth]{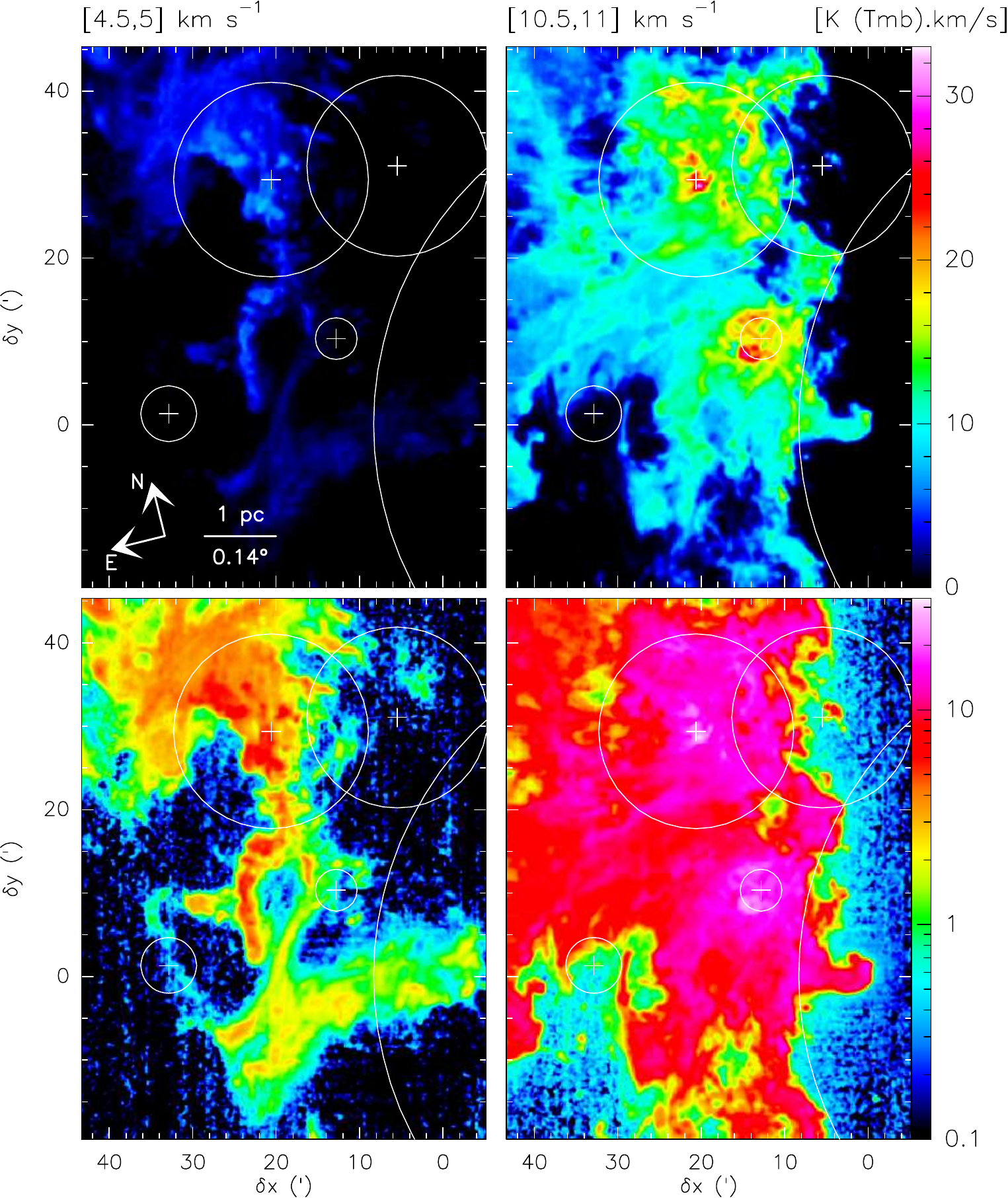}
    \caption{Spatial distribution of the \twCO{} \Jone{} emission
      integrated over two different 0.5\kms{} velocity ranges in linear
      (top row) and logarithmic (bottom row) color scales.}
    \label{fig:halo}
  \end{figure}}

\newcommand{\FigMaskedSpectraFluxAv}{%
  \begin{figure*}
    \centering %
    \includegraphics[width=0.21\linewidth]{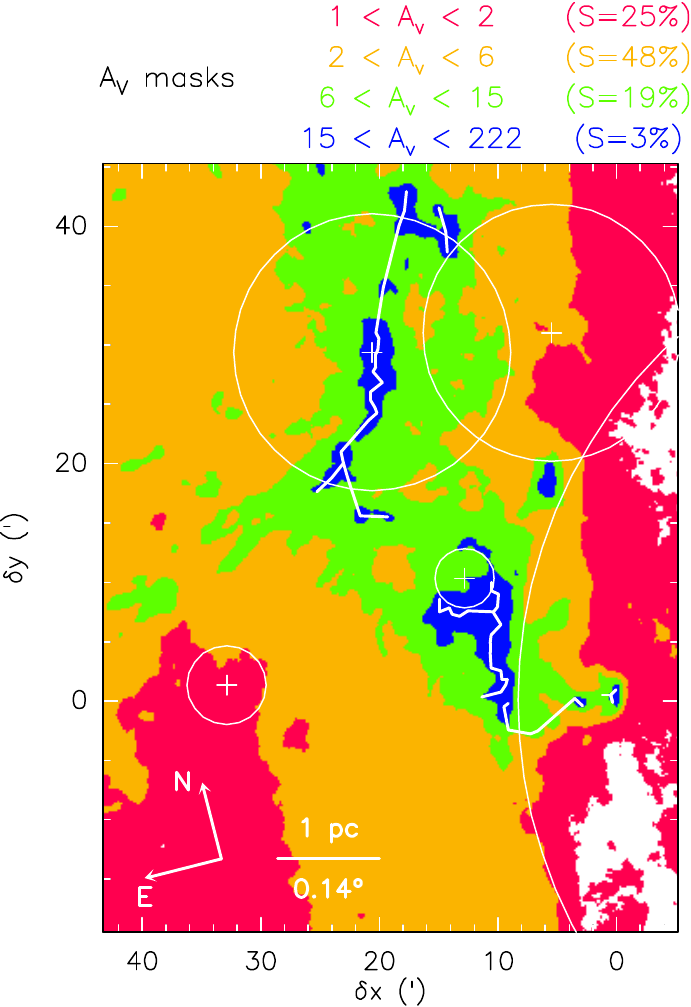}
    \hfill
    \includegraphics[width=0.75\linewidth]{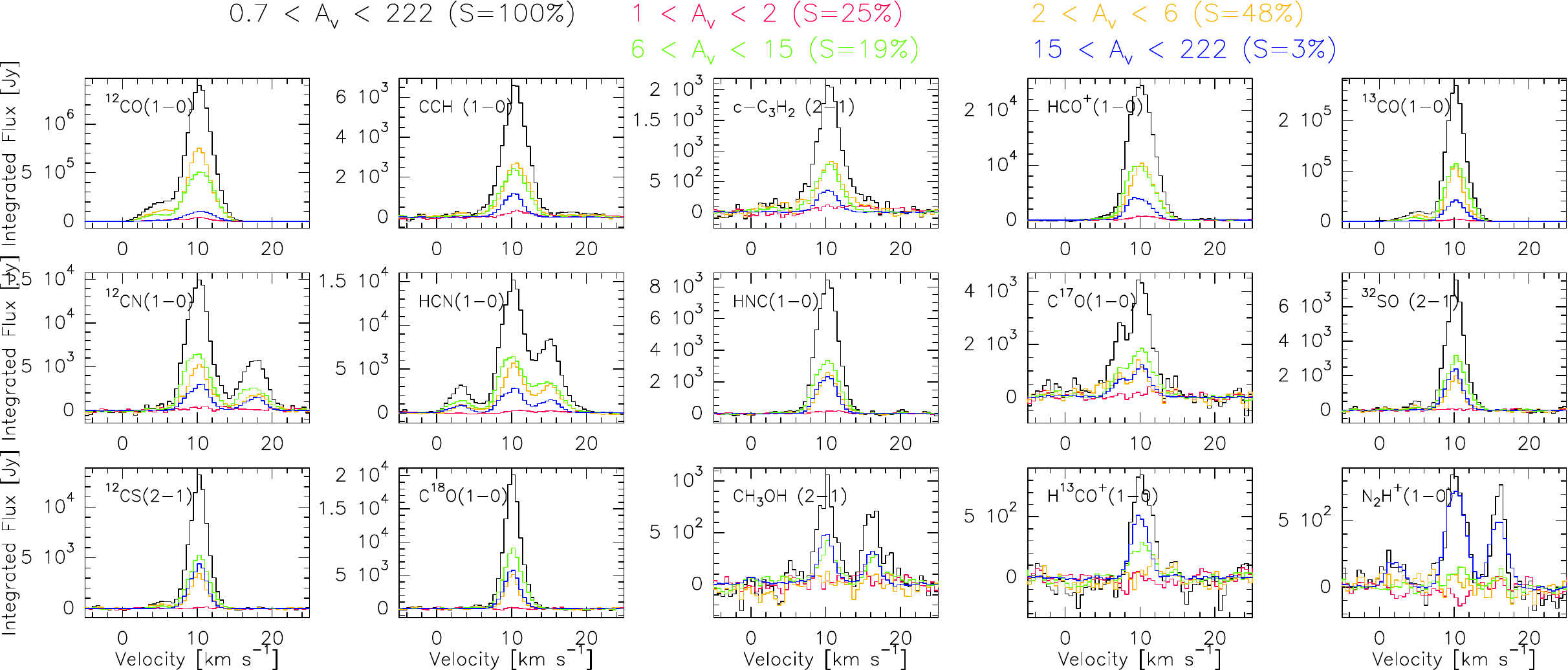}
    \caption{\textbf{Left:} Spatial distribution of the four following
      masks: $1 \le \Av < 2$ in red, $2 \le \Av < 6$ in orange, $6 \le \Av
      < 15$ in green, and $15\le \Av < 222$ in blue. The percentages in the
      legend list the fraction of the surface contained in the different
      masks. \textbf{Right:} Flux integrated over the masks as a function
      of velocity. The spectra of different color show the evolution of the
      flux in each line as a function of the mask used: All pixels observed
      in black, all pixels with $1 \le \Av < 2$ in red, $2 \le \Av < 6$ in
      orange, all pixels with $6 \le \Av < 15$ in green, and all pixels
      with $15\le \Av < 222$ in blue.}
    \label{fig:spec:mask:flux:Av}
  \end{figure*}}

\newcommand{\FigLineFluxAv}{%
  \begin{figure}
    \centering %
    \includegraphics[width=0.8\linewidth]{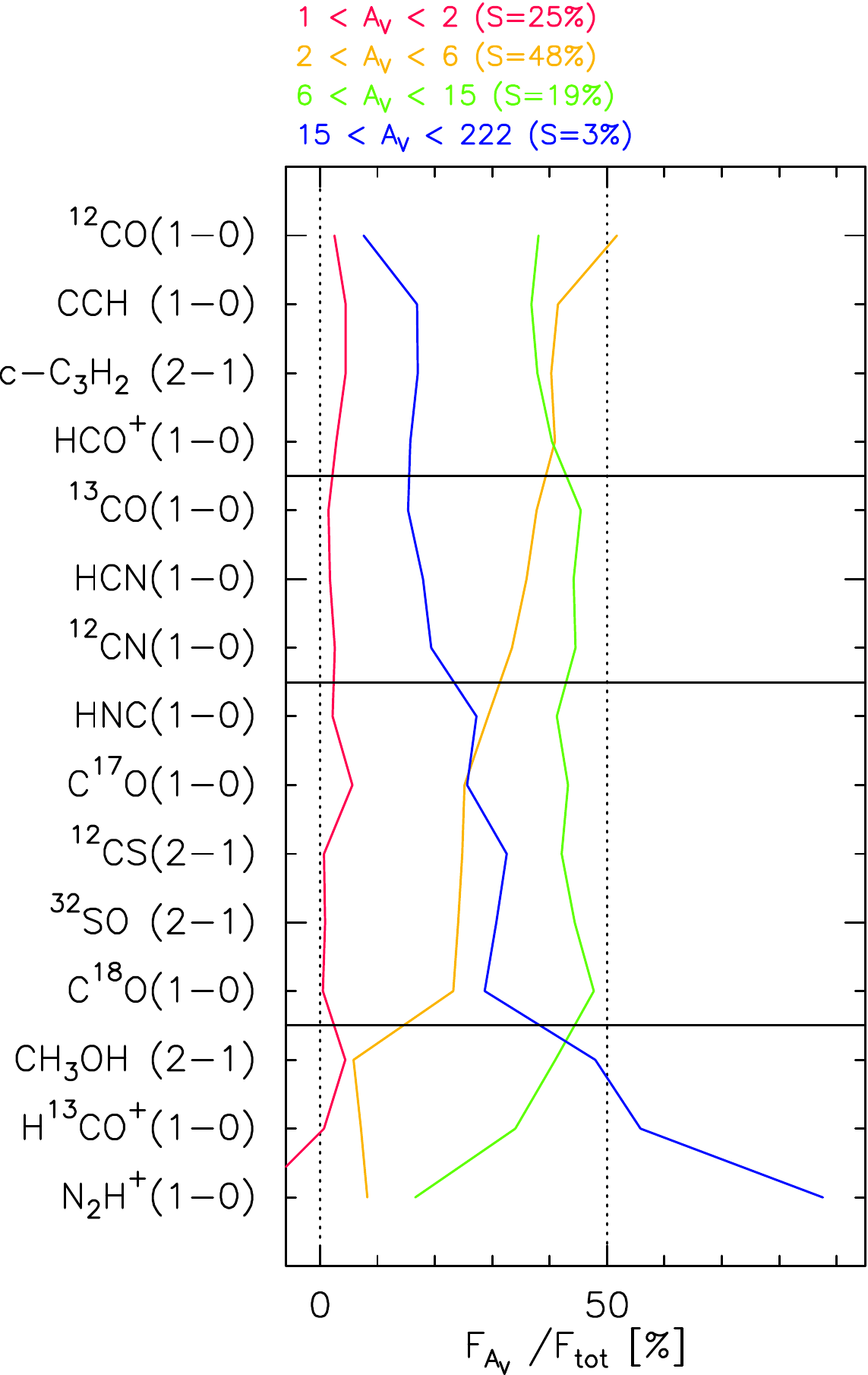}
    \caption{For each line, flux integrated over each of the four \Av{}
      mask divided by the flux computed over the observed field of
      view. All fluxes are computed between 9 and 12\kms{}. The black
      horizontal lines define the groups of lines described in
      Section~\ref{sec:flux:av}.}
    \label{fig:line:F:Av}
  \end{figure}}

\newcommand{\FigMaskedSpectraFluxTdust}{%
  \begin{figure*}
    \centering %
    \includegraphics[width=0.21\linewidth]{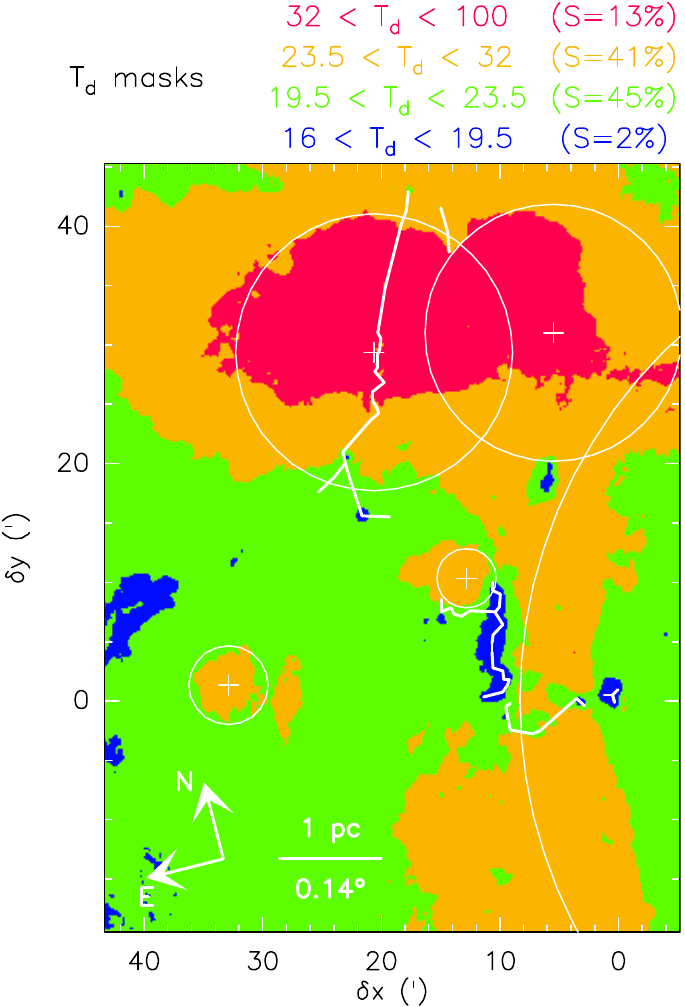}
    \hfill
    \includegraphics[width=0.75\linewidth]{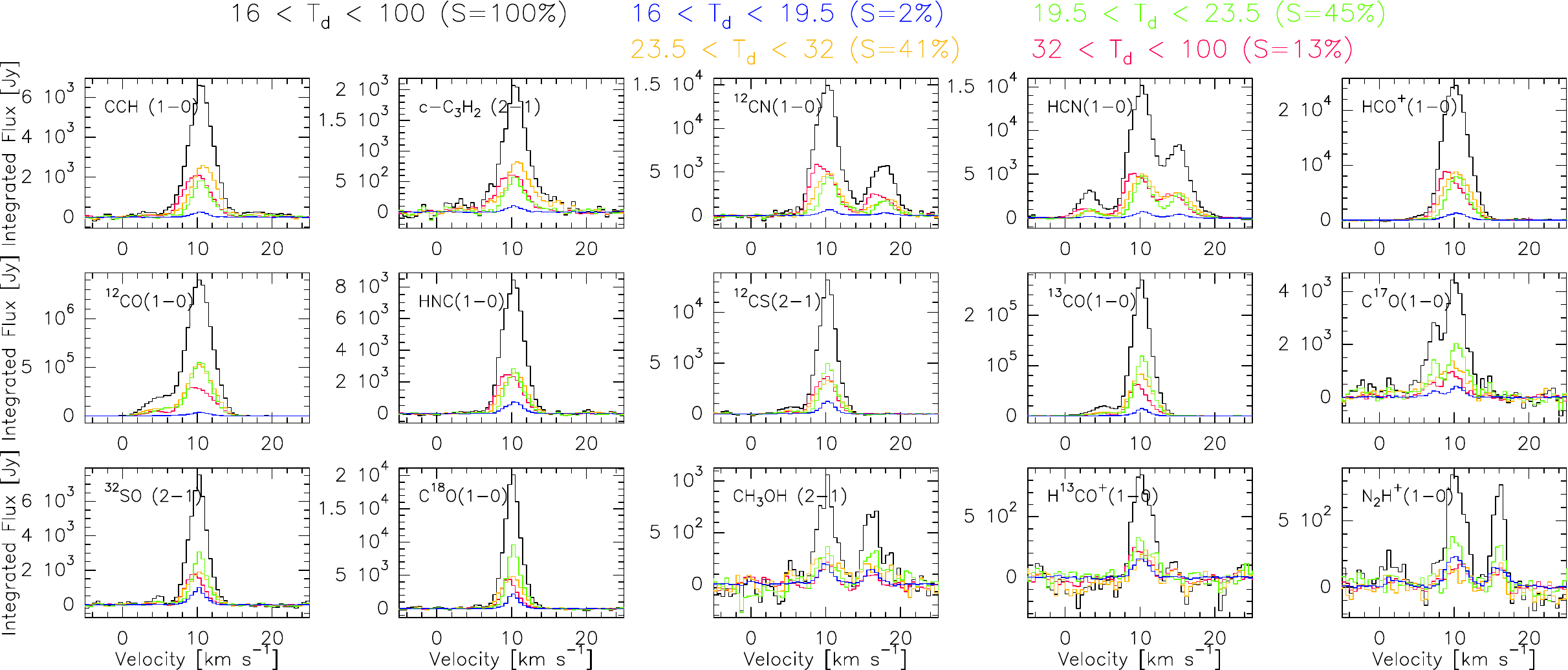}
    \caption{\textbf{Left:} Spatial distribution of the four following
      masks: $16 \le \Td < 19.5\K$ in blue, $19.5 \le \Td < 23.5\K$ in
      green, $23.5 \le \Td < 32\K$ in orange, and $32 \le \Td < 100\K$ in
      red. The percentages in the legend list the fraction of the surface
      contained in the different masks. \textbf{Right:} Flux integrated
      over the masks as a function of velocity. The spectrum of different
      colors the evolution of the flux for the same line as a function of
      the used mask: All pixels observed in black, all pixels with $16 \le
      \Td < 19.5\K$ in blue, all pixels with $19.5 \le \Td < 23.5\K$ in
      green, all pixels with $23.5 \le \Td < 32\K$ in orange, and all
      pixels with $32 \le \Td < 100\K$ in red.}
    \label{fig:spec:mask:flux:tdust}
  \end{figure*}}

\newcommand{\FigLineFluxTdust}{%
  \begin{figure}
    \centering %
    \includegraphics[width=0.8\linewidth]{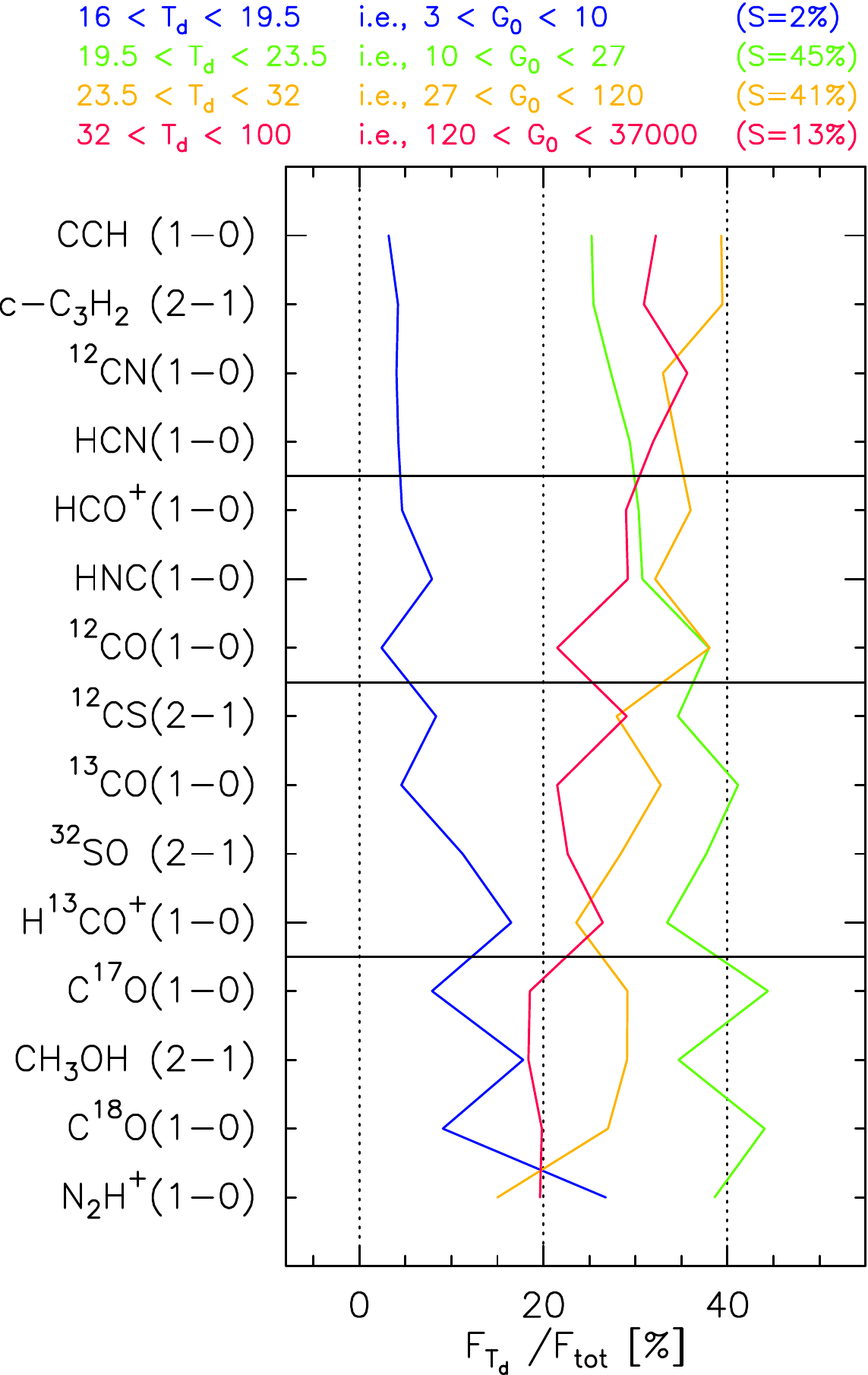}
    \caption{For each line, flux integrated over each of the four \Td{}
      masks divided by the flux integrated over the observed field of
      view. All fluxes are computed between 9 and 12\kms{}. The black
      horizontal lines define the groups of lines described in
      Section~\ref{sec:flux:tdust}.}
    \label{fig:line:F:Tdust}
  \end{figure}}

\newcommand{\FigAvVsArea}{%
  \begin{figure*}
    \centering %
    \includegraphics[width=\linewidth]{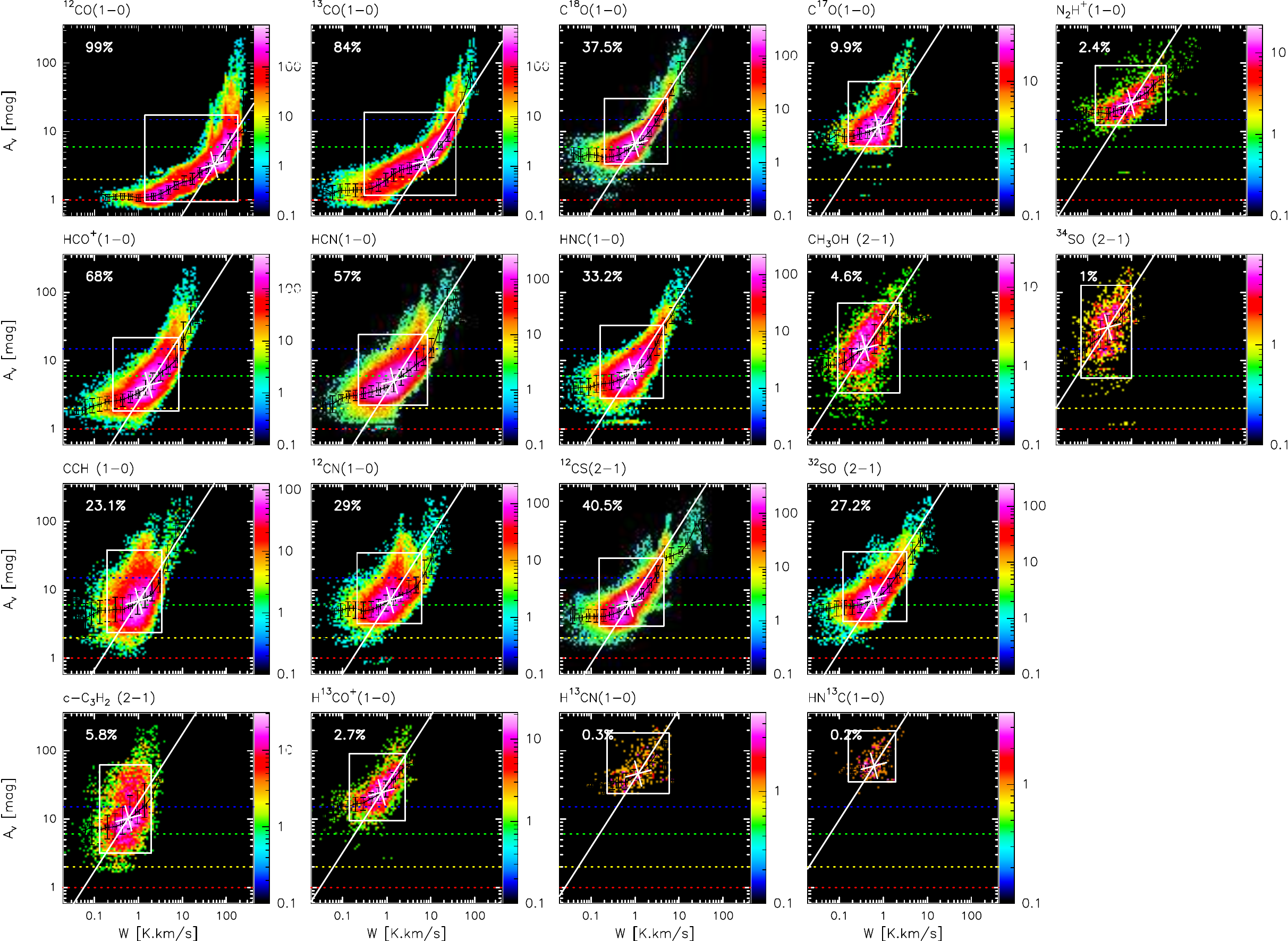}
    \caption{Joint distributions of the visual extinction as a function of
      the line integrated intensity for a selection of the detected
      lines. The percentage in the top left corner indicates the surface
      over which the joint distributions can be reliably computed. These
      distributions contains both the global trend for the bulk of the gas
      and extreme behavior at low and high visual extinctions. The number
      of sightlines falling in a given 2D bin of the distribution is
      color-coded using a logarithmic scale to emphasize the extreme
      behavior (in particular the dense cores) that occupy a small
      fraction of the observed field of view. In contrast, the white
      rectangle displays the region of the distribution where 90\% of the
      points are located: 2.5\% of the points are outside this rectangle on
      each side. This allows us to define more robust global trends for the
      bulk of the gas. The white point shows the median of the two
      marginalized distributions.  A line of unit slope, \ie, a linear
      relationship between visual extinction and the line integrated
      intensities, is overlaid as the white plain line going through the
      white cross. The black points show the median values of all data
      points falling in an interval regularly sampled of the logarithm of
      the line integrated intensity. The black error bars show the range of
      values where 50\% of the points in the current bin are located. This
      allows us to ask whether molecular lines are a good tracers of the
      visual extinction.  All these parameters are listed in
      Table~\ref{tab:corr:area:av}. The red, orange, green, and blue
      horizontal dashed lines show the visual extinction limits (1, 2, 6,
      and 15 magnitudes, respectively) used in the masks of
      Fig.~\ref{fig:spec:mask:flux:Av}. This enables us to visualize the
      amount of well detected pixels that falls in each of the masks for
      each line.}
    \label{fig:av:vs:area}
  \end{figure*}}

\newcommand{\FigAreaOverAv}{%
  \begin{figure*}[p]
    \centering %
    \includegraphics[width=0.825\linewidth]{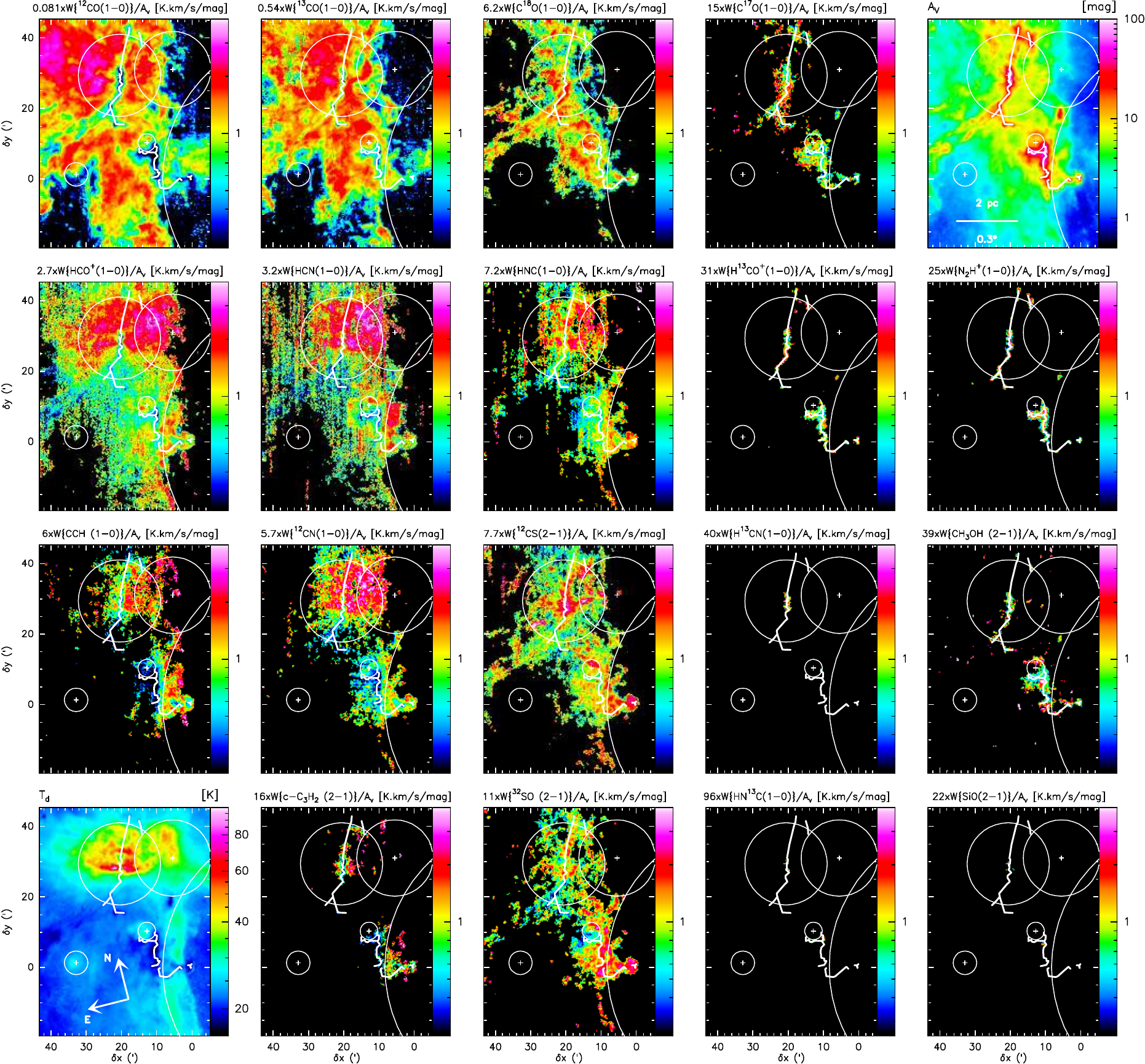}
    \caption{Spatial distribution of the ratio of the line integrated
      intensity to the visual extinction for some of the detected lines in
      the 3mm band, plus the dust temperature (bottom left panel) and the
      visual extinction (top right corner). The ratios are normalized by
      their median value that hence appears as 1 on the color look-up
      table. The color scale shows ratio values between 0.25 and 4 times
      the median value for all the ratio panels.}
    \label{fig:area:over:av}
  \end{figure*}}

\newcommand{\FigCorrTwCO}{%
  \begin{figure*}
    \centering %
    \includegraphics[width=0.9\linewidth]{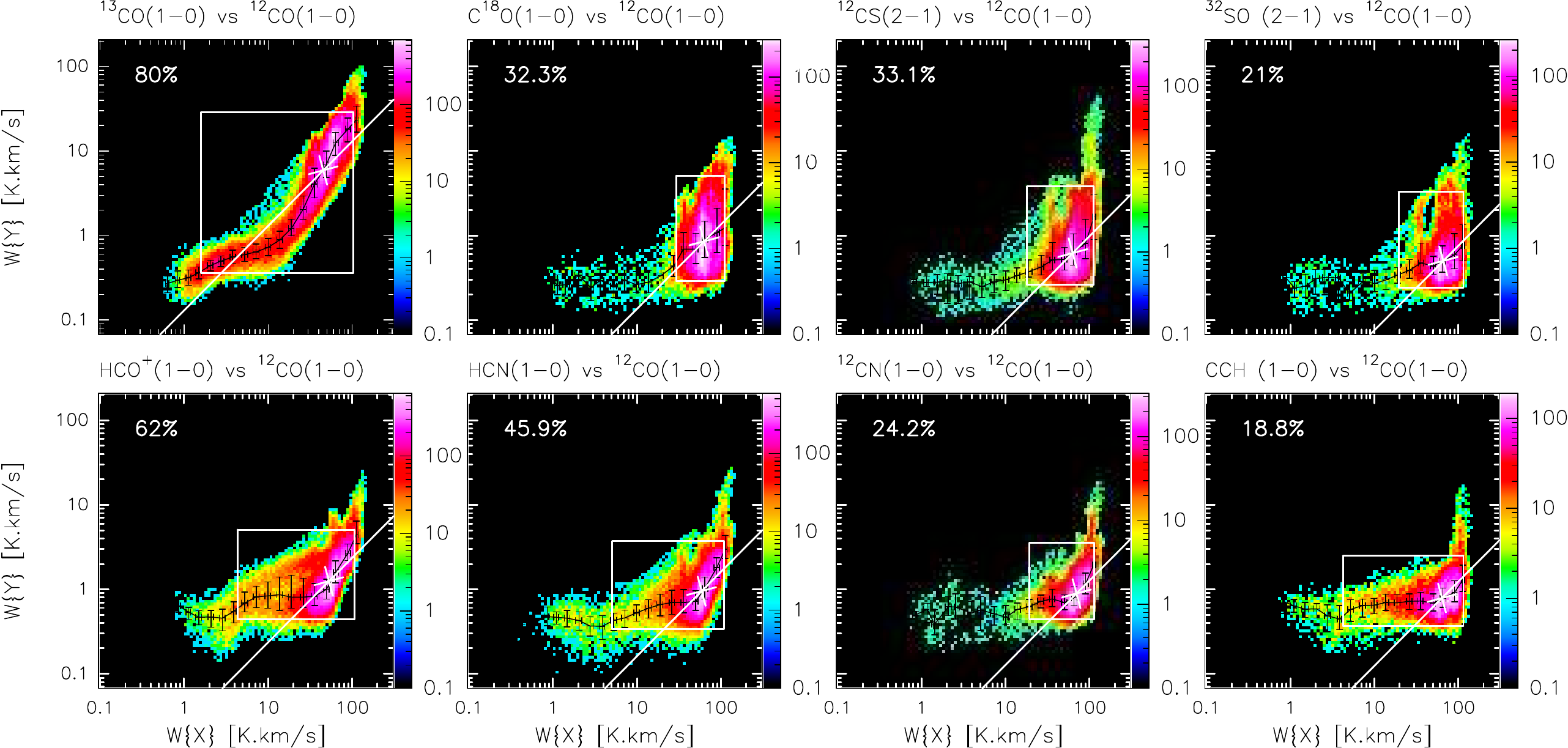}
    \caption{Joint distributions of various line intensities integrated
      over $[9,12\kms]$. Markers have the same signification as in
      Fig.~\ref{fig:av:vs:area}. The $x$-axis is always \W\cbrace{\twCO
        \Jone}.}
    \label{fig:area:vs:12co10}
  \end{figure*}}

\newcommand{\FigCorrThCO}{%
  \begin{figure*}
    \centering %
    \includegraphics[width=0.9\linewidth]{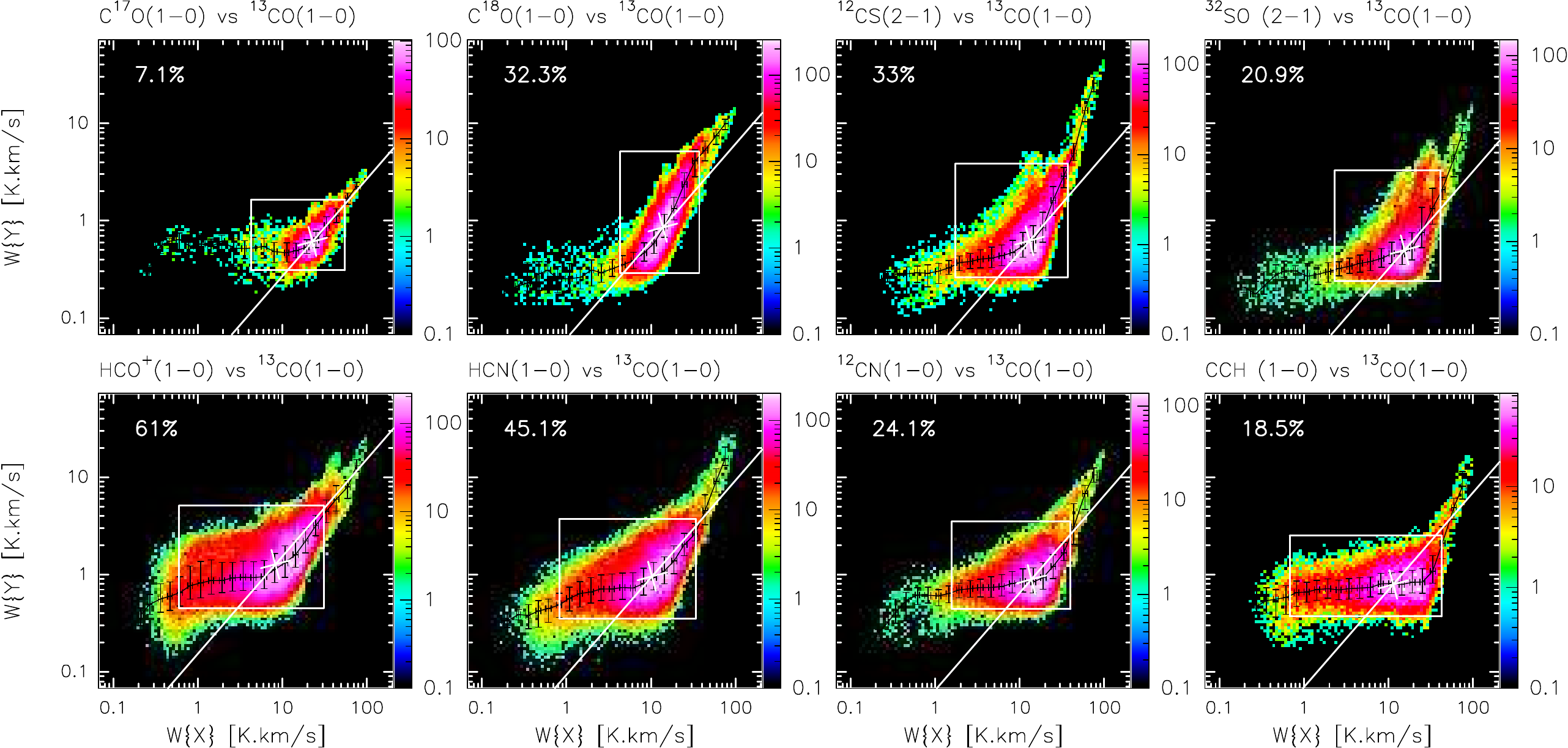}
    \caption{Same as Fig.~\ref{fig:area:vs:12co10}, except that the
    $x$-axis is now \W\cbrace{\thCO \Jone}.}
    \label{fig:area:vs:13co10}
  \end{figure*}}

\newcommand{\FigRatioTwCO}{%
  \begin{figure*}[p]
    \centering %
    \includegraphics[width=0.9\linewidth]{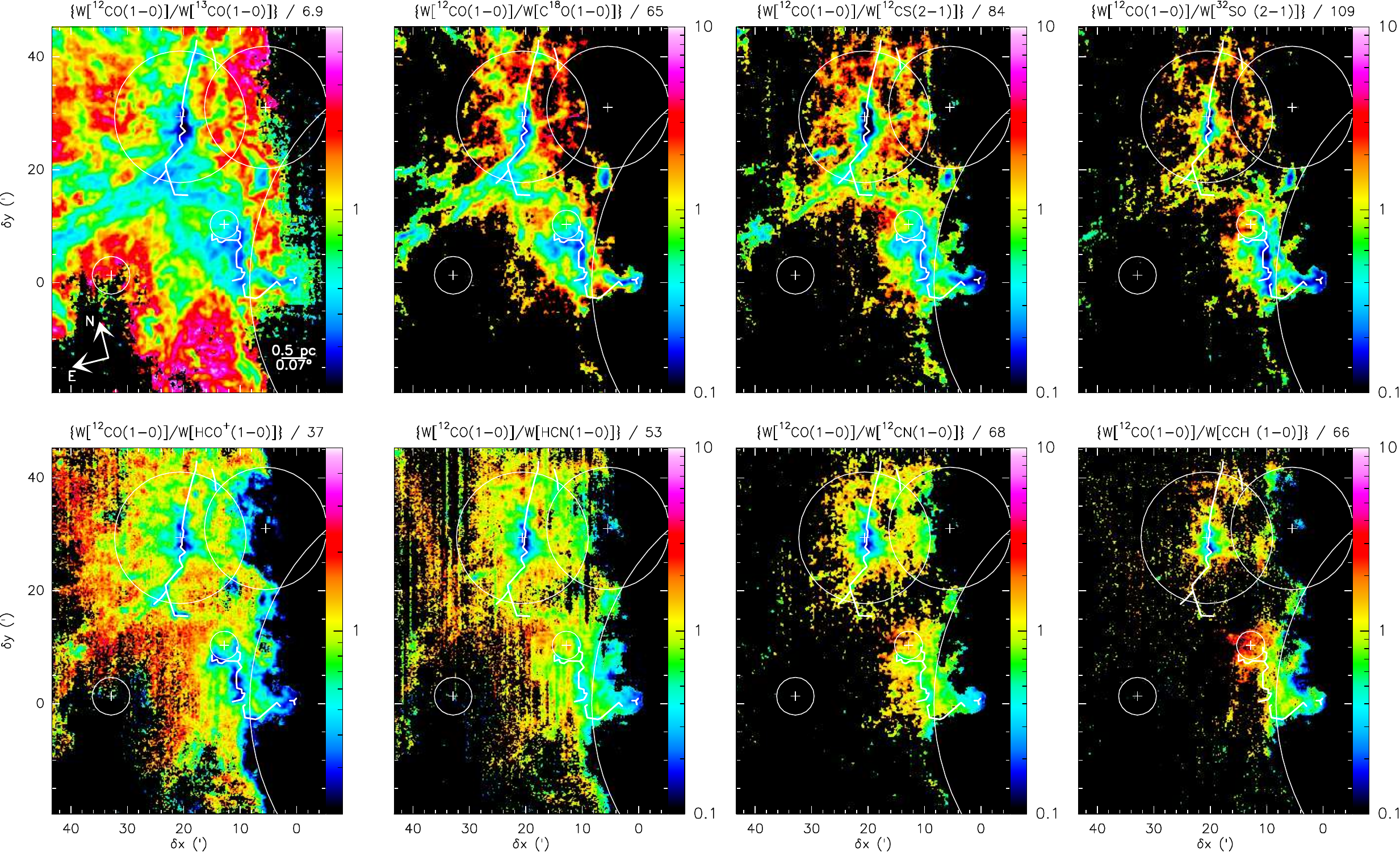}
    \caption{Spatial distribution of the ratios of line intensity
      integrated over $[9,12\kms]$. The ratios are normalized by the median
      value of the ratio. The numerator is always \W\cbrace{\twCO
        \Jone}. The color scale shows ratio values between 0.1 and 10 for
      all the ratio panels, except the \twCO/\thCO{} and \twCO/\HCOp{}
      panels where the color scale goes from 0.2 to 5.}
    \label{fig:area:over:12co10}
  \end{figure*}}

\newcommand{\FigRatioThCO}{%
  \begin{figure*}[p]
    \centering %
    \includegraphics[width=0.9\linewidth]{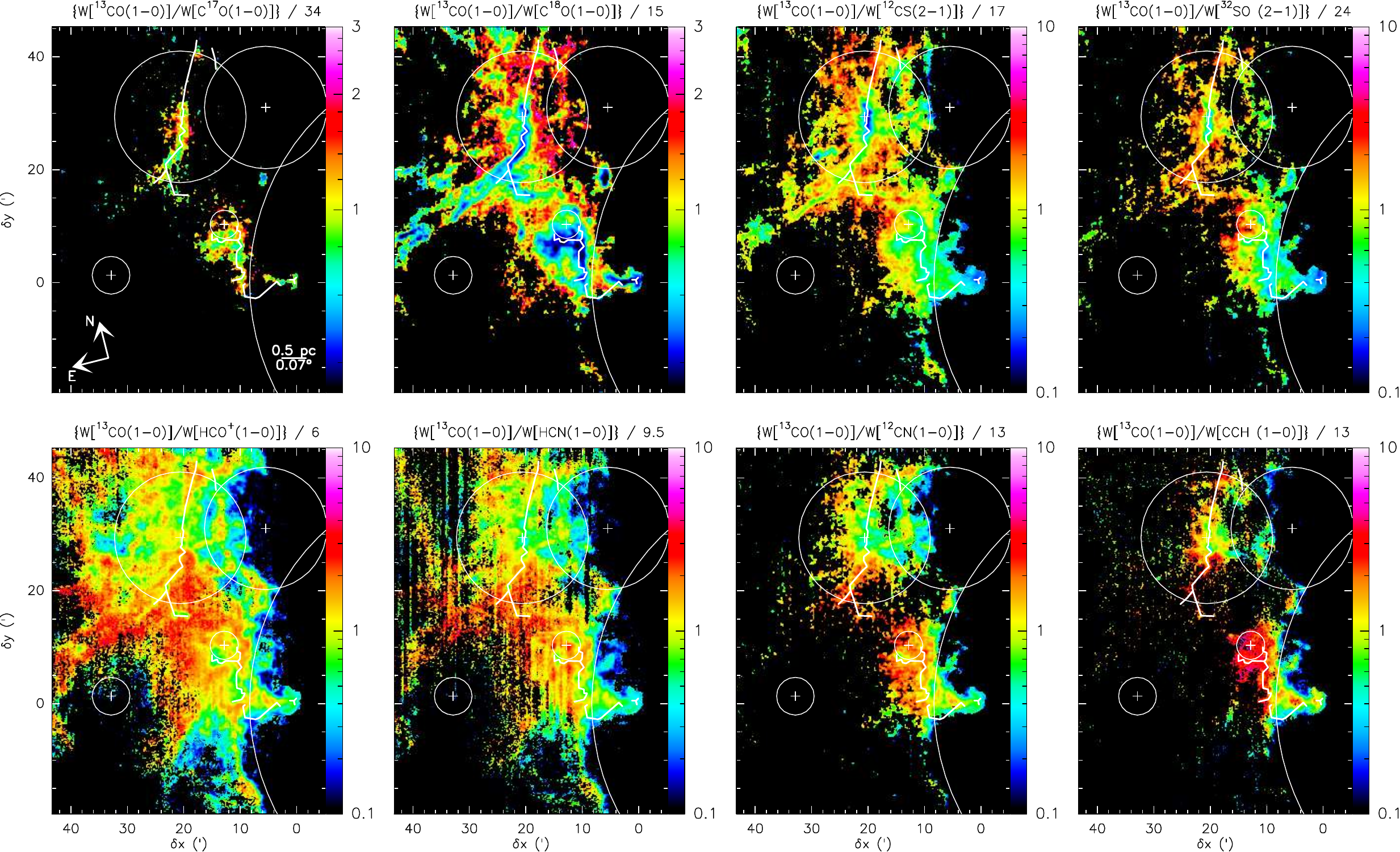}
    \caption{Same as Fig.~\ref{fig:area:over:12co10}, except that the
      numerator is now \W\cbrace{\thCO \Jone}. The color scale shows
      ratio values between 0.1 and 10 for all the ratio panels, except the
      \thCO/\CseO{} and \thCO/\CseO{} panels where the color scale goes
      from 0.33 to 3.}
    \label{fig:area:over:13co10}
  \end{figure*}}

\newcommand{\FigAvVsRatioTwCO}{%
  \begin{figure*}[p]
    \centering %
    \includegraphics[width=0.9\linewidth]{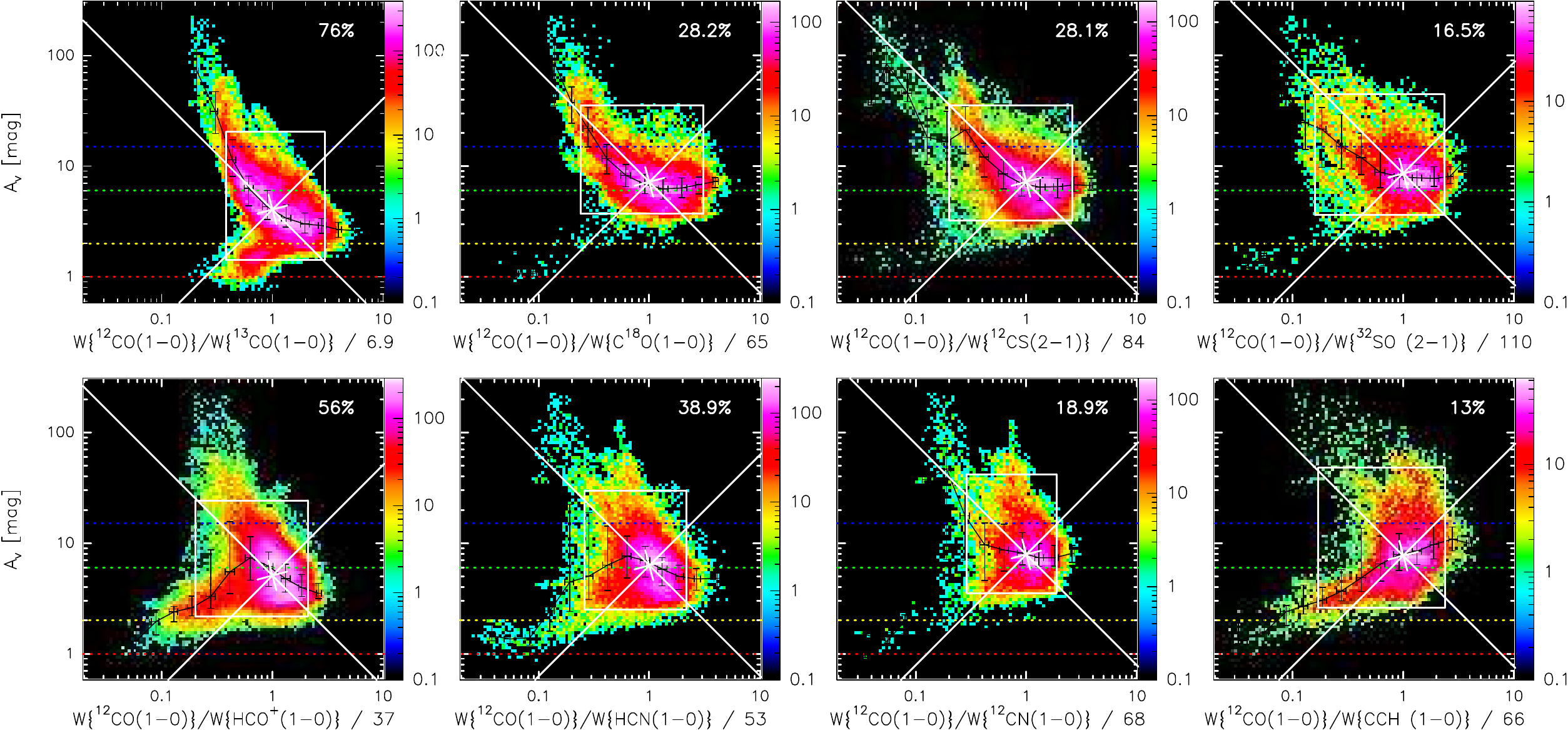}
    \caption{Joint distributions of the visual extinction as a function of
      the ratios of line integrated intensities. The ratios are normalized
      by their median value. The ratio numerator is always here
      \W\cbrace{\twCO \Jone}. Markers have the same signification as in
      Fig.~\ref{fig:av:vs:area}.}
    \label{fig:av:vs:ratio:12co10}
  \end{figure*}}

\newcommand{\FigAvVsRatioThCO}{%
  \begin{figure*}[p]
    \centering %
    \includegraphics[width=0.9\linewidth]{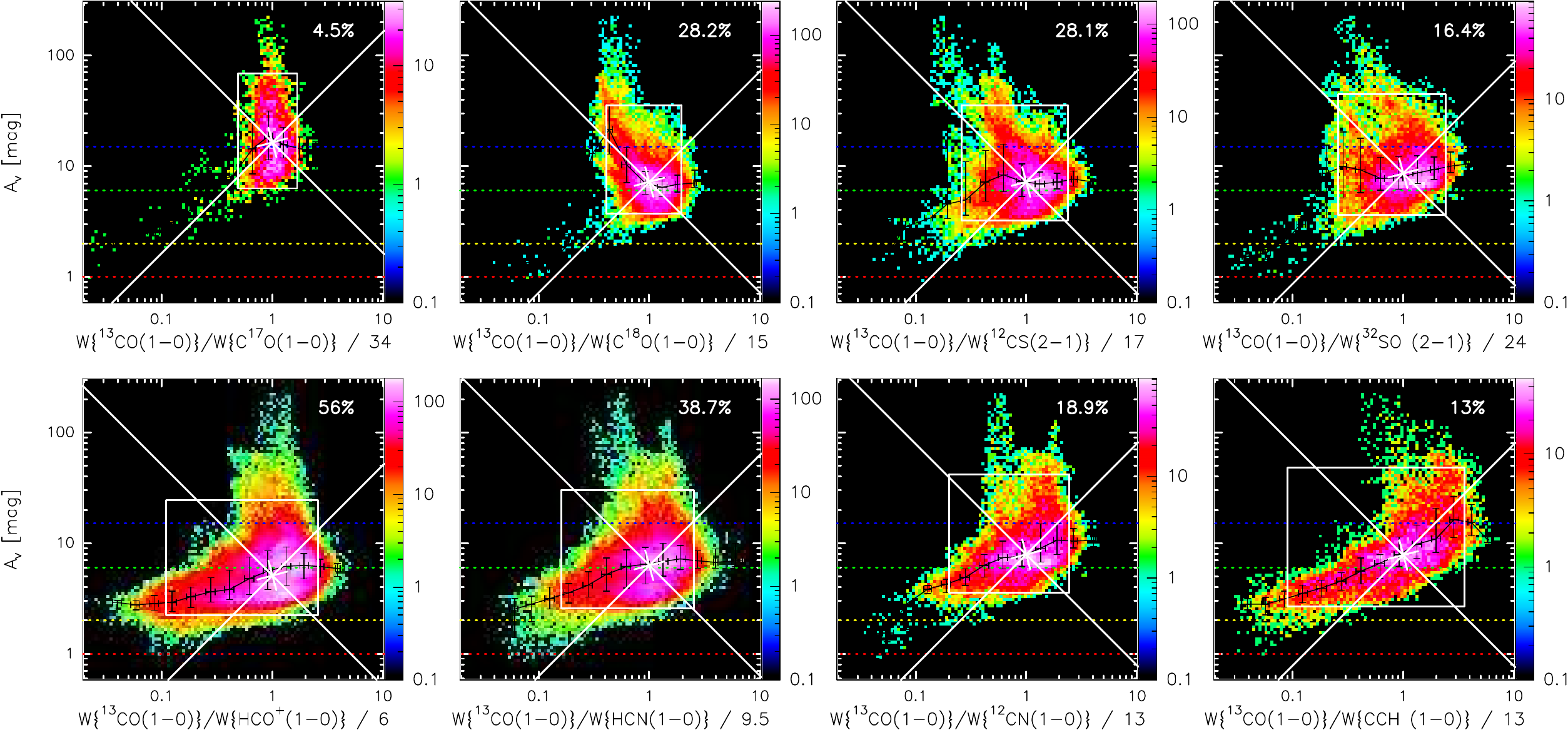}
    \caption{Same as Fig.~\ref{fig:av:vs:ratio:12co10}, except that the
      ratio numerator is now \W\cbrace{\thCO \Jone}.}
    \label{fig:av:vs:ratio:13co10}
  \end{figure*}}

\newcommand{\FigCorrDense}{%
  \begin{figure*}[p]
    \centering %
    \includegraphics[width=0.77\linewidth]{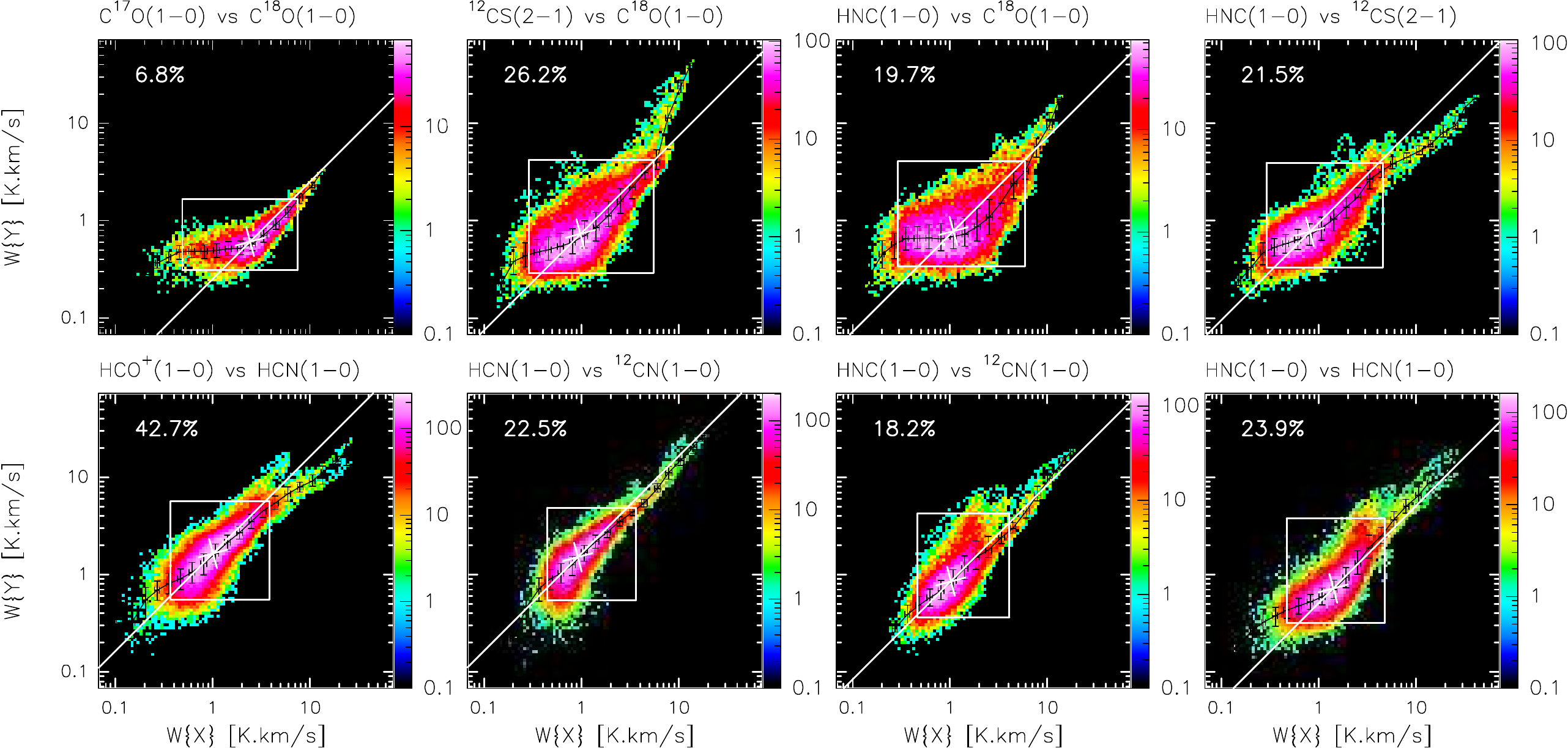}
    \caption{Same as Fig.~\ref{fig:area:vs:12co10}, except that there is no
      common $x$-axis in this figure.}
    \label{fig:area:vs:dense}
  \end{figure*}}

\newcommand{\FigRatioDense}{%
  \begin{figure*}[p]
    \centering %
    \includegraphics[width=0.77\linewidth]{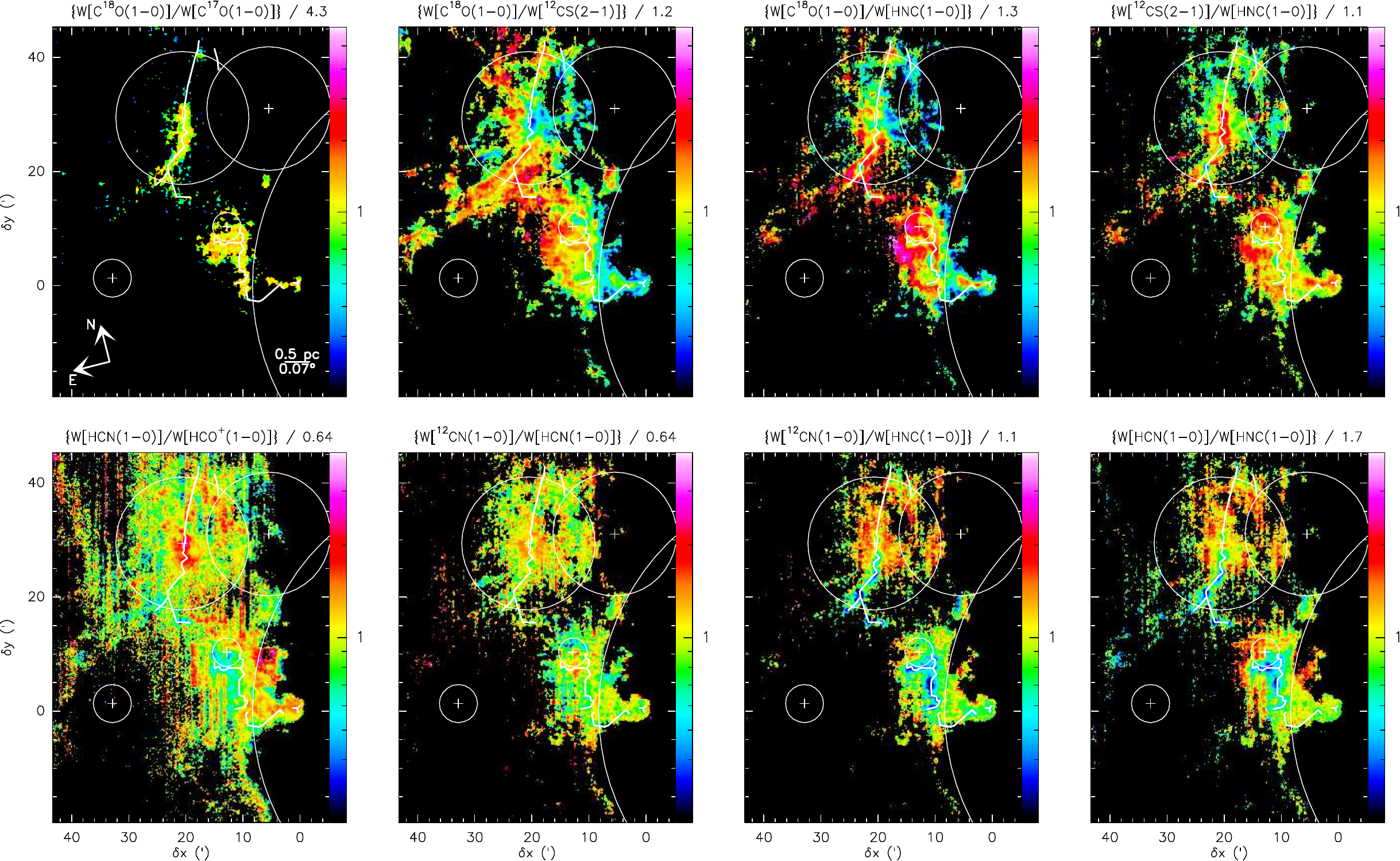}
    \caption{Same as Fig.~\ref{fig:area:over:12co10}, except that there is
      no common numerator in this figure.}
    \label{fig:area:over:dense}
  \end{figure*}}

\newcommand{\FigAvVsRatioDense}{%
  \begin{figure*}[p]
    \centering %
    \includegraphics[width=0.77\linewidth]{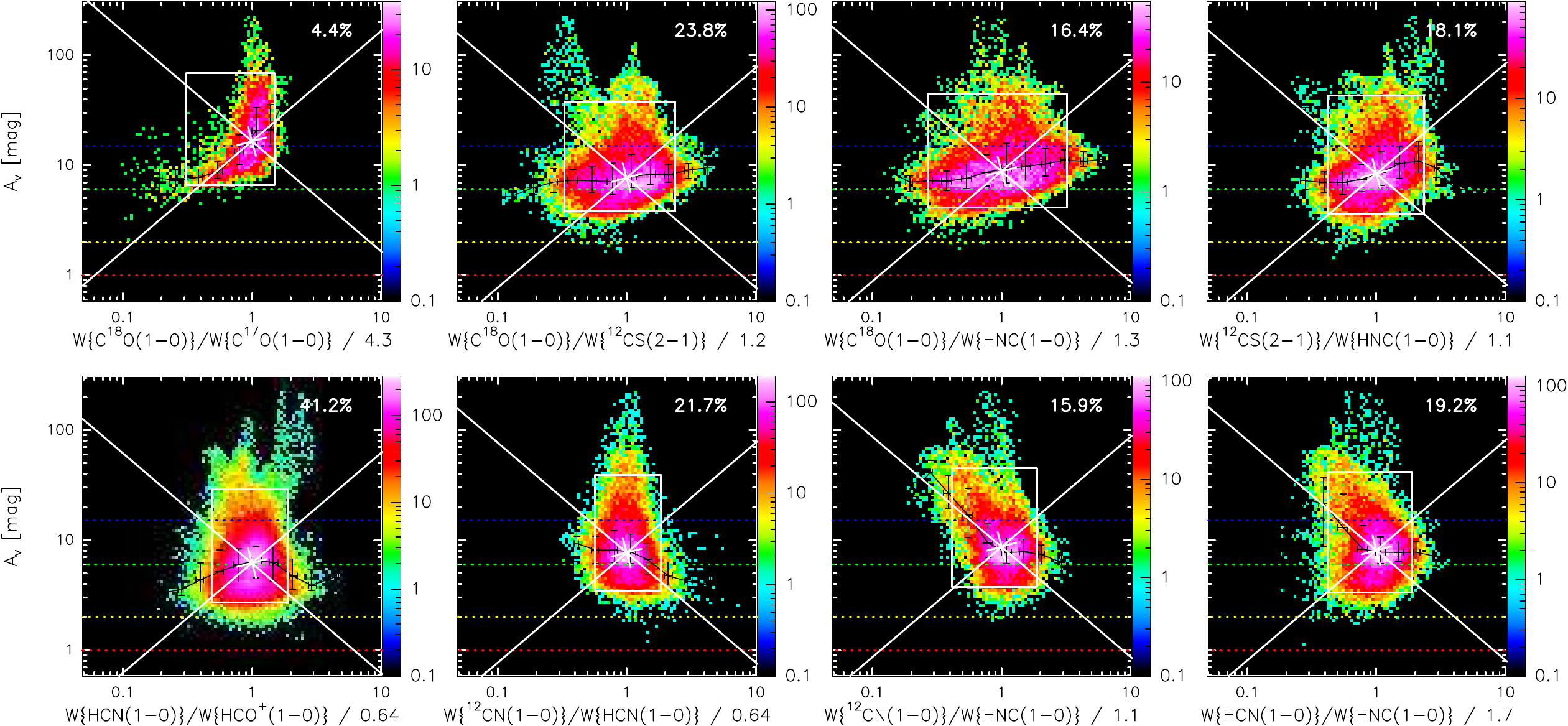}
    \caption{Same as Fig.~\ref{fig:area:over:12co10}, except that there is
      no common ratio numerator in this figure.}
    \label{fig:av:vs:ratio:dense}
  \end{figure*}}

\newcommand{\FigEnvironment}{%
  \begin{figure}
    \centering %
    \includegraphics[width=\linewidth]{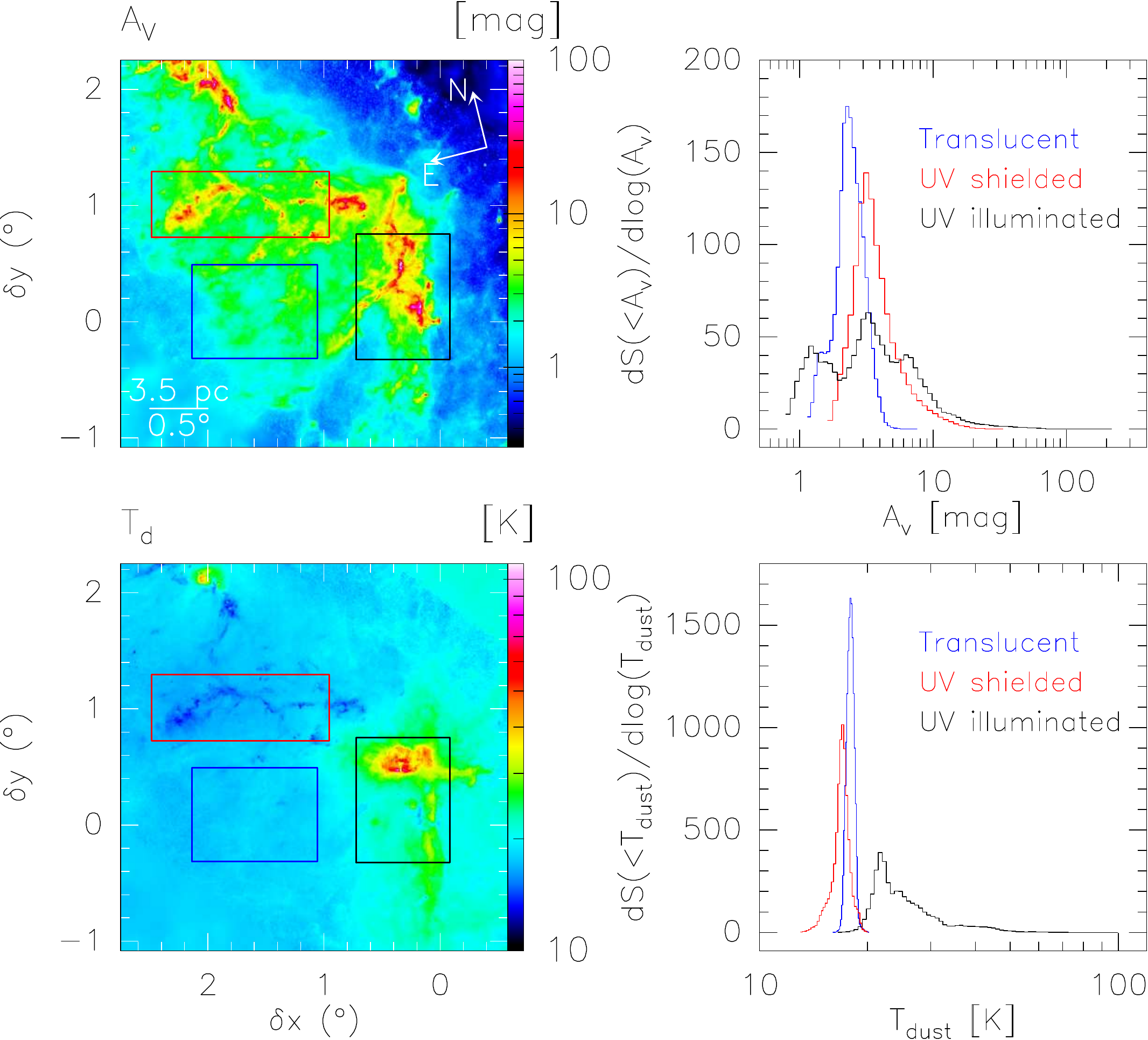}
    \caption{\textbf{Left panels:} Spatial distribution of the visual
      extinction and of the dust temperature. \textbf{Right panels:}
      Probability distribution functions of the visual extinction and the
      dust temperature for the regions inside the black, blue, and red
      rectangles, respectively.}
    \label{fig:dust:env}
  \end{figure}}

\newcommand{\FigTwCOprop}{%
  \begin{figure}
    \centering %
    \includegraphics[width=\linewidth]{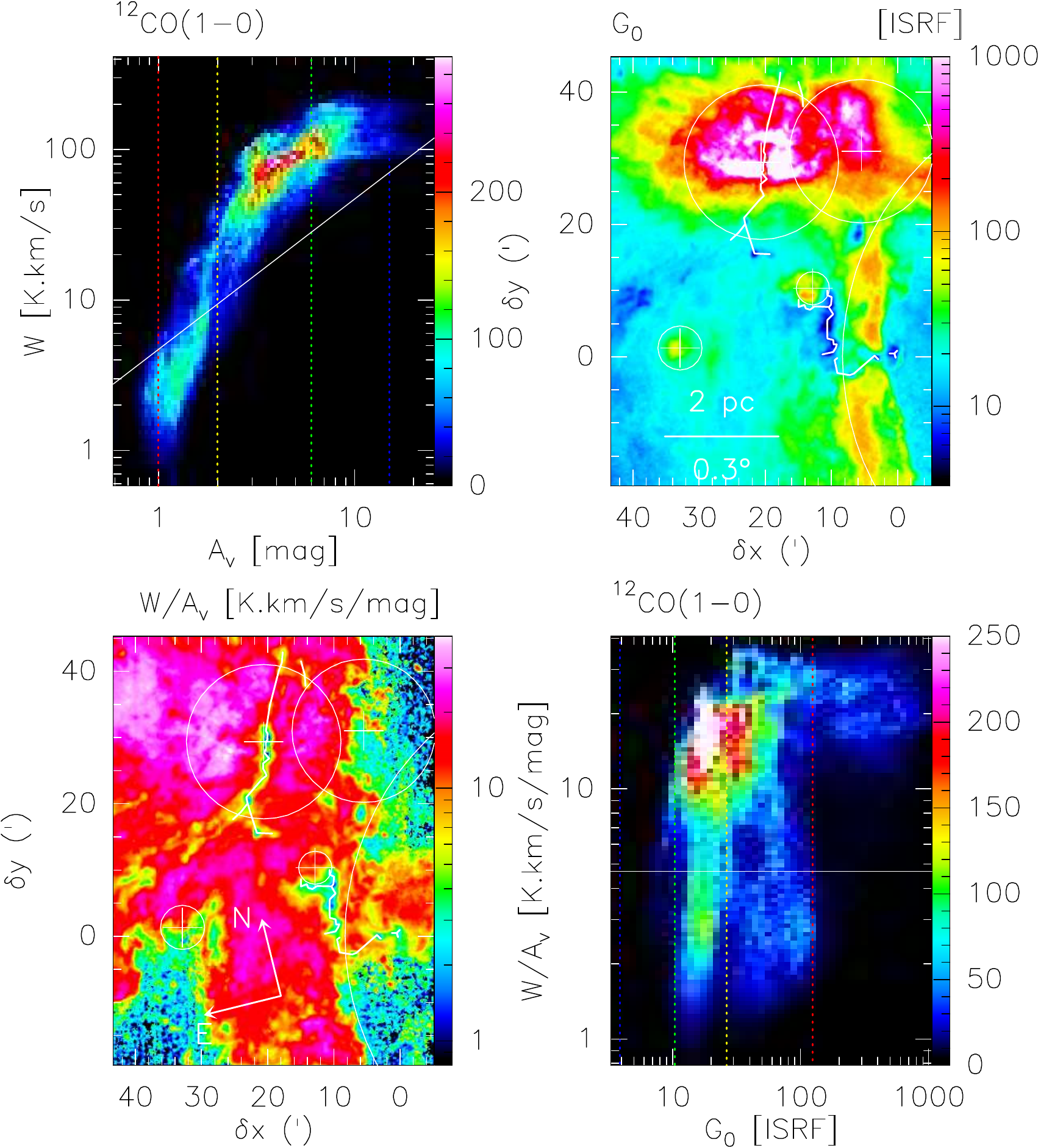}
    \caption{\textbf{Top, left panel:} Joint distribution of the \twCO{}
      \Jone{} line integrated intensity as a function of the visual
      extinction. The number of sightlines falling in a given 2D bin of the
      distribution is color-coded using a linear scale to emphasize the
      bi-modal nature of the distribution. The white line shows the
      location of points that follows $N(\HH) = \Xco\,\W$. The red, orange,
      green, and blue vertical dashed lines show the visual extinction
      limits (1, 2, 6, and 15 magnitudes, respectively) used in the masks
      of Fig.~\ref{fig:spec:mask:flux:Av}. \textbf{Top, right panel:}
      Spatial distribution of the far UV illumination in units of the
      ISRF~\citep{habing68}. \textbf{Bottom, left panel:} Ratio of the
      \twCO{} \Jone{} line integrated intensity to the visual extinction.
      \textbf{Bottom, right panel:} Joint distribution of this ratio as a
      function of the far UV illumination. The horizontal white line
      corresponds to the standard value of the \Xco{} factor. The blue,
      green, orange, and red vertical dashed lines show the far UV
      illumination limits (4, 10, 26, and 120, respectively), which
      corresponds to the dust temperature limits used in the masks of
      Fig.~\ref{fig:spec:mask:flux:tdust}. The color scales of the two
      images show the same ranges as the axes of the bottom right joint
      distribution.}
    \label{fig:12co10:prop}
  \end{figure}}

\newcommand{\FigRatioComparison}{%
  \begin{figure*}
    \centering 
    \includegraphics[width=\linewidth]{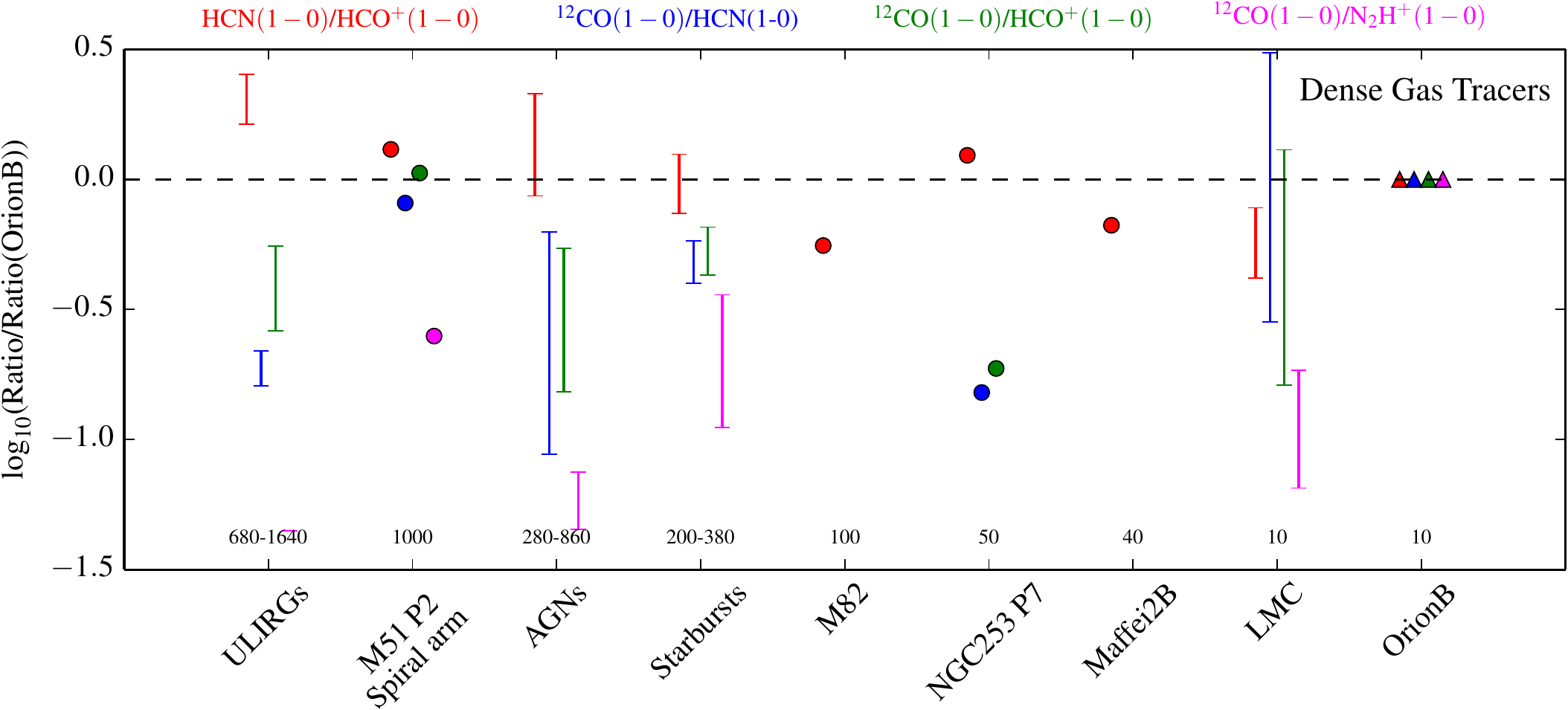}
    \includegraphics[width=\linewidth]{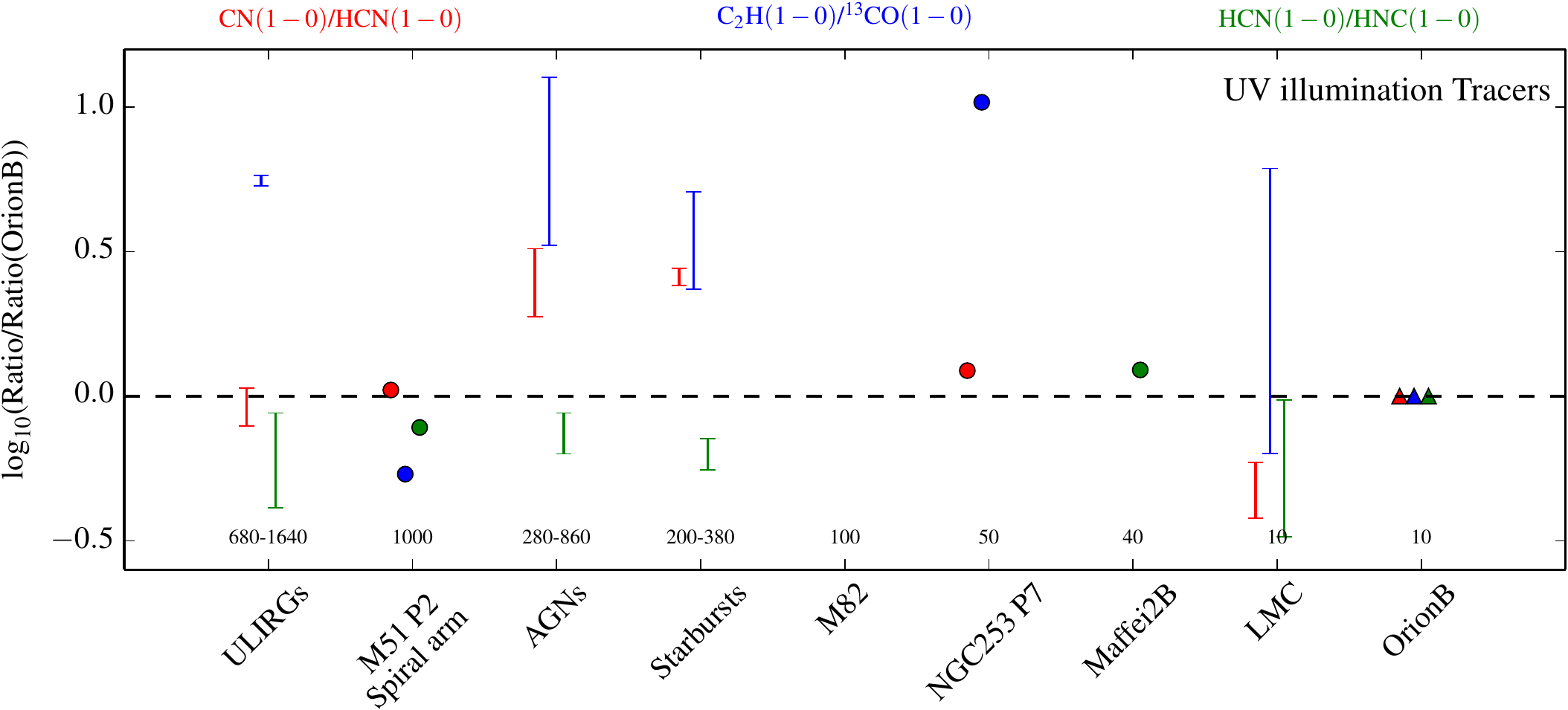}
    \caption{Line ratios observed towards nearby galaxies and the Orion\,B
      molecular cloud. The upper and lower panels include line ratios which
      are proposed dense gas and UV-illumination tracers, respectively.  A
      point is plotted for individual measures, while a range of values is
      given for the ULIRG, AGN, and starburst samples, which contain
      measures for several galaxies. The ULIRGs include Arp\,220 and
      Mrk\,231, while AGNs include M51, NGC\,1068 and NGC\,7469, and the
      starburst galaxies include M83, NGC\,253 and M82. The sources are
      sorted by decreasing spatial resolution, which is given at the bottom
      of the figure for each source. The Orion\,B values correspond to the
      ratios of the lines integrated over the full region to simulate a
      10\pc{} spatial resolution allowing a better comparison with the
      other galaxies.}
    \label{fig:ratio:comparison}
  \end{figure*}}

\newcommand{\TabObservations}{%
  \begin{table*}
    \caption{Observation parameters.}
    \centering
    \small
    \begin{tabular}{ccrcccrcccccc}
      \hline
      \hline
      Species & Transition & Frequency & Setup  & \Feff{} & \Beff{} & \Tsys{} & Beam\tablefootmark{a} & Vel. res.\tablefootmark{b} & Int. Time\tablefootmark{c} & Noise\tablefootmark{d} \\
              &            & \GHz{}    &        &         &         & K       & $''$                  & \kms{}                     & hr                         & K \\
      \hline
      \twCO{}       & \J{1}{0}           & 115.271202 & 110/USB & 0.95 & 0.78 & 287 & 22.5/31 & 0.51/0.5 & 40.4/62.4 & 0.49/0.18 \\
      \CN{}         & \J{1}{0}           & 113.490970 & 110/USB & 0.95 & 0.79 & 188 & 22.8/31 & 0.52/0.5 & 40.4/62.4 & 0.31/0.11 \\
      \CseO{}       & \J{1}{0}           & 112.358982 & 110/USB & 0.95 & 0.79 & 167 & 23.1/31 & 0.52/0.5 & 40.4/62.4 & 0.27/0.10 \\
      \thCO{}       & \J{1}{0}           & 110.201354 & 110/USB & 0.95 & 0.79 & 118 & 23.5/31 & 0.53/0.5 & 40.4/62.4 & 0.17/0.07 \\
      \CeiO{}       & \J{1}{0}           & 109.782173 & 110/USB & 0.95 & 0.79 & 114 & 23.6/31 & 0.53/0.5 & 40.4/62.4 & 0.17/0.07 \\
      \ttSO{}       & \T{2}{3}{1}{2}     &  99.299870 & 110/LSB & 0.95 & 0.80 &  95 & 26.1/31 & 0.59/0.5 & 40.4/62.4 & 0.13/0.06 \\
      \twCS{}       & \J{2}{1}           &  97.980953 & 110/LSB & 0.95 & 0.80 &  90 & 26.5/31 & 0.60/0.5 & 40.4/62.4 & 0.12/0.06 \\
      \methanol{}-A & \T{2}{0}{1}{0}     &  96.741375 & 110/LSB & 0.95 & 0.81 &  93 & 26.8/31 & 0.60/0.5 & 40.4/62.4 & 0.11/0.06 \\
      \methanol{}-E & \T{2}{1}{1}{1}     &  96.739362 & 110/LSB & 0.95 & 0.81 &  93 & 26.8/31 & 0.60/0.5 & 40.4/62.4 & 0.11/0.06 \\
      \NNHp{}       & \J{1}{0}           &  93.173764 & 110/LSB & 0.95 & 0.81 & 100 & 27.8/31 & 0.63/0.5 & 40.4/62.4 & 0.13/0.07 \\
      \HNC{}        & \J{1}{0}           &  90.663568 & 102/LSB & 0.95 & 0.81 & 115 & 28.6/31 & 0.64/0.5 & 44.9/70.5 & 0.12/0.08 \\
      \HCOp{}       & \J{1}{0}           &  89.188525 & 102/LSB & 0.95 & 0.81 & 130 & 29.1/31 & 0.66/0.5 & 44.9/70.5 & 0.13/0.09 \\
      \HCN{}        & \J{1}{0}           &  88.631848 & 102/LSB & 0.95 & 0.81 & 124 & 29.3/31 & 0.66/0.5 & 44.9/70.5 & 0.12/0.09 \\
      \CCH{}        & \J{1}{0}           &  87.316898 & 102/LSB & 0.95 & 0.82 & 132 & 29.7/31 & 0.67/0.5 & 44.9/70.5 & 0.15/0.11 \\
      \HNthC{}      & \J{1}{0}           &  87.090825 & 102/LSB & 0.95 & 0.81 & 137 & 28.6/31 & 0.64/0.5 & 44.9/70.5 & 0.14/0.11 \\
      \SiO{}        & \J{2}{1}           &  86.846960 & 102/LSB & 0.95 & 0.82 & 142 & 29.9/31 & 0.67/0.5 & 44.9/70.5 & 0.14/0.11 \\
      \HthCOp{}     & \J{1}{0}           &  86.754288 & 102/LSB & 0.95 & 0.81 & 136 & 28.6/31 & 0.64/0.5 & 44.9/70.5 & 0.14/0.10 \\
      \HthCN{}      & \J{1}{0}           &  86.340184 & 102/LSB & 0.95 & 0.81 & 136 & 28.6/31 & 0.64/0.5 & 44.9/70.5 & 0.13/0.10 \\
      \cCCCHH{}     & \T{2}{1,2}{1}{0,1} &  85.338893 & 102/LSB & 0.95 & 0.82 & 123 & 30.4/31 & 0.69/0.5 & 44.9/70.5 & 0.12/0.10 \\
      \hline
    \end{tabular}
    \tablefoot{%
      \tablefoottext{a}{Listed as natural/smoothed resolution.}%
      \tablefoottext{b}{Listed as natural/oversampled channel spacing.}
      \tablefoottext{c}{Listed as on-source time/telescope time.}
      \tablefoottext{d}{Listed as measured on the natural/resampled
        resolution cubes.}}
    \label{tab:obs}
  \end{table*}}

\newcommand{\FigTsysVsFreq}{%
  \begin{figure*}
    \centering %
    \includegraphics[width=\linewidth]{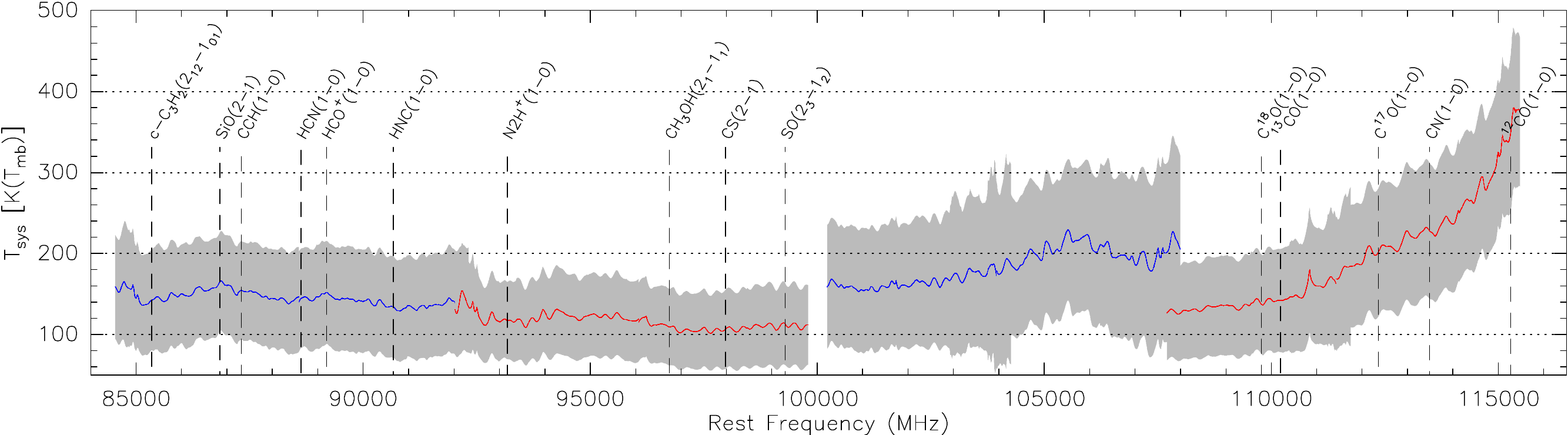}
    \caption{\Tsys{} as a function of the frequency. The
      solid lines display the average \Tsys{} and the grey shaded
      backgrounds show the $\pm3\sigma$ interval at each frequency. The
      blue and red colors present the instantaneous 8\GHz{}-bandwidth of
      the lower and upper sidebands covered by the two tunings. The
      vertical dashed lines show the frequencies of the brightest lines
      studied here.}
    \label{fig:tsys:vs:freq}
  \end{figure*}}

\newcommand{\FigNoise}{%
  \begin{figure}
    \centering %
    \includegraphics[width=\linewidth]{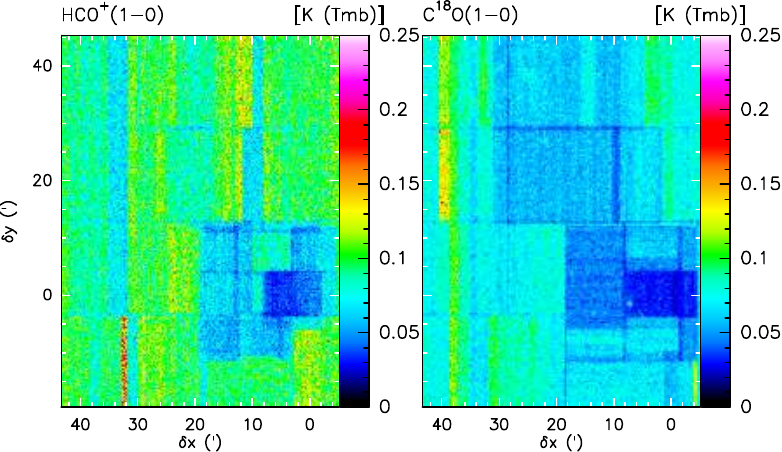}
    \caption{Typical spatial distribution of the RMS noise for two lines,
      each one belonging to one of the two tuning setups.}
    \label{fig:noise}
  \end{figure}}

\newcommand{\TabAvMaskedMean}{%
  \begin{table*}[p]
    \caption{Line averaged intensities in $\mKkms$ and in percentage of the total
      flux inside the four \Av{} mask regions.}
    \centering{} %
    \tiny{%
      \begin{tabular}{llccccc} 
      \hline 
      \hline 
      Species     & Transition         & $0\le\Av<222$     & $1\le\Av<2$       & $2\le\Av<6$       & $6\le\Av<15$      & $15\le\Av<222$    \\ 
      \hline 
      \twCO{}     & \J{1}{0}           & $33000$ $(100\%)$ & $3300$ $(9.9\%)$  & $36000$ $(110\%)$ & $65000$ $(200\%)$ & $77000$ $(230\%)$ \\ 
      \thCO{}     & \J{1}{0}           & $6200$ $(100\%)$  & $350$ $(5.7\%)$   & $4900$ $(79\%)$   & $15000$ $(230\%)$ & $29000$ $(470\%)$ \\ 
      \HCOp{}     & \J{1}{0}           & $980$ $(100\%)$   & $110$ $(11\%)$    & $840$ $(85\%)$    & $2000$ $(210\%)$  & $4700$ $(480\%)$  \\ 
      \HCN{}      & \J{1}{0}           & $590$ $(100\%)$   & $40$ $(6.9\%)$    & $440$ $(75\%)$    & $1300$ $(230\%)$  & $3200$ $(550\%)$  \\ 
      \CeiO{}     & \J{1}{0}           & $430$ $(100\%)$   & $8.4$ $(1.9\%)$   & $210$ $(49\%)$    & $1100$ $(240\%)$  & $3800$ $(880\%)$  \\ 
      \twCS{}     & \J{2}{1}           & $370$ $(100\%)$   & $9.9$ $(2.7\%)$   & $190$ $(52\%)$    & $800$ $(220\%)$   & $3700$ $(990\%)$  \\ 
      \twCN{}     & \J{1}{0}           & $350$ $(100\%)$   & $36$ $(10\%)$     & $250$ $(70\%)$    & $810$ $(230\%)$   & $2100$ $(590\%)$  \\ 
      \HNC{}      & \J{1}{0}           & $310$ $(100\%)$   & $26$ $(8.4\%)$    & $190$ $(61\%)$    & $650$ $(210\%)$   & $2600$ $(830\%)$  \\ 
      \CCH{}      & \J{1}{0}           & $270$ $(100\%)$   & $47$ $(18\%)$     & $230$ $(87\%)$    & $510$ $(190\%)$   & $1400$ $(520\%)$  \\ 
      \ttSO{}     & \T{2}{3}{1}{2}     & $200$ $(100\%)$   & $6.8$ $(3.4\%)$   & $100$ $(50\%)$    & $460$ $(230\%)$   & $1900$ $(940\%)$  \\ 
      \CseO{}     & \J{1}{0}           & $100$ $(100\%)$   & $23$ $(22\%)$     & $53$ $(53\%)$     & $230$ $(220\%)$   & $800$ $(780\%)$   \\ 
      \cCCCHH{}   & \T{2}{1,2}{1}{0,1} & $87$ $(100\%)$    & $15$ $(18\%)$     & $73$ $(84\%)$     & $170$ $(190\%)$   & $450$ $(520\%)$   \\ 
      \HthCOp{}   & \J{1}{0}           & $33$ $(100\%)$    & $0.87$ $(2.6\%)$  & $4.9$ $(15\%)$    & $57$ $(170\%)$    & $560$ $(1700\%)$  \\ 
      \NNHp{}     & \J{1}{0}           & $29$ $(100\%)$    & $-13$ $(-45\%)$   & $5$ $(17\%)$      & $25$ $(85\%)$     & $780$ $(2700\%)$  \\ 
      \methanol{} & \J{2}{1}           & $27$ $(100\%)$    & $4.7$ $(17\%)$    & $3.3$ $(12\%)$    & $57$ $(210\%)$    & $390$ $(1500\%)$  \\ 
      \hline 
      \end{tabular}
      \tablefoot{%
        \tablefoottext{a}{The lines are sorted by decreasing value of their intensity.}}}
    \label{tab:av:mask:mean}
  \end{table*}}

\newcommand{\TabTdustMaskedMean}{%
  \begin{table*}[p]
    \caption{Line intensities averaged over $[9,12\kms]$ inside the four \Td{} mask regions in
      $\mKkms$ and in percentage of the intensity of the cold dust region.}
    \centering{} %
    \tiny{%
      \begin{tabular}{llccccc} 
      \hline 
      \hline 
      Species     & Transition         & $16\le\Td<100$    & $16\le\Td<19.5$   & $19.5\le\Td<23.5$ & $23.5\le\Td<32$   & $32\le\Td<100$    \\ 
      \hline 
      \twCO{}     & \J{1}{0}           & $33000$ $(100\%)$ & $44000$ $(130\%)$ & $28000$ $(85\%)$  & $31000$ $(94\%)$  & $56000$ $(170\%)$ \\ 
      \thCO{}     & \J{1}{0}           & $6200$ $(100\%)$  & $15000$ $(250\%)$ & $5800$ $(92\%)$   & $5000$ $(81\%)$   & $10000$ $(170\%)$ \\ 
      \HCOp{}     & \J{1}{0}           & $980$ $(100\%)$   & $2500$ $(250\%)$  & $670$ $(68\%)$    & $870$ $(89\%)$    & $2200$ $(220\%)$  \\ 
      \HCN{}      & \J{1}{0}           & $590$ $(100\%)$   & $1400$ $(230\%)$  & $390$ $(66\%)$    & $500$ $(85\%)$    & $1500$ $(250\%)$  \\ 
      \CeiO{}     & \J{1}{0}           & $430$ $(100\%)$   & $2100$ $(490\%)$  & $420$ $(99\%)$    & $290$ $(67\%)$    & $660$ $(150\%)$   \\ 
      \twCS{}     & \J{2}{1}           & $370$ $(100\%)$   & $1700$ $(450\%)$  & $290$ $(78\%)$    & $250$ $(69\%)$    & $830$ $(230\%)$   \\ 
      \twCN{}     & \J{1}{0}           & $350$ $(100\%)$   & $770$ $(220\%)$   & $220$ $(61\%)$    & $290$ $(81\%)$    & $970$ $(280\%)$   \\ 
      \HNC{}      & \J{1}{0}           & $310$ $(100\%)$   & $1300$ $(430\%)$  & $210$ $(69\%)$    & $240$ $(79\%)$    & $690$ $(230\%)$   \\ 
      \CCH{}      & \J{1}{0}           & $270$ $(100\%)$   & $460$ $(170\%)$   & $150$ $(57\%)$    & $260$ $(97\%)$    & $670$ $(250\%)$   \\ 
      \ttSO{}     & \T{2}{3}{1}{2}     & $200$ $(100\%)$   & $1200$ $(610\%)$  & $170$ $(84\%)$    & $140$ $(70\%)$    & $350$ $(180\%)$   \\ 
      \CseO{}     & \J{1}{0}           & $100$ $(100\%)$   & $440$ $(430\%)$   & $100$ $(99\%)$    & $73$ $(72\%)$     & $150$ $(140\%)$   \\ 
      \cCCCHH{}   & \T{2}{1,2}{1}{0,1} & $87$ $(100\%)$    & $200$ $(230\%)$   & $50$ $(57\%)$     & $85$ $(97\%)$     & $210$ $(240\%)$   \\ 
      \HthCOp{}   & \J{1}{0}           & $33$ $(100\%)$    & $290$ $(890\%)$   & $25$ $(75\%)$     & $19$ $(58\%)$     & $67$ $(210\%)$    \\ 
      \NNHp{}     & \J{1}{0}           & $29$ $(100\%)$    & $420$ $(1500\%)$  & $25$ $(86\%)$     & $11$ $(37\%)$     & $44$ $(150\%)$    \\ 
      \methanol{} & \J{2}{1}           & $27$ $(100\%)$    & $260$ $(970\%)$   & $21$ $(78\%)$     & $19$ $(72\%)$     & $38$ $(140\%)$    \\ 
      \hline 
      \end{tabular}
      \tablefoot{%
        \tablefoottext{a}{The lines are sorted by decreasing value of their intensity.}}}
    \label{tab:tdust:mask:mean}
  \end{table*}}

\newcommand{\FigMaskedSpectraMeanAv}{%
  \begin{figure*}[p]
    \centering %
    \includegraphics[width=0.21\linewidth]{orionb-f06a.pdf}
    \hfill
    \includegraphics[width=0.76\linewidth]{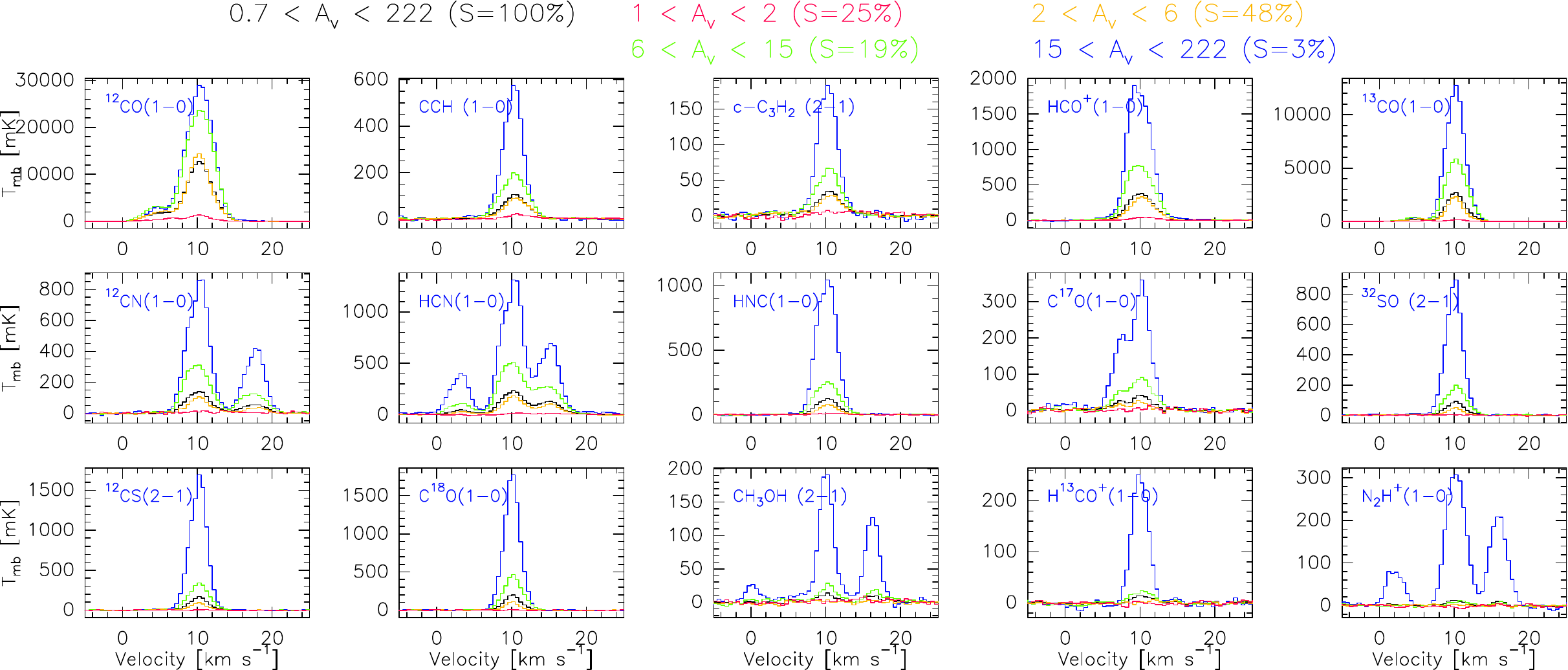}
    \caption{Same as Fig.~\ref{fig:spec:mask:flux:Av}, except that the
      spectra show the temperature intensity averaged over the different
      masks.}
    \label{fig:spec:mask:mean:Av}
  \end{figure*}}

\newcommand{\FigLineAreaAv}{%
  \begin{figure*}[p]
    \centering %
    \includegraphics[width=0.4\linewidth,angle=90]{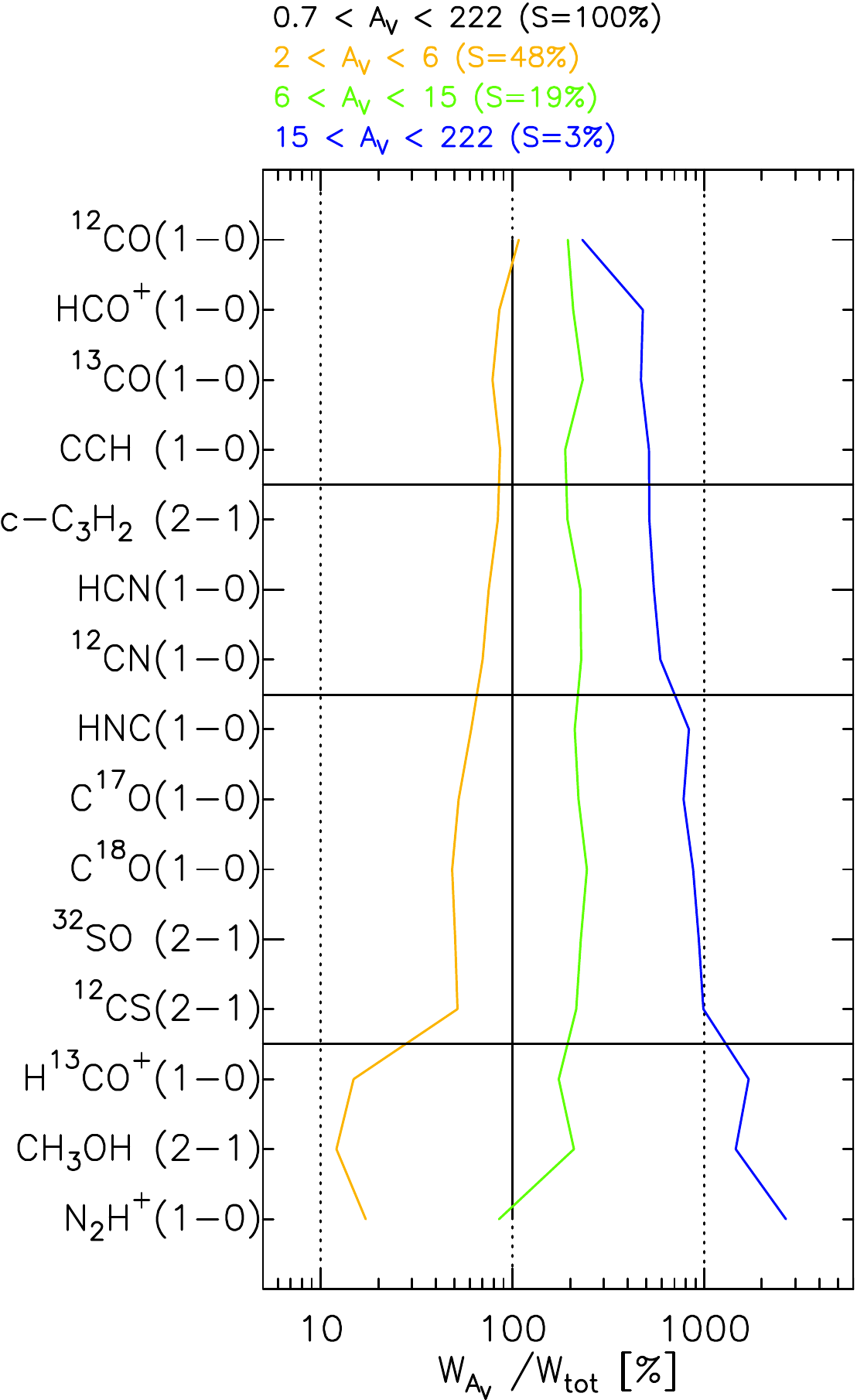}
    \caption{For each transition, line integrated intensity computed over
      the same velocity range divided by the line integrated intensity
      computed over the surface where $\Av \ge 15$. All line integrated
      intensities are computed between 9 and 12\kms{}. The black vertical
      lines define the groups of lines described in
      Section~\ref{sec:flux:av}.}
    \label{fig:line:W:Av}
  \end{figure*}}

\newcommand{\FigMaskedSpectraMeanTdust}{%
  \begin{figure*}[p]
    \centering %
    \includegraphics[width=0.21\linewidth]{orionb-f08a.pdf}
    \hfill
    \includegraphics[width=0.76\linewidth]{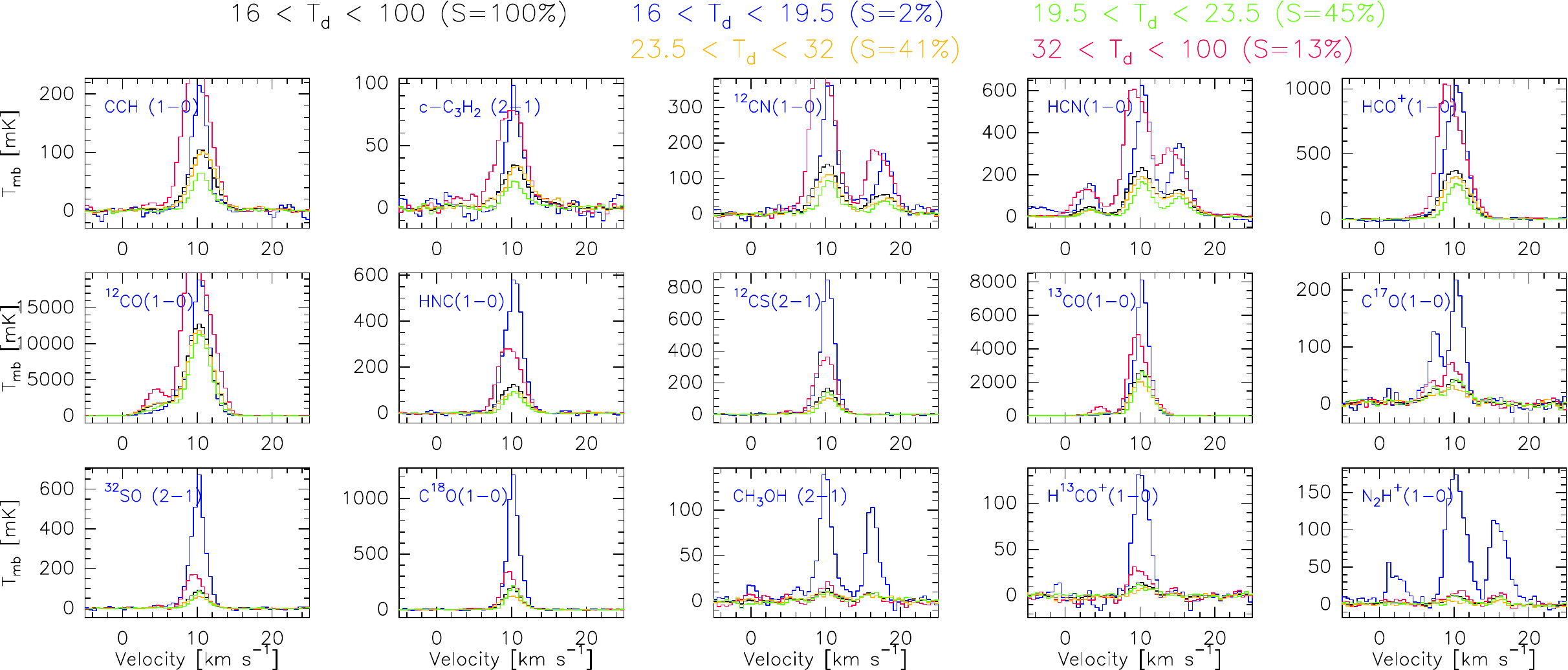}
    \caption{Same as Fig.~\ref{fig:spec:mask:flux:tdust}, except that the
      spectra show the temperature intensity averaged over the different
      masks.}
    \label{fig:spec:mask:mean:tdust}
  \end{figure*}}

\newcommand{\FigLineAreaTdust}{%
  \begin{figure*}[p]
    \centering %
    \includegraphics[width=0.4\linewidth,angle=90]{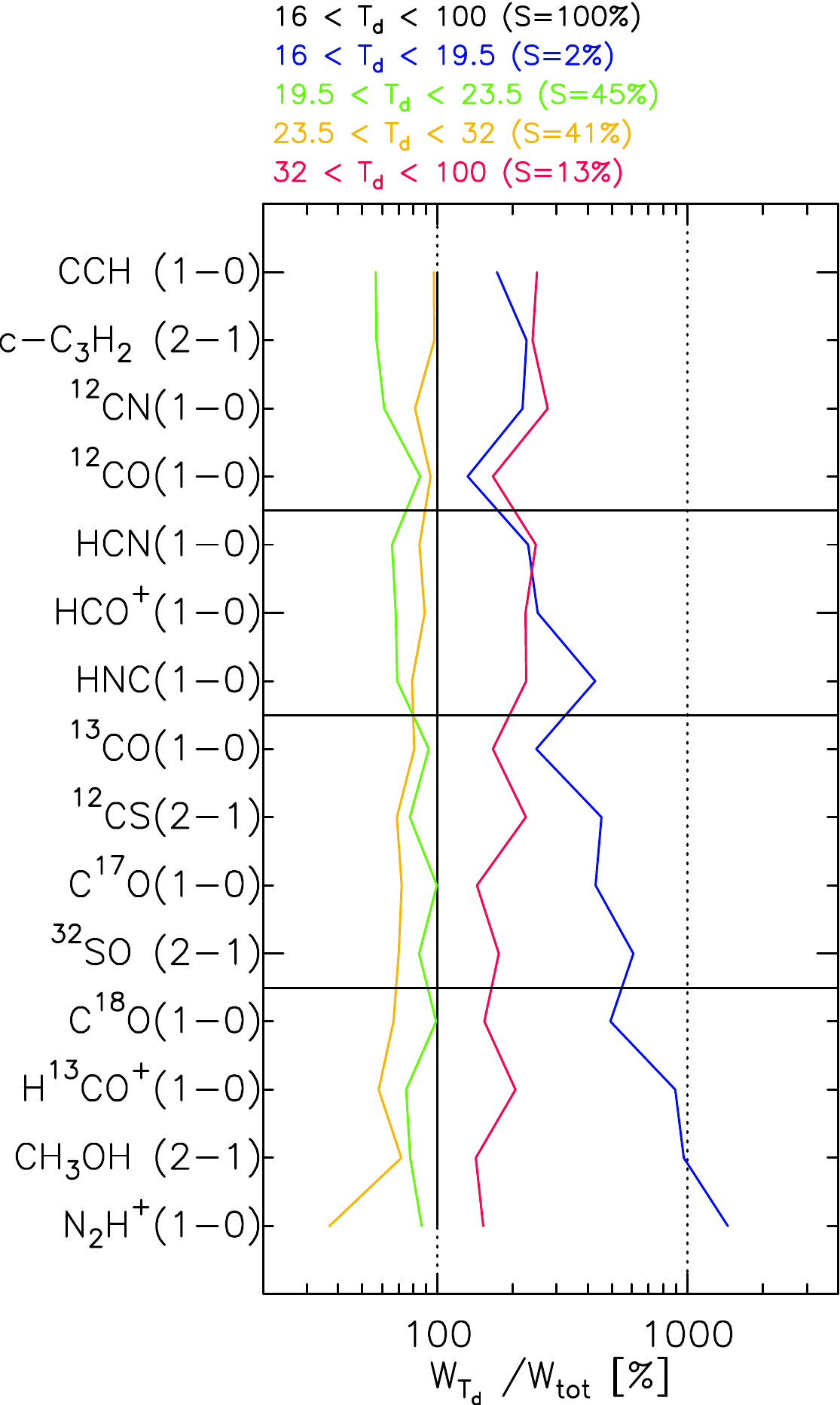}
    \caption{For each transition, line integrated intensity divided by the
      line integrated intensity computed over the surface where $\Td <
      20\K$. All line integrated intensities are computed between 9 and
      12\kms{}. The black vertical lines define the groups of lines
      described in Section~\ref{sec:flux:tdust}.}
    \label{fig:line:W:Tdust}
  \end{figure*}}

\begin{document} 

\title{The anatomy of the Orion\,B Giant Molecular Cloud:\\
  A local template for studies of nearby galaxies\thanks{Based on
    observations carried out at the IRAM-30m single-dish telescope. IRAM is
    supported by INSU/CNRS (France), MPG (Germany) and IGN (Spain).}}
\titlerunning{The anatomy of the Orion\,B Giant Molecular Cloud}

\author{Jérôme Pety \inst{\ref{IRAM},\ref{LERMA}} \and %
  Viviana V. Guzm\'an \inst{\ref{CFA}} \and %
  Jan H. Orkisz \inst{\ref{IRAM},\ref{LERMA}} \and %
  Harvey S. Liszt \inst{\ref{NRAO}} \and %
  Maryvonne Gerin \inst{\ref{LERMA},\ref{LERMA6}} \and \\ %
  Emeric Bron \inst{\ref{LERMA},\ref{LERMA6}} \and %
  Sébastien Bardeau \inst{\ref{IRAM}} \and %
  Javier R. Goicoechea \inst{\ref{ICMM}} \and %
  Pierre Gratier \inst{\ref{LAB}} \and %
  Franck Le Petit \inst{\ref{LERMA},\ref{LERMA6}} \and \\%
  François Levrier \inst{\ref{LERMA},\ref{LERMA6}} \and %
  Karin I. \"Oberg \inst{\ref{CFA}} \and %
  Evelyne Roueff \inst{\ref{LERMA},\ref{LERMA6}} \and %
  Albrecht Sievers \inst{\ref{IRAM-SPAIN}} %
}

\institute{%
  IRAM, 300 rue de la Piscine, 38406 Saint Martin d'Hères,
  France \label{IRAM} %
  \and LERMA, Observatoire de Paris, CNRS UMR8112, Ecole Normale
  Supérieure, PSL research university 24 Rue Lhomond, 75231 Paris cedex 05,
  France \label{LERMA} %
  \and Harvard-Smithsonian Center for Astrophysics, 60 Garden Street,
  Cambridge, MA, 02138, USA \label{CFA} %
  \and National Radio Astronomy Observatory, 520 Edgemont Road,
  Charlottesville, VA, 22903, USA \label{NRAO} %
  \and Sorbonne Universit\'es, UPMC Univ. Paris 06, UMR8112, LERMA, F-75005
  Paris, France \label{LERMA6} %
  \and Univ. Bordeaux, LAB, UMR 5804, 33270, Floirac, France \label{LAB} %
  \and ICMM, Consejo Superior de Investigaciones Cientificas (CSIC).
  E-28049. Madrid, Spain \label{ICMM} %
  \and IRAM, Avenida Divina Pastora, 7, Núcleo Central, E-18012 Granada,
  España \label{IRAM-SPAIN}%
}

\date{}

\abstract
{Molecular lines and line ratios are commonly used to infer properties of
  extra-galactic star forming regions. The new generation of millimeter
  receivers turns every observation nearly into a line survey. To fully
  exploit this technical advance in extra-galactic studies requires
  detailed bench-marking of available line diagnostics.}
{We aim to develop the Orion\,B Giant Molecular Cloud (GMC) as a local
  template for interpreting extra-galactic molecular line observations.}
{We use the wide-band receiver at the IRAM-30m to spatially and spectrally
  resolve the Orion\,B GMC. The observations cover almost 1 square degree
  at $26''$ resolution with a bandwidth of 32\GHz{} from 84 to 116\GHz{} in
  only two tunings. Among the mapped spectral lines are the \twCO, \thCO,
  \CeiO, \CseO, \HCN, \HNC, \twCN, \CCH, \HCOp, \NNHp{} \Jone, and \twCS{},
  \ttSO{}, \SiO{}, \cCCCHH{}, \methanol{} \Jtwo{} transitions.}
{We introduce the molecular anatomy of the Orion\,B GMC, including
  relations between line intensities and gas column density or far-UV
  radiation fields, and correlations between selected line and line
  ratios. We also obtain a dust-traced gas mass that is less than about one
  third the CO-traced mass, using the standard \Xco{} conversion
  factor. The presence of overluminous CO can be traced back to the
  dependence of the CO intensity on UV illumination. As a matter of fact,
  while most lines show some dependence on the UV radiation field, CN and
  \CCH{} are the most sensitive. Moreover dense cloud cores are almost
  exclusively traced by \NNHp{}. Other traditional high density tracers,
  such as \HCN{}\Jone, are also easily detected in extended translucent
  regions at a typical density of $\sim500\,\HH\pccm$. In general, we find
  no straightforward relation between line critical density and the
  fraction of the line luminosity coming from dense gas regions.}
{Our initial findings demonstrate that the relations between line (ratio)
  intensities and environment in GMCs are more complicated than often
  assumed. Sensitivity (\ie, the molecular column density), excitation, and
  above all chemistry contribute to the observed line intensity
  distributions, and they must be considered together when developing the
  next generation of extra-galactic molecular line diagnostics of mass,
  density, temperature and radiation field.}


\maketitle{} %

\TabStars{} %
\FigRGB{} %

\section{Introduction}
\label{sec:intro}

The star formation process from interstellar gas raises many outstanding
questions.  For instance, what is the relative role of micro-physics and
galactic environment on the star formation efficiency? More precisely, what
is the role of magnetic field, gravity, turbulence~\citep[see
\eg{},][]{hennebelle11,hennebelle13}, on one hand, and of external
pressure, position in galactic arm/interarms, streaming
motions~\citep[\eg{},][]{meidt13,hughes13b}, on the other hand?  How does
feedback from {\sc{H\,ii}} region expansions and supernovae limit the star
formation efficiency~\citep{kim13}? What are the key dynamical parameters
controlling star formation: Mach number, virial parameter, amount of energy
in solenoidal/compressive modes of the
turbulence~\citep{federrath12,federrath13}? What is the amount of CO-dark
molecular gas and does it bias the global estimation of the mass of the
molecular reservoir at cloud scales~\citep{wolfire10,liszt12}?  What is the
amount of diffuse $(\sim 100-500\pccm, 80\K)$ vs. dense $(\sim10^4\pccm,
10\K)$ gas in a GMC?  In other words, what is the fraction of star-forming
dense gas~\citep{lada10,lada12,lada13}?

All these questions also arise in extra-galactic studies with the
additional difficulty that GMCs are unresolved at the typically achieved
angular resolution ($1''$ corresponds to 15\pc{} for a 3\Mpc{} distant
galaxy). It is therefore crucial to first understand how the average
spectra of molecular lines relate to actual physical properties when the
line emission is spatially resolved. By mapping a significant fraction of a
GMC at a spatial resolution of $\sim50\mpc$ and a spectral resolution of
$0.6\kms$, we address some of the following issues: What linear resolution
must be achieved on a GMC to correctly derive its global properties
including star formation rate and efficiency~\citep{leroy16}? For instance,
are usual extra-galactic line tracers of the various molecular cloud
density regimes reliable~\citep{bigiel16}?  Do we get a more accurate
estimate of the mass by resolving the emission?  More generally, can we
derive empirical laws that link tracer properties averaged over a GMC to
its internal star forming activity?

With the advent of wide-bandwidth receivers associated to high resolution
spectrometers, any observation now simultaneously delivers emission from
many different tracers. Moreover, the increased sensitivity makes it
possible to cover large fields of view. The possibility to map many
different lines in many different environments allows us to start answering
the questions presented above. The essence of the ORION-B (Outstanding
Radio Imaging of OrioN B, PI: J.~Pety) project is to recast the science
questions of star formation in a statistical way. Wide-field hyper-spectral
mapping of Orion\,B is used to obtain an accurate 3D description of the
molecular structure in a Giant Molecular Cloud, a key for defining chemical
probes of the star formation activity in more distant Galactic and
extragalactic sources.

About thirty 3\mm{} lines are detected in only two frequency tunings with
the same sensitive radio single-dish telescope at a typical resolution of
$26''$ over almost 1 square degree.  The field of view ($5.6\times7.5\pc$)
would fall in a single resolution element of a map of the Orion\,B
molecular cloud observed at 3\mm{} with a telescope of similar diameter as
the IRAM-30m from the Small or Large Magellanic Clouds. The spectra
averaged over the field of view would then represent the spectra of
Orion\,B as seen by an alien from the Magellanic Clouds.  Conversely, our
imaging experiment allows us to reveal the detailed anatomy of a molecular
emission that is usually hidden behind these mean spectra in nearby galaxy
studies.  The south-western edge of the Orion\,B molecular cloud
(a.k.a. Barnard\,33 or Lynds 1630) represents an ideal laboratory for this
kind of study. It forms both low-mass and massive stars. It contains
regions of triggered or spontaneous star formation, photon-dominated
regions and UV-shielded cold gas, all in a single source.

In companion papers, \citet{gratier16} study a Principal Component Analysis
of the same dataset to understand the main correlations that exist between
the different lines. \citet{orkisz16} quantify the fractions of turbulent
energy that are associated to the solenoidal/compressive modes, and they
relate these values to the star formation efficiency in Orion\,B. In this
paper, we present the observational results of the ORION-B project,
focusing on the mean properties of this GMC and evaluate the diagnostic
power of commonly used line tracers and ratios.

We present the targeted field of view, as well as the observations and data
reduction process in Section~\ref{sec:orionb}. Typical properties, such as
UV-illumination, mean line profiles, CO-traced, dust-traced and virial
masses, are computed in Section~\ref{sec:mean:prop}. In
Section~\ref{sec:flux}, we investigate the fraction of flux arising in
different gas regimes for each line. In Section~\ref{sec:Av:vs:W}, we
compare the visual extinction map with the line integrated intensities and
compute the luminosity per proton of the different line tracers. The
properties of various line ratios are discussed in
Section~\ref{sec:line:ratios}.  A discussion is presented in
Section~\ref{sec:discussion}, focusing on possible biases introduced by the
characteristics of the observed field of view, and whether the \HCOp, \HCN,
and \HNC{} \Jone{} lines are good tracers of dense gas.  We end the
discussion by comparing the observed line ratios in Orion\,B with
extra-galactic observation results. Section~\ref{sec:conclusion} summarizes
the results and concludes the paper.

\section{The Orion\,B Giant Molecular Cloud}
\label{sec:orionb}

\TabOrionB{} %

\subsection{Targeted field of view}
\label{sec:RGB}

Figure~\ref{fig:rgb} displays a composite image of the \twCO{} (blue),
\thCO{} (green), \CeiO{} (red) \Jone{} peak-intensity main-beam
temperatures. It shows the south-western edge of the Orion\,B molecular
cloud. This region samples the interaction of the molecular cloud with at
least 4 \Hii{} regions. First, \object{$\sigma$Ori} is an O9.5V star that
illuminates the western edge of the \object{Orion\,B} cloud. It creates the
\object{IC\,434} nebula from which the \object{Horsehead} pillar
emerges. Second, \object{NGC\,2023} and \object{NGC\,2024} are two younger
\Hii{} regions embedded in the Orion\,B molecular cloud, powered by B1.5V
(\object{HD\,37903}) or late O, early B (IRS2b) stars,
respectively~\citep{bik03}.  NGC\,2024 covers 20 arcminutes at the northern
edge of the mapped field of view, while NGC\,2023 is situated approximately
halfway between IRS2b and the Horsehead. The B5V \object{HD\,38087} star
creates the \object{IC\,435} nebula at the south-eastern edge of the field
of view. Finally, one of the 3 Orion\,Belt stars, the O9.71b star
\object{Alnitak} (a.k.a. $\zeta$Ori), falls in the observed field of
view. Table~\ref{tab:stars} lists the characteristics of these exciting
stars. To guide the eye, we overlaid on the right panel of
Fig.~\ref{fig:rgb} crosses at the position of the main exciting stars, and
circles at the approximate boundaries of the different \Hii{}
regions. These visual markers will be used throughout the paper.

\subsection{IRAM-30m observations}
\label{sec:observations}

The observations were taken with the IRAM-30m telescope in four observing
runs: August 2013, December 2013, August 2014, and November 2014 during
133~hours in total (telescope time) under average summer weather (6\mm{}
median water vapor) and good winter weather (3\mm{} median water
vapor). During all these runs, we observed with a combination of the 3\mm{}
sideband separated EMIR receivers and the Fourier transform spectrometers,
which yields a total bandwidth of $\sim 32\GHz$ per tuning (\ie{}, $\sim
8\GHz$ per sideband and per polarization) at a channel spacing of 195\kHz{}
or $0.5-0.7\kms$. The two tuned frequencies were 102.519\GHz{} and
110.000\GHz{} at the 6.25\MHz{} intermediate frequency of the upper
sideband, resulting in local oscillator frequencies of 96.269 and
103.750\GHz, respectively. This allowed us to observe nearly the entire
3\mm{} band from 84.5 to 115.5\GHz{}.

We used the on-the-fly scanning strategy with a dump time of 0.25 seconds
and a scanning speed of $17''/$s to ensure a sampling of 5 dumps per beam
along the scanning direction at the $21.2''$ resolution reached at the
highest observed frequency, \ie{}, 116\GHz. We covered the full field of
view ($\sim 0.9$ square degrees) with $\sim103$ tiles of $\sim 110''\times
1000''$ size. The rectangular tiles had a position angle of $14\deg$ in the
Equatorial J2000 frame to adapt the mapping strategy to the global
morphology of the Western edge of the Orion\,B molecular cloud. These tiles
were covered with rasters along their long axis (almost the Dec
direction). The separation between two successive rasters was $\lambda/2D =
8.46''$ to ensure Nyquist sampling perpendicular to the scanning
direction. The scanning direction was reversed at the end of each line
(zigzag mode). This implied a tongue and groove shape at the bottom and top
part of each tile. We thus overlapped by $30''$ the top and bottom edges of
the tiles to ensure a correct sampling. On the other hand, the left and
right edges of the tiles were adjusted to avoid any overlap, \ie{}, to
maximize the overall scanning speed. The field of view was covered only
once by the telescope, except for the tiles observed in the worst
conditions (low elevation and/or bad weather) that were repeated once.

The calibration parameters (including the system temperature) were measured
every 15 minutes. The pointing was checked every two hours and the focus
every 4 hours. Following~\citet{mangum07} and~\citet{pety08}, we used the
optimum position switching strategy. A common off reference position was
observed during 11 seconds every 59 seconds with the following repeated
sequence OFF-OTF-OTF-OFF. No reference position completely devoid of
\twCO{} \Jone{} emission could be localized in the close neighborhood of
the Orion\,B western edge. As this reference position is subtracted to
every OTF spectrum in order to remove the common atmospheric contribution,
the presence of signal in the reference position results in a spurious
negative contribution to the signal everywhere in the final cube. Searching
for a reference position farther away in the hope that it is devoid of
signal would degrade the quality of the baseline because the atmospheric
contribution would vary from the OTF spectra to the reference position.  We
thus tested several nearby potential reference positions using the
frequency-switched observing mode that does not require a reference
position. This is possible because the observed lines have narrow
linewidth. We then selected the nearest position that has the minimum line
integrated emission in \twCO{} \Jone{}. Offsets of this position are
$(-500'',-500'')$ with respect to the projection center given in
Table~\ref{tab:l1630}. The \twCO{}, and \thCO{} \Jone{} peak intensities at
this position are $\sim 1$ and 0.05\K, respectively. The correction of the
negative contribution from the reference position to the final cube
requires a good observation of the reference position. We therefore
observed this reference position using the frequency-switched observing
mode in both tunings, a few minutes per observing session.

\subsection{IRAM-30m data reduction}
\label{sec:reduction}

Data reduction was carried out using the \texttt{GILDAS}\footnote{See
  \texttt{http://www.iram.fr/IRAMFR/GILDAS} for more information about the
  GILDAS softwares~\citep{pety05a}.}\texttt{/CLASS} software. The data were
first calibrated to the \Tas{} scale using the chopper-wheel
method~\citep{penzias73}. The data were then converted to main-beam
temperatures (\Tmb{}) using the forward and main-beam efficiencies (\Feff{}
and \Beff{}) listed in Table~\ref{tab:obs}. The \Beff{} values are derived
from the Ruze's formula
\begin{equation}
  \Beff{}(\lambda) =
  \Beff^{0}\,\exp\cbrace{-\paren{\frac{4\pi\,\sigma}{\lambda}}^2}
\end{equation}
\begin{equation}
  \mbox{with} \quad
  \Beff^{0} = 0.863,
  \quad \mbox{and} \quad
  \sigma = 65.6\mum,
\end{equation}
where $\lambda$ presents the wavelength dependence\footnote{The values of
  $\Beff^{0}$ and $\sigma$ can be found at
  \texttt{http://www.iram.es/IRAMES/mainWiki/Iram30mEfficiencies}.}. The
resulting amplitude accuracy is $\sim10\%$. A 12 to 20\MHz{}-wide subset of
the spectra was first extracted around each line rest frequency. We
computed the observed noise level after subtracting a first order baseline
from every spectrum, excluding the velocity range from 0 to 18\kms{} LSR,
where the gas emits for all observed lines, except \twCO{} and \HCOp{}
\Jone{} for which the excluded velocity range was increased from -5 to
20\kms{} LSR. A systematic comparison of this noise value with the
theoretical noise computed from the system temperature, the integration
time, and the channel width, allowed us to filter out outlier spectra
(typically 3\% of the data).

To correct for the negative contribution from the reference position to the
final cube, we first averaged all the observations of the reference
position 1) to increase the signal-to-noise ratio of the measured profiles,
and 2) to decrease the influence of potential calibration errors. Signal in
the reference position was only detected for the \twCO{} and \thCO{}
\Jone{} lines. The correction was thus applied only for these two
lines. The averaged spectra at these frequencies were fitted by a
combination of Gaussians after baseline subtraction, in order to avoid
adding supplementary noise in the final cube. This fit was then added to
every on-the-fly spectrum.

The spectra were then gridded into a data cube through a convolution with a
Gaussian kernel of $\emr{FWHM}\sim1/3$ of the IRAM-30m telescope beamwidth
at the rest line frequency. To facilitate comparison of the different line
cubes, we used the same spatial (pixels of $9''$ size) and spectral (80
channels spaced by 0.5\kms) grid. The position-position-velocity cubes were
finally smoothed at the common angular resolution of $31''$ to avoid
resolution effects.

\subsection{Map of visual extinction and dust temperature from Herschel and
  Planck data}
\label{sec:ancillary}

In this paper, we will observationally check the potential of line
intensities and of ratios of line intensities to characterize physical
properties of the emitting gas. Ancillary data are thus needed to deliver
independent estimates of these physical properties. We will use recent dust
continuum observations to provide estimates of the column density of
material and of the far UV illumination.

After combining the Herschel Gould Belt Survey~\citep{andre10,schneider13}
and Planck observations~\citep{tauber11} in the direction of Orion\,B,
\citet{lombardi14a} fitted the spectral energy distribution to yield a map
of dust temperature and a map of dust opacity at 850\mum{}
(\tauk). \citet{hollenbach91} indicates that the equilibrium dust
temperature at the slab surface of a 1D Photo-Dissociation Region (PDR) is
linked to the incident far UV field, \Gz{} at $\Av = 0$, through
\begin{equation}
  \label{eq:G0}
  \Td = 12.2\,\Gz[0.2]\K,
\end{equation}
where the \Gz{} value is given in units of the local interstellar radiation
field~\citep[ISRF,][]{habing68}. We will invert this equation to give an
approximate value of the far UV illumination. This value is likely a lower
limit to the actual \Gz{} in most of the mapped region. Indeed, it is the
far UV field at the surface of the PDR, while there are embedded \Hii{}
regions in the field of view. However, \citet{abergel02} estimates a
typical $\Gz \sim 100$ for the western edge of L\,1630, which is a large
scale edge-on PDR. Using Eq.~\ref{eq:G0}, this value is compatible with the
typical dust temperature fitted towards this edge, \ie, about 30\K{}.

\citet{lombardi14a} compared the obtained 850\mum{} opacity map to an
extinction map in the K band (\Ak) of the region. A linear fit of the
scatter diagram of \Ak{} and \tauk{} give for Orion\,B $\Ak = 3460\,\tauk$
(\citet{lombardi14a} name this factor $\gamma$). They used a value of
$\Ak/\Av = 0.112$ from \citet{rieke85}. However, this value, including
their estimated $\Rv \simeq 3.1$, is not based on observations towards
Orion stars. \citet{cardelli89} measured the properties of dust absorption
\Ebv{} and \Rv{} towards two stars of our field of view. Using their
parametrization, we yield $\Ak/\Av = 0.1254$ for $\Rv = 4.11$ towards
HD\,37903, and $\Ak/\Av = 0.1335$ for $\Rv = 5.30$ towards HD\,38087. We
here take an average of both values, $\Ak/\Av = 0.13$, \ie, a 20\% larger
value than \citet{rieke85}. We therefore have
\begin{equation}
  \Av = 2.7\e{4}\,\tauk\magn.
\end{equation}
The dust properties (both the temperature and visual extinction) are
measured at an angular resolution of $36''$.

\subsection{Noise properties, data size, percentage of signal channels,
  line integrated intensities}
\label{sec:data:cube}

The median noise levels (computed on the cubes that have 0.5\kms{} channel
spacing and $31''$ angular resolution) range from 100 to 180\mK{} (\Tmb{})
depending of the observed frequency. Details can be found in
Appendix~\ref{sec:noise}.
The reduced data cube amounts to about 160\,000 images of $325\times435$
pixels or 84\,GB of uncompressed data. It would make a movie of 1h50m at 24
images per second. However, about 99.5\% of the channels show mostly noise
because of the limited sensitivity of our observation. The 0.5\% of the
bandwidth where clear signal is detected includes the emission from low $J$
lines of CO, \HCOp{}, HCN, HNC, CN, CS, SO, \CCH, \cCCCHH, \NNHp{},
\methanol{}, SiO, and some of their isotopologues, in particular, CO
isotopologues (see Table~\ref{tab:luminosity} and Fig.~\ref{fig:area}).

Most of this paper will study the properties of the line integrated
intensity defined as $W = \int T(v)\,dv$. To produce reliable spectral line
maps, we used all the pixels matching two conditions: 1) its own
signal-to-noise ratio is larger than 4, and 2) the signal-to-noise ratio of
at least 25\% of its neighbors are larger than 4. Residual striping may be
seen along the vertical scanning direction in particular at low intensity
on the images of line integrated intensities. Indeed, baselining corrects
for the striping to first order. Hence residual striping is more visible
for faint lines and/or lines for which the overlapping of the velocity and
the hyperfine structures require the definition of wider baselining
windows, \eg{}, for the \HCN{} ground state transition. Finally, we used
two different flavors of the integrated intensity.  On one hand, we will
use the line intensity integrated over the full line profiles when we aim
at studying the gas properties along the full line of sight. This will
happen, for instance, when we will compute the CO-traced mass in
Section~\ref{sec:mass} and the correlations between the column density of
material along the full line of sight and the line integrated intensity in
Section~\ref{sec:NH:vs:W}. On the other hand, lines are detected over
different velocity ranges. Using the same velocity range for all lines,
\eg, $[-2,+18\kms]$, will result in noisy integrated intensities for
tracers that have the narrowest lines. In contrast, adapting the velocity
range to each line could bias the results. We thus adopted a compromise for
sections where we can restrict our investigations to the bulk of the gas:
We computed the line integrated intensity over the velocity range where the
core of the line can be found for each species and transition over the
measured field of view. This velocity range is $[9,12\kms]$.

\section{Mean properties}
\label{sec:mean:prop}

\FigAreaIma{} %

From this section on, we will only study the properties of the \Jone{} line
for the CO isotopologues (\twCO, \thCO, \CeiO, and \CseO), \HCOp, \HCN,
\HNC, and their \chem{^{13}C} isotopologues, \twCN, \CCH, and \NNHp{}, as
well as the \Jtwo{} transition for \twCS{}, \ttSO, \methanol, and SiO.

\subsection{Geometry, spatial dynamic, typical visual extinction, dust
  temperature, and far UV illumination}
\label{sec:typical}

Table~\ref{tab:l1630} lists the typical properties of the observed field of
view. At a typical distance of $\sim400\pc$~\citep{menten07,schlafly14},
the mapped field of view corresponds to $5.6 \times 7.5\pc$. This
corresponds to a surface of $43\pc^2$. Assuming that the depth along the
line of sight is similar to the dimension projected on the plane of sky, we
get a volume equal to the surface at the power 3/2, or $280\pc^3$.

The angular resolution ranges from 22.5 to $30.4''$ at 3\mm{} while the
typical 30m position accuracy is $\sim 2''$. All the cubes were smoothed to
$31''$ angular resolution, \ie, $60\mpc$ or $\sim 10^5\au$. We will thus
explore a maximum spatial dynamic range of 125 for all the lines.

The visual extinction ranges from 0.7 to 222\magn{} with a mean value of
4.7\magn{}. This is associated to a range of \twCO{} \Jone{} integrated
intensity from 0 to 288\Kkms{} with a mean value of 61\Kkms. In other
words, the field of view contains all kind of gas from diffuse without CO
emission to highly visually extinct with bright CO emission, but most of
the gas is in the higher end of the translucent regime $(2 \la \Av \la 6)$.

The SED-fitted dust temperature along the line of sight ranges from 16 to
99\K{} with a mean value of 26\K{}. This translates into a typical far UV
illumination, \Gz, ranging from 4 to $3.6\e{4}$ using the Inter-Stellar
Radiation Field (ISRF) definition by~\citet{habing68}. The \Gz{} mean value
is 45. This confirms that the observed field of view is on average strongly
far UV illuminated by the different massive exciting stars listed in
Table~\ref{tab:stars} (see Section~\ref{sec:RGB}).

\subsection{Distribution of line integrated intensities}
\label{sec:moment:1}

Figure~\ref{fig:area} presents the spatial distribution of the line
integrated intensities. We also added the spatial distribution of the dust
temperature at the bottom left panel and the visual extinction at the top
right panel to give reference points on the underlying nature of the gas
that emits each line tracer (see Section~\ref{sec:ancillary}).

The spatial distributions of the molecular lines presented here are
different. The \Jone{} line of the CO isotopologues themselves show a very
different behavior. The line of the rarer isotopologue, \CseO{}, has a
spatial distribution that is similar to that of the \NNHp{} \Jone{} line,
which is a known tracer of the cold and dense regions in molecular
clouds~\citep{bergin07}. Indeed, \NNHp{} and \CseO{} are seen only towards
lines of sight of high extinction ($\Av>20-30$). The \Jone{} line emission
of the slightly more abundant isotopologue \CeiO{} is more extended and
clearly traces the dense and cold filaments of the cloud. Moreover, the
\CeiO{} \Jone{} emission is similar to the extinction map shown in
Fig.~\ref{fig:area}, a property consistent with the known linear
correlation of \CeiO{} \Jone{} integrated intensity with the visual
extinction~\citep{frerking82}. The \Jone{} emission of the second most
abundant CO isotopologue, \thCO{}, traces gas in the extended envelope
surrounding the filaments traced by the \CeiO{} \Jone{} emission.  The
emission of the main CO isotolopogue no longer traces the dense gas and it
is largely dominated by the extended and more diffuse or translucent gas
because it is then strongly saturated.

The \HCOp{}, HCN and HNC \Jone{} lines are usually considered to be good
tracers of dense molecular gas because of their high spontaneous emission
rates and large critical densities. Among these three species, the \HNC{}
\Jone{} map bears the closest resemblance with the \CeiO{} \Jone{}
map. Emission in the ground state lines of \HCOp{} and \HCN{} exhibits a
more extended component and it looks more like the \thCO{} \Jone{} map.
All three lines as well as CN, present bright emission towards
high-extinction lines of sight. Their emission also seems to trace the
edges of the \Hii{} regions. A clear difference between CN and the other
N-bearing species and \HCOp{} is the larger contrast between the warmer
northern region, near NGC\,2024, and the cooler southern region near the
Horsehead. In contrast to their main isotopologue, the \chem{^{13}C}
isotopologue of \HCOp, \HCN, and \HNC{} are only clearly detected towards
the dense cores.  The methanol emission is slightly more extended than the
\NNHp{} \Jone{} emission but it is clearly seeded by the dense cores as
traced by \NNHp{}.


The emission of the sulfur-bearing species, in particular the \twCS{}
\Jtwo{} line, has similar spatial distributions as that of the \CeiO{}
\Jone{} line. Finally, the SiO \Jtwo{} line is only detected at the
position of two previously known outflows. The first one is located at the
South-West of NGC\,2023 around the class-0 NGC\,2023 mm1 protostars located
at \radec{05}{41}{24.9} {-02}{18}{09}~\citep[J2000,][]{sandell99}. The
second one is located on both sides of the FIR5 young stellar object
located at
\radec{05}{41}{44.6}{-01}{55}{38}~\citep[J2000,][]{richer90,chernin96},
near the center of NGC\,2024.  This confirms that SiO is before all a shock
tracer.

\FigMeanSpectra{} %
\TabLuminosity{} %

\subsection{Mean line profiles over the observed field of view}
\label{sec:mean:spectra}

\FigTwoComponents{} %
\FigHalo{} %

Figure~\ref{fig:mean:spec} shows the spectra of the main detected lines
averaged over the mapped field of view. Several spectra show multiple
components for different reasons. First, the multi-peak nature of the
\CN{}, \HCN{}, \CseO{} and \NNHp{} ground state lines is a consequence of
the resolved hyperfine structure of these transitions. Second, the faintest
spectra (\eg, \HNthC{}) are detected at low signal-to-noise ratio, implying
a noisy profile. Finally, the western side of the Orion\,B cloud displays
two velocity components: The main one around 10.3\kms{} and a satellite
one, ten times fainter, around 4.9\kms{}. The two components have similar
linewidth $(3-4\kms)$ and they overlap between 5 and
9.5\kms. Figure~\ref{fig:two:components} displays the fit for these two
components on the \twCO{} and \thCO{} \Jone{} spectra averaged over the
field of view. The values of the \twCO/\thCO{} integrated line intensity
ratios are 6.4 and 11.8 for the main and the satellite components
respectively. The difference in line ratios suggest that the satellite
velocity component corresponds to lower column density material (see
Section~\ref{sec:line:ratios:CO}). The satellite component is barely
detected in \CeiO{}, \HCOp{}, and \twCS{}, and it stays undetected for the
other lines. Figure~\ref{fig:halo} shows the \twCO{} \Jone{} emission in
two channel maps belonging to the two velocity components, both in linear
and logarithmic color scales. The use of a logarithmic transfer function
shows that bright emission is surrounded by a halo of faint emission. This
shows that the fainter velocity component still covers a large fraction of
the observed surface.

We now argue that both velocity components along the line of sight are
associated with the Orion\,B Giant Molecular Cloud. Figure~\ref{fig:halo}
shows that the spatial distribution of both components overlap on most of
the observed field, and that they are in close interaction with the massive
stars of known distance, listed in Table~\ref{tab:stars}. Furthermore, the
3-dimensional structure of interstellar extinction has been studied
by~\citet{lallement14} and~\citet{green15} using differential reddening of
stars at known distances. Towards Orion\,B, the reddening steeply increases
between 300 and 500\pc{}, and most importantly, there is no significant
reddening detected at closer or larger distances in this direction of the
sky (Lallement, priv. comm.). These results are in excellent agreement with
the distance determination of Orion through maser
parallax~\citep{menten07}.  Overall the 3D structure of the Orion clouds is
complex and could extend over several tens of parsec along the line of
sight, a dimension comparable to the projected size on the plane of the
sky.

Table~\ref{tab:luminosity} lists the integrated line intensities, \W{}, and
luminosities, $L$, computed as
{\tiny
  \begin{equation*}
    \frac{L}{\Lsol} = 3.4\times10^{-8} 
                      \paren{\frac{3\mm}{\lambda}}^3 \paren{\frac{\W}{1\Kkms}}
                      \paren{\frac{D}{400\pc}} \paren{\frac{\Omega}{1'\times1'}},
  \end{equation*}}
where $k_\emr{bolt}$ is Boltzmann constant, $\lambda$ the line rest
wavelength, $D$ the source distance, and $\Omega$ the field-of-view
angle. The dynamic range of reliable integrated intensity is about
2\,400. Moreover, the typical intensity ratios of the \Jone{} lines would
be $\twCO/\thCO = 6.7$, $\thCO/\CeiO = 16.7$, $\CeiO/\twCO = 0.9\%$, and
$\HCOp/\twCO \sim \HCN/\twCO \sim 3\%$.  \twCN, \CeiO{}, \twCS{}, \CCH{},
\HNC{} emit about 1\% of the \twCO{} intensity. The low-J lines of \ttSO{},
\CseO{}, \cCCCHH{}, \NNHp{}, \HthCN{}, \HthCOp{} are up to 25 times fainter
than the previous family, exemplified by \CeiO. The \HNthC{} and \SiO{}
integrated intensity can not be reliably measured.

\subsection{CO-traced, dust-traced, and virial-traced mass and densities}
\label{sec:mass}

In this section, we will compute the typical gas mass and densities using
three common approaches: 1) the \twCO{} \Jone{} luminosity, 2) the dust
continuum luminosity, and 3) the virial theorem. Table~\ref{tab:l1630}
lists the found values.

The direct sum of the pixel intensity over the mapped field of view and
between the $[-2,+18]\kms$ velocity range indicates that the data cube
contains a total CO luminosity of $\sim 2\,500\Kkmspcpc$. Using the
standard CO-to-\HH{} conversion factor, $\Xco = 2.0 \times
10^{20}\pscmpKkms$ or $4.35\Msol\pc^{-2}/(\Kkms{})$~\citep[this includes
the factor 1.36 to account for the presence of helium,][]{bolatto13}, this
corresponds to a gas mass of $1.1\e{4}\Msol$. The total surface covered was
0.86 square degree, \ie{}, $43\pc^2$. The mean intensity and mean surface
density are $61\Kkms$ and $260\Msol\pc^{-2}$, respectively. The associated
column density of gas is about $10^{22}\,\HH\pscm$. This in turn gives a
typical volume density of $40\Msol\pc^{-3}$ or 590\,\HH\pccm{}.

Using the Gould Belt Survey data and its SED fits, we can derive values for
the same quantities from dust far infrared emission. To do this, we first
used in section~\ref{sec:ancillary} a value different from the standard one
for the conversion factor from \Ak{} to \Av{} because 1) this is an
observational quantity that can be measured relatively easily, and 2) we
mainly deal with molecular gas, while the standard value is derived in
diffuse gas. While this value depends on the optical properties (grain
composition, grain shapes, and size distribution, which leads to the
extinction curve) of the dust in Orion\,B, it is independent of any
assumption about the gas properties. On the other hand, to derive the dust
traced mass, we also need to use a value for the \NH/\Av{} ratio. Assuming
that the dependency of this ratio on the dust optical properties is only a
second order effect, this ratio mainly depends on the gas-to-dust ratio,
\ie, on how many grains there are per unit mass of gas. We thus use the
standard value, $\NH/\Av = 1.8 \times 10^{21}\,\H\pscm\magn^{-1}$, for this
ratio. This directly leads to a dust-traced mass of the mapped field of
view of $\sim3\,900\Msol$, a mean surface density of $92\Msol\pspc$ or
$4\times10^{21}\,\HH\pscm$, and a mean volume density of $14\Msol\pcpc$ or
$210\,\HH\pccm$.

The column density of neutral atomic hydrogen measured by integrating
across profiles of the 21\cm{} \Hi{} line taken by the LAB all-sky \Hi{}
survey~\citep{kalberla05} is
\begin{equation*}
  N(\Hi) = 1.823 \times 10^{18} \pscm  \int T_B dv \approx (1.8 \pm 0.2)\times10^{21}\pscm 
\end{equation*}
in the optically thin limit. This corresponds to approximately $1.0\magn$
of visual extinction using the usual conversion $\NH/\Ebv = 5.8 \times
10^{21}\pscm\,\magn^{-1}$ derived by \citet{bohlin78} and $\Rv = \Ebv/\Av =
3.1$.  The total expected foreground gas contribution for a source at a
distance of 400\pc{} is $\NH= 1.2 \times 10^{21}\pscm$ for a local mean gas
density $\langle n(\H)\rangle = 1.15 \pccm$~\citep{spitzer78} corresponding
to $\Av = 0.75\magn$ using the same conversion from column density to
extinction. The minimum value of the visual extinction across the observed
field of view, (\ie, 0.7) is therefore in good agreement with the expected
contribution of diffuse material along the line of sight. As the mean
visual extinction is 4.7\magn, correcting for this diffuse component would
result in decreasing the molecular part of the dust-traced mass and
densities by less than 20\%. We choose to consider this difference
negligible, \ie{}, to consider that all the dust-traced mass refers to gas
where hydrogen is molecular.

Following~\citet{solomon87} and~\citet{bolatto13}, we can also compute a
mass assuming that turbulent pressure and gravity are in virial
equilibrium. \citet{bolatto13} indicate that the virial mass, \Mvir{}, is
given by
\begin{equation}
  \Mvir = f \, \R \, \dv^2,
\end{equation}
where \R{} is the projected radius of the measured field of view, \dv{} is
the 1D velocity dispersion (full width at half maximum of a Gaussian
divided by 2.35), and $f$ a factor that takes into account projection
effects. This factor depends on the assumed density profile of the GMC. For
a spherical volume density distribution with a power-law index $k$, \ie,
\begin{equation}
  \rho(r) \propto r^{-k}, 
\end{equation}
$f$ is 1\,160, 1\,040, and $700\unit{\Msol\pc^{-1}(\kms)^{-2}}$, for $k=0$,
1, and 2, respectively. In our case, $\R \simeq 0.5\,\sqrt{5.6 \times 7.5}
= 3.3$, and $\dv \simeq 3.9/2.35 = 1.7\kms,$ when we only take into account
the Gaussian fit of the main velocity component around 10.5\kms. We thus
obtain a virial mass between 6\,200 and 9\,500\Msol.

We find that, contrary to expectations, the CO-traced mass is typically 3
times the dust-traced mass, and that the virial mass is lower than the
CO-traced mass but it is much higher than the dust-traced mass. Throughout
the paper, we will propose that this discrepancy is related to the strong
far UV illumination of the mapped field of view (see
Section~\ref{sec:typical}).  In the meantime, we will take an average
between the CO-traced and dust-traced mass and densities when we will need
an order of magnitude estimate for these quantities.

\section{Fraction of line fluxes from different gas regimes}
\label{sec:flux}

\TabAvMaskProp{} %
\TabAvMaskedFlux{} %
\FigMaskedSpectraFluxAv{} %
\FigLineFluxAv{} %

In this section, we will explore which fraction of the line fluxes comes
from more or less dense gas, and from more or less far UV illuminated gas.

\subsection{Flux profiles over different \Av{} ranges}
\label{sec:flux:av}

We chose 4 ranges of \Av{}, representing diffuse $(1\le\Av<2)$, and
translucent $(2\le\Av<6)$ gas, the environment of filaments ($6\le\Av<15$),
and dense gas ($15\le\Av$). Table~\ref{tab:av:mask:prop} lists the physical
properties of the different regions based on their \twCO{} \Jone{} and far
infrared emission. While the different regions have by construction
increasing values of their mean visual extinction (1.4, 4, 9, and 29,
respectively), they present similar mean dust temperature and far UV
illumination. As expected the minimum dust temperature decreases when the
range of visual extinction increases. In contrast, the maximum dust
temperatures, and thus far UV illuminations, are also found in the masks of
highest visual extinctions. This is related to the presence of very dense
(probably cold) molecular gas in front of young massive stars that excite
\Hii{} regions (see, \eg, the dark filament in front of IRS2 that excites
the NGC\,2024 nebula). This could also be due to the presence of embedded
heating sources.

Contrary to standard expectations, the dust and CO-traced mass are similar
for diffuse and dense regions, while they differ by a factor 3 mostly in
the translucent gas and filament environment. Moreover, both the dust and
CO-traced matter indicate that about 50\% of the gas lies in diffuse and
translucent gas. Dense cores $(15\le\Av)$ represent between 10 and 20\% of
the mass but only 3\% of the surface and 0.6\% of the volume. The sum of
the volume fractions of the 4 regions only amounts to 55\% because of the
simplified way the volumes are computed $(V=S^{3/2})$. This implies that
volume densities can only be interpreted as typical values. Finally, the
volume densities increase from $\sim100$ to $7\,300\,\HH\pccm$ for diffuse
and dense gas, respectively. Translucent gas and the environment of
filaments have typical density values of $\sim 500$ and $1\,500\,\HH\pccm$,
respectively. The volume density increases by a factor $3-5$ from each gas
regime to the next. We will use this fact to statistically identify
high/low \Av{} lines of sight with high/low density gas, respectively.

Figure~\ref{fig:spec:mask:flux:Av} presents the \Av{} masks and it displays
the flux profiles integrated over regions of different extinction ranges.
To better quantify the different behavior of the fluxes integrated over
these different regions, Table~\ref{tab:av:mask:flux} and
Fig.~\ref{fig:line:F:Av} present, for each line, the percentage of the
total line flux that comes from the different \Av{} masked regions
$(F_{\Av}/F_\emr{tot})$. In all cases, the fluxes are integrated in the
$[9, 12\kms]$ velocity range. The lines were sorted by increasing value of
the $F(1\le\Av<2)+F(2\le\Av<6)$ ratio. This value represents for each line
the flux coming from both diffuse and translucent gas. The layout of the
panels in Fig.~\ref{fig:spec:mask:flux:Av} also follows this order. We can
group the lines in 4 categories depending on how the line flux is divided
between regions of very low ($1\le\Av<2$), low ($2\le\Av<6$), intermediate
($6\le\Av<15$), or high $(15\le\Av)$ visual extinction.

In the first category of lines, the regions of low and intermediate visual
extinctions contribute more than $\sim45\%$ of the total flux, and regions
of high visual extinction contributes less than $\sim20\%$ of the flux.  In
this category, the total flux is predominantly coming from translucent
lines of sight $(2\le\Av<6)$. This is the case of the \Jone{} lines of
\twCO, \HCOp{}, \CCH{}, and the \cCCCHH{} (\T{2}{1,2}{1}{0,1}) line. From
these species, \twCO{} is the one with the largest contribution (55\%) from
diffuse and translucent gas ($\Av\le6$).

In the 2nd category, the total flux is now predominantly coming from
regions of intermediate visual extinction $(45\%$ coming from
$6\le\Av<15)$. But the diffuse and translucent gas still contributes for a
similar fraction $(35-40\%)$ of the total flux, and dense gas do not
contribute more than 20\% of the total flux. This is the case of the
\Jone{} lines of \thCO{}, HCN, and CN.

In the third category, the flux comes predominantly from regions of
intermediate visual extinction as in the 2nd category. But the regions of
low and high visual extinctions both contribute for similar fractions of
the total flux (around 30\%). The \Jone{} lines of HNC, \CeiO{}, \CseO{},
and the lines of the sulfur species, namely the \Jtwo{} line of \twCS, and
\ttSO{}, belong to this category.

The \Jone{} lines of \NNHp{} and \HthCOp{}, as well as the \Jtwo{} lines of
\methanol{} form the last category. In this one, the flux is predominantly
coming from the regions of high visual extinctions $(15\le\Av)$. These
lines all present a small surface filling factor $(\la5\%)$ and negligible
contribution from the translucent and diffuse gas. In this category,
\NNHp{} plays a special role. This is the only easily mapped line, where
the flux is completely dominated (at 88\%) by regions of high visual
extinctions, probably dense cores.

\subsection{Flux profiles over different \Td{} ranges}
\label{sec:flux:tdust}

\TabTdustMaskProp{} %
\TabTdustMaskedFlux{} %
\FigMaskedSpectraFluxTdust{} %
\FigLineFluxTdust{} %

We chose 4 ranges of \Td, representing cold dust $(16 \le \Td < 19.5\K)$
that corresponds to gas that is shielded from the UV field (\eg, the dense
cores), lukewarm dust $(19.5 \le \Td < 23.5\K)$, warm dust $(23.5 \le \Td <
32\K)$, and hot dust $(32 \le \Td < 100\K)$. It is clear that this fitted
dust temperature is biased toward the presence of warm dust because the
dust emissivity increases rapidly with the temperature in the far
infrared. Hence, cool dense gas is probably present along the line of sight
of highest extinction, even though the fitted dust temperature is
relatively high.

We here use the dust temperature as a proxy for the typical far UV
illumination along the line of sight (see Section~\ref{sec:ancillary}). In
fact, using Eq.~\ref{eq:G0}, we obtain that the mean far UV illumination is
9, 18, 50, and 400 for the cold, lukewarm, warm, and hot dust masks,
respectively. This implies very different kinds of PDRs present along the
line of sight. Moreover, the 4 masks of dust temperature display a
morphology very different from that of the masks of visual extinction. Only
the dense cores in Horsehead and near NGC\,2023 are clearly delineated in
both families of masks, while the intermediate density filamentary
structure and diffuse/translucent gas are present in all 4 masks of dust
temperature. Instead, the morphology of these temperature masks coincides
well with the boundaries of the different \Hii{} regions. We thus interpret
the cold, lukewarm, warm, and hot dust masks as very low, low, medium, and
high far UV illumination masks.

The field of view is dominated by intermediate far UV illumination PDRs
(83\% of the surface have a \Gz{} between 10 and 120). Less than 2\% of the
lines of sight have $\Gz \le 10$ and about 15\% have $\Gz > 120$.
Moreover, the CO-traced mass is 0.86, 2.6, 3.1, and 3.5 times the
dust-traced mass in the cold, lukewarm, warm, and hot dust regions. Similar
ratios are found for the volume densities. This confirms that the
discrepancy between CO and dust traced mass is linked to the enhanced far
UV illumination of the South-Western edge of Orion\,B.

Figure~\ref{fig:spec:mask:flux:tdust} presents the \Td{} masks and it
displays the flux profiles integrated over regions of different far UV
illumination.  To better quantify the different behavior of the fluxes
integrated over these different regions, Table~\ref{tab:tdust:mask:flux}
and Fig.~\ref{fig:line:F:Tdust} presents, for each line, the percentage of
the total line flux that comes from the different \Td{} masked regions
$100\,(F_{\Td}/F_\emr{tot})$. In all cases, the fluxes are integrated in
the $[9, 12\kms]$ velocity range. The lines were sorted by decreasing
distance between the sum of the flux coming from the highest far UV
illumination regions (red and orange masks) and the sum of the flux coming
from the lowest far UV illumination regions (green and blue masks).  The
layout of the panels in Fig.~\ref{fig:spec:mask:flux:tdust} also follows
this order. While oscillations on the 4 individual curves of
Fig.~\ref{fig:line:F:Tdust} are present, the general tendency is that the
percent of flux coming from the highest far UV illuminated regions
decreases from top to bottom. We can thus group the lines in 4 categories
depending on whether the line flux comes predominantly from the very low,
low, medium, or high far UV illumination regions.

In the first category, the regions of high far UV illumination $(\Gz>27)$
contribute about 70\% of the total line flux and the region of very low
illumination $(\Gz<10)$ contributes less than 5\%. High $(\Gz>120)$ and
medium $(27<\Gz<120)$ illumination regions contribute about equally to the
total flux. The fundamental lines of the \CCH{}, \cCCCHH{}, \twCN{}, and
\HCN{} belong to this category.

In the second category, containing the \HCOp{}, \HCN{} and \twCO{} \Jone{}
lines, the line flux comes predominantly $(65-75\%)$ comes from
intermediate far UV illumination regions $(10<\Gz<120)$. The highest
illumination region still contributes for $20-30\%$ of the total flux,
while lowest illumination region contributes for less than 10\%.

In the third category, the flux comes first from the region where
$10<\Gz<27$. Quantitatively, this is the category where the flux coming
from $\Gz<27$ starts to dominates compared to medium and intermediate
illumination regions. The \Jone{} line of \thCO{} and \HthCOp{}, as well as
the \Jtwo{} line of \twCS{}, and \ttSO{} belong to this category.

In the last category, the flux coming from regions where $\Gz<27$
contributes between 52 and 66\% of the total flux. This contains the
\methanol{} \Jtwo{}, and the \Jone{} line of the rarest CO isotopologues
and \NNHp{}.

\section{Molecular low-J lines as a probe of the column density}
\label{sec:NH:vs:W}

\subsection{Visual extinction vs. line integrated intensities}
\label{sec:Av:vs:W}

\FigAvVsArea{} %

Figure~\ref{fig:av:vs:area} presents the joint distributions of the visual
extinction and line integrated intensities for the studied molecular
tracers. As the visual extinction is proportional to the amount of matter
along the line of sight, it is desirable to make a comparison of all the
matter traced by the molecules along this line of sight. Hence, the line
profiles are integrated over the full velocity range where the line is
measured, \ie{}, not just integrated anymore on the [9,12\kms] velocity
range. The visual extinctions are defined over the full field of view. In
contrast, the line integrated intensities are well defined for only a
fraction of the field of view. The joint distributions were thus only
computed where the line integrated intensities are well defined (the
criteria can be found in Section~\ref{sec:data:cube}). For each
distribution, the additional statistics (in particular for the visual
extinction) are computed on this fraction of
points. Table~\ref{tab:corr:area:av} lists these statistics.

The first obvious trend in Fig.~\ref{fig:av:vs:area} is the global
correlation between the visual extinction, \Av, and the line integrated
intensities, \W. This correlation is clearly visualized through the
comparison of the black curves, which show the typical behavior of the
variations of \Av{} as a function of \W{}, with the white lines that
represent a linear relation between these two quantities. While the lines
are often overly bright with respect to the white line at low extinction,
and their integrated intensity sometimes saturate at high visual
extinction, a correlation is clearly present for a large fraction of the
measured lines of sight between these two regimes. More precisely, the
\CeiO{}, and \HNC{} \Jone{} lines are the best tracers of the visual
extinctions when the integrated line intensity is above 1\Kkms{}. Indeed,
there is an excellent agreement between the black line and the white line
when \W{} is above the intensity median value, and the scatter is low
around these curves for both transitions. The \CeiO{}, and \HNC{} \Jone{}
lines are followed by \ttSO{} and \twCS{} \Jtwo{} lines. But these start to
show a second twofold behavior at high visual extinction, a fraction of the
pixels showing a saturation of the line integrated intensity at high visual
extinction. This saturation branch is amplified for the \HCN{}, \twCN{},
and \CCH{} \Jone{} lines. The \thCO{} \Jone{} line is also a good tracer of
the visual extinction, as it has a clearly monotonic (though non-linear)
relationship with low scatter from $\Av \sim 2$ to $\Av \sim 20$.

The second trend concerns the visual extinction thresholds at which the
lines become clearly detected. Lines that are detected over a smaller
fraction of the mapped field of view show up at a higher \Av{} than lines
with a more extended spatial distribution. Moreover, this threshold
behavior is amplified when the position-position-velocity cubes are not
smoothed at a common angular resolution in the first place. We will
emphasize two particular examples. First, the \NNHp{} \Jone{} line has a
surface filling factor of 2.4\% and it is detected at a median visual
extinction of 26, while the filling factor of the \HCOp{} \Jone{} is 68\%
and this line is detected at a median visual extinction of 4.4, close to
the median visual extinction at which \twCO{} \Jone{} is emitted. Second,
this \Av-threshold behavior is also clear for the suite of CO
isotopologues, where \twCO{} and \thCO{} \Jone{} are detected at visual
extinctions even lower than 1, while \CeiO{} and \CseO{} \Jone{} are mostly
detected for visual extinctions above 3 and 6, respectively. The obvious
explanation is related to detection limits. Rarer isotopologues produce
weaker lines per unit column density, hence require a larger total gas
column density to produce a signal above the detection threshold.

However, the \Av{}-flat asymptotes at low values of the integrated
intensities are also evidence that it is not just a detection problem.
Indeed, a linear relation is expected between the visual extinction and the
integrated intensity at low values, \ie, in the optically thin regime.  The
linear trend should thus just be interrupted at the detection threshold. In
contrast, there is a \Av{}-threshold above which the species starts to
emit. This is corroborated by the fact that intensity ratios do not match
the values expected from the known carbon isotope ratios, even at low
visual extinction where optical depth effects are negligible. The
\Av-thresholds could either be explained by chemical or dynamical
reasons. Turbulent mixing between the phases of the ISM or the existence of
dense but diffuse globulettes at the edge of \Hii{} regions belong to the
latter category. In the former category, we have selective chemistry.

In summary, the \Jone{} or \Jtwo{} lines of molecular tracers are to first
order sensitive to different range of visual extinction when detected at a
similar noise level. In addition, they are overall well correlated with the
amount of matter along the line of sight. This behavior will be quantified
in another paper about the Principal Component Analysis of the
dataset~\citep{gratier16}.

\TabCorrAreaAv{} %
\FigAreaOverAv{} %
\clearpage{} %

\subsection{Tracer luminosities per proton}
\label{sec:luminosity:per:proton}

Figure~\ref{fig:area:over:av} shows the spatial distribution of the ratio
of the line integrated intensity to the visual extinction. The panels show
these ratios for the molecular tracers ordered in the same way as the
figure displaying the line integrated intensities (Fig.~\ref{fig:area}). We
also added the spatial distribution of the visual extinction and dust
temperature as the top right and bottom left panels, respectively, for
reference. The intensity ratios are normalized by their median values and
the intensities are displayed using a logarithmic scale symmetrically
stretched around 1. This eases the visualization of departure of the ratio
by a multiplicative factor, \eg{}, 1/2 and 2. The luminosity per proton is
easily computed by dividing the $\W/\Av$ ratios by the standard value of
$\NH/\Av = 1.8 \times 10^{21}\,\H\pscm/\magn$.

The $\W(\twCO)/\Av$ and $\W(\thCO)/\Av$ present a similar pattern, \ie{}, a
luminosity per proton higher than the median value in translucent gas and
lower in dense gas. The luminosity per proton decreases again at the very
edge of the molecular cloud. The $\W(\CeiO)/\Av$ ratio shows less variation
by a factor 2 to 3. The $\W(\HCOp)/\Av$, $\W(\HCN)/\Av$, $\W(\HNC)/\Av$,
and $\W(\twCN)/\Av$ ratios show maxima associated with the Orion\,B Eastern
edge and with the NGC\,2024 \Hii{} bubble. The dark filament in front of
NGC\,2024 delineates the frontier between ratios higher/lower than the
median.  This can be interpreted as an excitation effect due to higher
electron density or an abundance effect. The West/East asymmetry of the
ratio is more marked for the $\W(\CCH)/\Av$ ratio, in particular around the
NGC\,2023 region.  The $\W(\twCS)/\Av$ shows a specific pattern with a
maximum at the center of NGC\,2024 and a minimum between NGC\,2024 and
NGC\,2023.

\subsection{Typical abundances}
\label{sec:typical:abundances}

\TabPseudoAbundance{}%

As the low-J molecular lines are overall well correlated to the column
density of molecular gas, the luminosities per proton could in principle be
used to estimate the abundance of the different species. To do this, we
computed the column density of each species, $N_\emr{species}$, that is
required to produce an integrated intensity of 1\Kkms{} assuming that the
gas is at local thermal equilibrium. The values of $N_\emr{species}$ vary
by less than 20\% when the temperature increases from 20 to 30\K{}. Typical
abundances with respect to the proton number can then be computed with
\begin{equation}
  [\emr{species}] = \frac{\W}{\Av} \, \frac{\Av}{\NH} \, \frac{N_\emr{species}}{1\Kkms}.
\end{equation}
Table~\ref{tab:abundance} lists the minimum, maximum, and median values of
the so-called abundances for each line. The deduced abundances are
reasonable for all the studied lines except \twCO{} \Jone{}, which delivers
abundances too low by one order of magnitude, because this line is highly
optically thick.

\section{Line ratios as tracers of different physico-chemical regimes}
\label{sec:line:ratios}

\FigCorrTwCO{} %
\FigCorrThCO{} %

Line intensity ratios are commonly used to study the physical and chemical
properties of the gas in different environments. The advantage of using
line ratios instead of absolute line intensities is that it allows to
remove calibration uncertainties (when lines are observed
simultaneously). It then is easier to compare from source to source. Line
ratios may also remove excitation effects and bring forward actual chemical
variations. Our knowledge of the chemistry of the gas then allows us to use
line ratios to constrain the physical properties of the gas.

An important basic property we wish to determine easily from observations
is the density of the gas. Forming dense gas is a required step to form
stars, and the availability to form dense gas may regulate star formation
efficiency~\citep{lada13}. Line ratios of HCN and \HCOp{} with respect to
\twCO{} and \thCO{} are commonly used to trace the fraction of dense gas in
galactic and extragalactic GMCs~\citep[\eg,][]{lada12}. This is because
\twCO{} and \thCO{} can be excited at low densities ($\sim10^2\pccm$)
compared to HCN and \HCOp{}, which are expected to be excited only at high
densities ($\sim10^5\pccm$). Indeed, the HCN/\twCO{} ratio is observed to
be well correlated with the star formation efficiency, traced by IR/HCN in
M51~\citep[\eg,][]{bigiel16}.

In this section we first show line ratios involving the brightest detected
lines, \ie{}, \twCO{} and \thCO{} \Jone{}, and we conclude with a few other
interesting ratios. For this, we discuss 2D-histograms of the ratio
denominator vs. the ratio numerator, the spatial distribution of the
ratios, and the 2D-histogram of the ratio vs. the visual extinction. This
will allow us to study the correlations present before computing the ratio,
to visually assess the correlations of the line ratios with different kinds
of regions, and to quantitatively study potentially remaining correlations
with the visual extinction.

\vspace*{-0.1cm}

\subsection{Ratios with respect to \twCO{} and \thCO{} \Jone{}}
\label{sec:line:ratios:CO}

The 2D-histograms shown in Fig.~\ref{fig:area:vs:12co10} display the
relation of the integrated intensity of different lines as a function of
the integrated intensity of the \twCO{} \Jone{} line. While the eye is
mainly caught by the saturation of the \twCO{} line, \ie{}, the fact that
other tracer intensity increases by a large factor when $\W(\twCO) \sim
100-200\Kkms$, most of the data follows a different trend.  The running
median and running interval containing 50\% of the data, materialized as
black points and error bars, indicate that most of the tracers have first a
relatively constant integrated intensity as $\W(\twCO)$ increases, and then
their integrated intensity is well correlated to $\W(\twCO)$. As shown by
the white rectangles that display the part of the 2D-histogram populated by
90\% of the points, most of the tracers only emit when the \twCO{} \Jone{}
line is already quite bright at $\sim 10-20\Kkms$. On the other hand, the
CCH, \HCN{}, \HCOp{}, and \thCO{} \Jone{} lines show a significant fraction
of the data at \twCO{} \Jone{} integrated intensity between 1 and
10\Kkms. The \thCO{} \Jone{} line has a specific behavior as it is
under-luminous with respect to a linear correlation going through the
median behavior at intermediate \twCO{} \Jone{} intensities $(5 \la
\W(\twCO) \la 30\Kkms)$.

Figure~\ref{fig:area:vs:13co10} shows the same 2D-histograms as before but
with respect to $\W(\thCO)$. In general, the same trends seen for \twCO{}
are seen for \thCO{}. The species integrated intensities have a relatively
constant or slightly increasing integrated intensity as $\W(\thCO)$
increases up to $\sim 10\Kkms$. Their intensity is then well correlated to
$\W(\thCO)$. The effects of the \thCO{} \Jone{} saturation are visible but
less pronounced than with respect to the \twCO{} \Jone{} line.

Figure~\ref{fig:area:over:12co10} presents the spatial distribution of the
line ratios involving \twCO{}. The ratios are normalized by their median
value to emphasize the symmetric departure of the ratios compared to the
general trend. The ratios all show minimum values in the dense regions
associated with NGC\,2024, NGC\,2023, and Horsehead. This probably reflects
the saturation of the \twCO{} emission in regions with the highest column
density. These regions are also the densest regions, which implies that the
molecular tracers are easily produced and excited. %

\FigRatioTwCO{} %
\FigRatioThCO{} %
\FigAvVsRatioTwCO{} %
\FigAvVsRatioThCO{} %
\clearpage{} %

\noindent{} As the \twCO{} line becomes saturated, the available energy
gets carried away by other \twCO{} transitions or other molecular
species. A West-East gradient is superimposed on the first pattern for the
ratios involving \HCOp{}, \HCN{}, \CCH{}, \twCN{}, and \HNC{} (not shown),
for both \twCO{} and \thCO{}.  The minimum values are obtained on the
Western edge, the maximum values in the Eastern diffuse region. This
pattern probably indicates a gradient in excitation and abundance in UV
illuminated regions for molecules sensitive to the far-UV
radiation. Finally an approximately circular structure around NGC\,2024
with a luminosity deficit in \CeiO, \twCS{} and marginally \ttSO{} probably
traces UV illuminated material near NGC\,2024.

Similar spatial behavior is also seen for the ratios with respect to
\thCO{} \Jone{} (see Fig. \ref{fig:area:over:13co10}), but slightly
attenuated because the saturation of the \thCO{} \Jone{} line is less
pronounced. The first pattern (minimum ratio values in regions of highest
density) for the ratios including \CeiO{}, \twCS, and \ttSO{} supports the
interpretation in terms of opacity for the densest/brightest regions. The
East-West pattern is even more pronounced for the \HCOp{} species with an
excess emission of the molecular tracers in the UV illuminated regions and
a deficit in the diffuse/translucent gas. This may be a combined effect of
lower heating, moderate density, and an increase of \thCO{} due to isotopic
fractionation $(\twCO + \thCp \longrightarrow \thCO +\twCp)$. Finally, the
East-West pattern does not reach the translucent regions on the Eastern
side for \CCH, and CN. In these cases, we mostly see the increase of the
line ratio in the high extinction gas, including the compressed Western
edge~\citep{schneider13}.

Finally, Fig.~\ref{fig:av:vs:ratio:12co10} shows the 2D-histograms of \Av{}
as a function of the line ratios involving \twCO{}. The line ratios have a
bimodal behavior relative to \Av{}, with values lower than the median
(marked by the white cross) found both for high and low \Av{} regions.
Values of the ratios higher than the median are associated with a small
range of visual extinctions, either the translucent ($2<\Av<6$) or
filamentary ($6<\Av<15$) gas. The bimodal trend is present in all ratios,
but more pronounced for those involving lines presenting an extended
emission ($\thCO/\twCO$ and $\HCOp/\twCO$).

The increasing branch (in \Av{}) is the dominating one for the ratios
involving \thCO{}, \CeiO{}, \twCS{}, \ttSO, and \twCN{}. This means that
the other decreasing branch, while existing, represents a small number of
points in our field of view. Low values of these ratios thus mostly point
to high density regions. The increasing branch is most likely a consequence
of the saturation of the \twCO{} \Jone{} line compared to the other lines
at large gas column densities. In addition, many molecular species will
become more abundant at large column densities, as they become shielded
from UV-radiation. Both effects will produce lower ratios at large
\Av{}. Because all tracers are correlated to first order to the column
density, one would expect to remove a correlation with \Av{} by taking the
ratio of two lines. However, a (anti-)correlation may remain. This is due
to the fact that the correlation between the integrated intensity of weaker
lines and $\W(\twCO)$ have a non-linear behavior, probably because these
lines have a lower opacity than the \twCO{} \Jone{} line for large gas
column densities.

In contrast, the decreasing branch dominates for the ratios involving the
\HCOp{}, \HCN{}, and \CCH{} \Jone{} lines. Low values of these ratios point
to the lower visual extinction range. This is clear for $\twCO/\CCH$, whose
running median almost monotonically increases from an \Av{} of $\sim 2$ to
10\magn{}. When the $\twCO\Jone/\HCOp\Jone$ and $\twCO\Jone/\HCN\Jone$
ratios increase, the running median of the visual extinction first
increases from values lower than $2\magn$ up to $\sim8\magn$, and it then
starts to decrease again. One interesting result is that all decreasing
branches sample values of the visual extinction as low as $1-2\magn$. This
is consistent with the fact that all the associated species are detected in
diffuse clouds through absorption against extra-galactic continuum
sources~\citep{lucas96,lucas00,liszt01}. This probably means that the
strongly polar species, \CCH{}, and \HCOp{}, reach a radiative regime where
they emit more efficiently than \twCO{}, the weak excitation
regime described by \citet{liszt16}.

Figure~\ref{fig:av:vs:ratio:13co10} shows the 2D-histograms of the visual
extinction as a function of the ratio involving \thCO{}. We see similar
global results as for \twCO{}, \ie{}, a globally constant visual extinction
at values of the ratio higher than the median, and an increasing and a
decreasing \Av{} branches when the ratio decreases below the median
value. In contrast with the ratios involving \twCO{} \Jone{}, the
decreasing \Av{} branch dominates for most of the ratios. Indeed, the
running median for all ratios but $\CeiO/\thCO$ decrease for low visual
extinctions. The decrease is even almost monotonic for ratios involving the
\HCOp{}, \CCH, \twCN, and \HCN{} \Jone{} lines. This behavior is consistent
with the fact that the correlation between the integrated intensity of the
different species and $\W(\thCO)$ is more linear. Taking the ratios thus
better removes the correlations with the gas column density.

\TabWyOverWx{} %

\FigCorrDense{} %
\FigRatioDense{} %
\FigAvVsRatioDense{} %
\clearpage{} %
\subsection{Various other line ratios}
\label{sec:line:ratios:dense}

Figure~\ref{fig:area:vs:dense} to~\ref{fig:av:vs:ratio:dense} show the same
plots as the previous section but for other line ratios including tracers
of the column density (\CeiO{}, \twCS{}, and \HNC{}), and other highly
studied ratios in extra-galactic observations (\HCOp{}, \HCN{}, \HNC{}, and
CN).

The 2D-histograms that display the relation of the integrated intensity of
\CseO{}, \HNC{}, and \twCS{} with respect to $\W(\CeiO)$ show two main
behaviors (see Fig.~\ref{fig:area:over:dense}). The \CseO{}, \HNC{}, and
\twCS{} \Jtwo{} lines are over-luminous compared to the \CeiO{} \Jone{}
line at low intensities. Their intensities then become linearly
correlated. In addition, the \twCS{} \Jtwo{} line becomes much brighter
than both the \CeiO{} and \HNC{} \Jone{} lines at high intensity
values. The integrated brigthnesses of the HCN-\HCOp{}, \HNC-\HCOp,
\HNC-\CN{}, and HNC-HCN \Jone{} line pairs are all linearly correlated. The
best correlation is found for the \HCN{}-\CN{} pair of lines.

The spatial patterns of the ratios shown in Fig.~\ref{fig:area:over:dense}
are less obvious to describe because the fraction over the field of view
where the two lines are detected at enough signal-to-noise ratio is
lower. The $\CeiO/\CseO$ ratio, shown in the upper left panel, is fairly
flat with no clear spatial pattern. The East-West pattern is particularly
marked on the $\CeiO/\twCS$ and $\HNC/\CeiO$ ratios. The $\twCS/\HNC$,
$\twCN/\HNC$, and $\HCN/\HNC$ ratios all show an approximately circular
structure around NGC\,2024 with a deficit of \HNC{} integrated
intensity. We relate the latter behavior to an isomerisation of \HNC{} into
\HCN{} when the gas temperature increases.

Computing the ratio for these lines removed almost any linear
(anti-)correlation with the visual extinction, except for the $\HCN/\HNC$
and $\CN/\HNC$ ratios (see Fig.~\ref{fig:av:vs:ratio:dense}). For the
filamentary gas (\Av{} between 6 and 15\magn) the ratio spans a large range
of values, up to one order of magnitude for $\CeiO/\twCS$ and $\CeiO/\HNC$.

\section{Discussion}
\label{sec:discussion}

\subsection{Typical line intensities in a strongly UV illuminated part of a
  GMC}
\label{sec:gmc:sampling}

\TabEnvironment{} %
\FigEnvironment{} %

The field of view sampled here is not a random 1 square degree part of any
GMC. The left panels of Fig.~\ref{fig:dust:env} show the spatial
distribution of the visual extinction and dust temperature over a much
larger fraction of the Orion\,B molecular cloud than the one presented in
this paper, which is shown as the black rectangle. The right panels compare
the Probability Distribution Functions (PDFs) of dust properties over our
field of view with the PDFs of two other regions with same surface
area. Table~\ref{tab:dust:env} lists the minimum, median, and maximum
values of the associated distributions, as well as their 5 and 95\%
quantiles. This clearly shows that three different kinds of environment
exist in the Orion\,B molecular cloud. First, the lowest dust temperatures
are associated with relatively high visual extinctions (red
rectangle). Second, the blue rectangle identifies a translucent region (all
pixels have $\Av \la 8$), associated with a typical dust temperature of
about 18\K. In both cases, the distribution of visual extinction and dust
temperature are single peaked with a narrow full width at half maximum. In
contrast, our field of view displays wide \Av{} and \Td{} distributions,
and it is associated with the highest dust temperatures with a median value
of $\sim24\K$ $(\Gz\sim30)$. The presence of high gas temperatures is
confirmed by the large \twCO{} \Jone{} peak temperatures that are lower
limits of the kinetic temperature~\citep{orkisz16}.  These properties are
associated with the presence of at least four \Hii{} regions (see
Section~\ref{sec:RGB}) that imply a large UV illumination (see the fourth
column of Table~\ref{tab:dust:env}). In particular, the minimum dust
temperature in our field (16.4\K) is rather high in Orion\,B compared to
the Taurus molecular cloud~\citep{marsh14}.

Table~\ref{tab:luminosity} indicates that, under these sampling conditions
and at the typical sensitivity achieved in studies of nearby galaxies
($3\sigma = 1\Kkms$), only the \Jone{} line of \twCO, \thCO, \HCOp{},
\HCN{} would easily be detected by a single-dish radio-telescope of
30m-diameter with a single-beam receiver. A 10 times better sensitivity
(100 longer integration) is required to detect the \Jone{} or \Jtwo{} lines
of HCN, CN, \CeiO, \twCS, \CCH, \HNC, \ttSO, \CseO, and \cCCCHH{}. Finally,
another order-of-magnitude increase of the sensitivity ($3\sigma =
0.01\Kkms$) would be needed to detect \NNHp{}, \methanol{}, \HthCOp, and
\HthCN. This means that detecting rare isotopologues of \HCOp{}, \HCN{}, or
\HNC{} in nearby galaxies is difficult to achieve with a single-dish
radio-telescope because of the dilution of the signal in the beam.

\subsection{On the influence of the UV field on the determination of
  molecular mass}
\label{sec:co:luminosity}

\FigTwCOprop{} %

The average visual extinction and CO integrated intensity for the observed
field of view are 4.7\magn{} and 61\Kkms{}. This turns into a $\Wco/\Av =
13.0\Kkms/\magn$, while the standard \Xco{} factor, \ie,
$2\times10^{20}\,\HH\pscm/(\Kkms)$, corresponds to $4.7\Kkms/\magn$ when we
assume a standard $\NH/\Av$ factor and fully molecular gas.  The \Hi{}
emission indicates that diffuse gas accounts for about 1 magnitude of
extinction towards Orion\,B (see Section~\ref{sec:mass}). Assuming that
contribution from atomic hydrogen to the mass is negligible towards the
mapped field of view overestimates the dust-traced molecular mass by 27\%,
increasing the discrepancy between the CO and dust-traced mass. Therefore,
we neglect this subtlety and we directly compare the CO and dust-traced
mass. We find that the CO-traced mass (and thus the associated surface and
volume density) is about 3 times higher than the dust-traced mass.

The origin of this discrepancy lies in the intense UV illumination of the
gas by massive stars. The bottom left panel of Fig.~\ref{fig:12co10:prop}
compares the spatial distribution of the CO integrated intensity per visual
extinction. The standard value of 4.7\Kkms/\magn{} corresponds to the
transition between yellow and green. Only diffuse gas or the UV shielded
dense gas have \Xco{} values close or lower than standard. This is
confirmed by the joint distribution of the \Wco{} and \Av{} where most of
the points (lines of sight with $2\la \Av \la 15$) lie above the white line
of slope 4.7\Kkms/\magn{}. When the visual extinction increases, the CO
intensity saturates. At the lowest visual extinction ($\Av \la 2$), CO is
destroyed into \chem{C^{+}}. The spatial distribution of \Gz{} clearly
shows that most of the gas lies in regions with $\Gz > 10$, the mean value
of \Gz{} being 45. Under such conditions, dust and gas are heated to higher
temperature than in the standard interstellar radiation field. In the
physical conditions of Orion\,B, the CO emission per \HH{} molecule is
increasing with the kinetic temperature, leading to a possible bias in the
mass determination. The bottom right panel of Fig.~\ref{fig:12co10:prop}
shows that the CO intensity per \Av{} increases with \Gz{}.

While it is tempting to conclude that only the CO traced mass is widely
overestimated in such conditions, the dust-traced mass is in fact also
underestimated, as indicated by the range of virial mass that we estimated
from the field of view size and CO linewidth (see
Table~\ref{tab:l1630}). It is known that using a single dust temperature to
fit the spectral energy distribution on a line of sight that contains dust
at different temperatures hides the presence of cold dust along the line of
sight, because the luminosity of dust increases extremely fast with its
temperature~\citep{shetty09}.

All in all, the typical volume density we infer for the regions, \ie{},
between 200 and $600\,\HH\pccm$, is typical of galactic Giant Molecular
Clouds~\citep{heyer15}. While the local values of $\NH/\Av$ and \Xco{} are
uncertain, and they could well be different from their standard values in
Orion\,B, we here wish to study Orion\,B as if it was observed from nearby
galaxies. In these studies, standard values are used when the metallicity
is similar to that of the Milky Way~\citep[\eg, see the PAWS
project][]{schinnerer13,pety13}. From a practical viewpoint, we thus
proceeded with standard values, knowing that the correct result is probably
in between the CO-traced and dust-traced masses, surface densities, and
volume densities.

Enhanced far UV fields heat large masses of CO gas that turns over-luminous
with respect to the standard \Xco{} factor, \ie, the average behavior of
the CO gas in our Milky way~\citep{bernard11}. This effect could compensate
for the presence of CO-dark gas~\citep{wolfire10,grenier15}, as proposed by
\citet{liszt12} with different observations.  Therefore the standard value
of \Xco{} may well be applicable to galaxies with a higher than average,
yet moderate, massive star formation rate. An easy check would be to
increase the size of the mapped field of view to the full Orion\,B cloud to
test at which scale the CO and dust-traced masses derived with standard
values of \Xco{} and the \NH/\Av{} ratio get reconciled to better than a
factor 3. Complementary \textsc{Cii} observations would also help to settle
this point~\citep{goicoechea15}.

\TabFillingFactor{} %
\TabRatioComparison{} %

\subsection{Dense gas tracers}
\label{sec:dense:gas:tracers}

The brightness of a molecular line depends on the column density of the
species, which is affected by the chemistry, and the excitation properties
of the line. These in turn depend on the physical properties of the gas
(density, temperature, ionization fraction, ...). Indeed, two conditions
must be satisfied for a line to be detected. First, the molecule has to
abundant enough, and second, the excitation conditions must be favorable
for the line to be excited and produce bright line emission. It is
therefore often assumed that lines with high critical densities, such as
the \Jone{} lines of HCN and \HCOp{}, are good tracers of dense gas because
these species are abundant and their emission is expected to be seen only
in regions where the density is high enough to excite the
line. Table~\ref{tab:filling:factor} lists the critical density of each
molecular line as well as the percentage of total flux that is emitted from
regions of intermediate $(6\le\Av<15)$ and high $(15\le\Av)$ visual
extinction, as measured in Section~\ref{sec:flux:av}. These two regimes are
representative of the gas arising in filaments and dense cores,
respectively. We do not find any clear correlation between the critical
density of the lines and the percentage of flux emitted.

For instance, the lines of the CO isotopologues have nearly equal critical
densities (\ie{}, similar excitation conditions) but the percentage of flux
coming from the densest regions $(n_\HH \sim 7\,300\pccm)$ varies from 8\%
for \twCO{} \Jone{} to 29\% for \CeiO{} \Jone. The higher fraction of flux
coming from high density regions for the rarer isotopologues is the result
of both sensitivity and chemistry. The intrinsic lower abundance of
\thCO{}, \CeiO{} and \CseO{} compared to the main isotope will result in
weaker emission for the rarer isotopologues everywhere in the cloud. Also,
in the UV-illuminated layers \twCO{} will survive longer than the CO
isotopologues due its capacity to self-shield (process known as selective
photo dissociation), although this effect is probably not resolved by the
observations.

Lines with much higher critical densities ($>10^{5}\pccm$) than CO \Jone{}
($\sim10^{3}\pccm$), such as \HCOp{}, \HCN{}, and \HNC{} \Jone{}, which are
expected to trace dense gas, emit only 16, 18, and 27\% of their total flux
in mapped Orion\,B regions of high density. Most of the emission then
arises in lower density gas. \citet{shirley15} extensively discusses the
relevance of the notion of critical density. He reminds that the critical
density is the density at which collisional deexcitation equals the net
radiative emission. He emphasizes that it is computed in the optically thin
limit, implying that it is only an upper limit in the presence of
photon-trapping. The fact that these lines can be excited and thus detected
in diffuse gas was in addition recently discussed by~\citet{liszt16}. While
their results can not be quantitatively applied to the observed field of
view because the line peak intensities are slightly outside the range of
applicability of the weak excitation regime, the underlying physical
process is still present. At the limit of detectability, \citet{liszt16}
showed that the intensity of low-energy rotational lines of strongly polar
molecules, such as \HCOp{} and \HNC, is proportional to the product of the
total gas density and the molecule column density, independent of the
critical density, as long as the line intensity does not increase above a
given value. This implies that for any given gas density there is a column
density that will produce an observable line intensity. In the observed
field of Orion\,B, we are in an intermediate radiative transfer regime
where the fundamental lines of \HCOp, \HCN, and \HNC{} can be excited in
regions of density much smaller than their critical densities.

The lines with the largest fraction of their emission arising in the
densest gas are \methanol{} and \NNHp{}. In particular, \NNHp{} \Jone{}
that presents a similarly high critical density as \HCOp, \HCN, and \HNC{}
\Jone, has the highest proportion of its flux coming from the densest
regions (88\%). Moreover, in contrast to all the other lines, only 17\% of
the total \NNHp{} \Jone{} flux is associated with regions of intermediate
\Av{} ($6\le\Av<15$) which have typical densities of $\sim
1\,500\pccm$. This shows that the \NNHp{} \Jone{} line is the best
molecular tracer of dense regions among the lines studied in this paper.

Despite the similar critical densities between \NNHp{} and \HCOp{}, their
behavior is completely different. First, \NNHp{} is detected over only
2.4\% of the observed field of view, while \HCOp{} emission covers 68\% of
the field. Second, the percentages of the flux coming from
dense/translucent regions are 88/8\% for \NNHp{}, and 15/41\% for
\HCOp{}. These differences can only be understood by their different
chemistry. \HCOp{} can be formed from ion-molecule reactions involving
C$^+$ and other cations, notably CH$^+_2$ and CH$^+_3$, in addition to the
protonation of CO. On the contrary, the sole reaction producing \NNHp{} is
the protonation of \NN{}. Furthermore, the destruction of \HCOp{} by
dissociative recombination with electrons produces CO while \NNHp{} can
react with CO to produce \HCOp{}. Hence, \NNHp{} only survives in regions
where the electron abundance is low (to prevent dissociative recombination)
and where CO is frozen on dust grains (to prevent proton transfer to CO),
\ie, in cold and dense cores.

In summary, species that have a chemical reason to only be present in dense
gas are the only really reliable high density tracers. More generally, the
knowledge of the chemical behavior is fundamental in understanding how
molecular species can be used to trace the different physical environments.

\subsection{Typical line ratios in Orion\,B and other galaxies}
\label{sec:typical:ratios}

\FigRatioComparison{} %

Molecular line ratios have the potential to be powerful probes of physical
properties related to star formation activity. Thanks to the current
observing capabilities, many unbiased line surveys have recently been made
towards nearby galaxies, and more than 27 species have been detected in the
3\mm{} band~\citep{meier12,salas14,
  watanabe14,aladro15,meier15,nishimura16}. Common line ratios include the
\HCN\Jone/\HCOp\Jone{}, \twCO\Jone/\HCN\Jone, \HCN\Jone/\HNC\Jone{} and the
CN\Jone/\HCN\Jone, which are proposed tracers of density, temperature and
radiation field, respectively.

Figure~\ref{fig:ratio:comparison} shows observed line ratios in nearby
galaxies and in the Orion\,B molecular cloud. The comparison includes line
ratios obtained with both single-dish telescopes and interferometers, and
galaxies with distances between 3.3 and 170\Mpc, in addition to the LMC,
the nearest external galaxy (50\kpc). The sample therefore spans two orders
of magnitude in spatial resolution (written at the bottom of the y-axis in
units of pc). To ease the comparison with Orion\,B, the ratios are
normalized by the Orion\,B ratios of the lines integrated over the full
observed field of view (\ie, at a resolution of 10\pc).

Line ratios observed in Orion\,B are, in general, comparable to what is
observed in nearby galaxies. Noticeably, line ratios (except \twCO/\NNHp)
in Orion\,B at a resolution of 10\pc{} are very similar to those observed
at a resolution of 1000\pc{} in the spiral arm of the famous whirlpool
galaxy (M\,51), a prototype of grand design spiral galaxy.

The \HCN{}/\HCOp{} ratio (shown in red in the upper panel) is assumed to
trace dense gas that will eventually form stars, and thus is often used to
trace the star formation activity in other galaxies. Both lines are among
the brightest lines observed and are thus easily detected. ULIRGs and AGNs
present the largest \HCN/\HCOp{} ratios, but there are no major differences
between the sources. The resolved spatial distribution in Orion\,B (see
Fig.~\ref{fig:area:over:dense}) shows that in the case of UV-illuminated
gas, the \HCN/\HCOp{} ratio does not trace the high density gas. Another
proposed tracer of dense gas is the inverse of the \twCO/HCN ratio (shown
in blue in the upper panel). This ratio is higher in Orion\,B, the LMC and
in the spiral arm of M51 than in the other galaxies. The \twCO/\HCOp{}
ratio behaves in a similar way. While the resolved spatial distribution in
Orion\,B (see Fig.~\ref{fig:area:over:12co10}) shows that these ratios
actually separate diffuse from dense gas, the quantitative analysis shows
that \HCOp{} and \HCN{} \Jone{} fluxes mostly trace densities around
$500-1500\,\HH\pccm$ instead of $\sim10^4\pccm$.

Better tracers of high-density gas are ratios involving \NNHp, such as
CO\Jone/\NNHp\Jone{} (shown in magenta), because \NNHp{} resides solely in
dense gas ($>10^4\pccm$), contrary to \HCN{} that can be present at lower
densities (see Section~\ref{sec:flux:av}).  Starbursts (including M\,51)
and ULIRGs forms many stars, requiring the presence of many dense cores,
and thus a high \NNHp{} \Jone{} brightness relative to \twCO{} \Jone{} that
traces the total reservoir of molecular gas. In contrast, Orion\,B has a
low star formation efficiency~\citep{lada10,megeath16}, probably implying a
low number of dense cores and thus a lower relative \NNHp{} brightness.
The low surface filling factor of \NNHp{} makes it difficult to detect at
high signal-to-noise ratio with single-dish telescopes in external
galaxies. Fortunately, the much better resolving power of NOEMA and ALMA
relieves this difficulty and \NNHp{} starts to be detected in nearby
galaxies.

The \HCN{}/\HNC{} is another popular line ratio measured in nearby galaxies
(shown in green in the bottom panel). Both species are abundant in cold
clouds, but at temperatures higher than about 30\K, \HNC{} starts to be
converted to \HCN{} through reactions with H~\citep[see,
\eg,][]{schilke92,graninger14}. The HCN/HNC ratio is thus increasing at
higher temperatures. The sources included in this comparison present
similar HCN/HNC ratios within a factor of 2. The spatial distribution shown
in Fig.~\ref{fig:area:over:dense} is consistent with the proposed
temperature dependence of this ratio, whose lowest values are found in the
cold filaments, and which correlates well with the map of the dust
temperature seen in Fig.~\ref{fig:area}.

Finally, the CN/HCN intensity ratio (shown in red in the bottom panel) is
expected to trace UV-illuminated gas because CN is a major product of HCN
photodissociation. Indeed, higher {\it abundance} ratios are found in PDRs
and XDRs compared to cold dark clouds~\citep{fuente05,baan08}. However, in
spatially resolved observations, there is no clear correlation of the
CN/HCN flux ratio with radiation field (see
Fig.~\ref{fig:area:over:dense}). Furthermore, in our comparison, AGNs and
starbursts present the largest CN/HCN ratios, and ULIRGs have similar
ratios to the Orion\,B cloud and the spiral arm of M51. This is
contradictory with CN/HCN being a radiation field tracer. In fact, the
ratios that involve the HCN \Jone{} line must be taken with care because
this line is known to be pumped by IR photons~\citep{aalto07}. This will
produce brighter HCN lines, and thus could explain the low CN/HCN ratio
observed in ULIRGs. A better tracer of radiation field is the
\CCH\Jone/\thCO\Jone{} ratio. Carbon chains, such as \CCH{} and \CCCHH,
have been observed to form efficiently in UV-illuminated
regions~\citep{pety05b,guzman15}. ULIRGs, AGNs and starburst galaxies
present the largest \CCH/\thCO{} ratios, consistent with their high star
formation activities. Moreover, the spatial distribution of the
\CCH/\thCO{} ratio (see Fig.~\ref{fig:area:over:13co10}) shows a clear
gradient between the illuminated edge (right) and the UV-shielded (left)
side of the Orion\,B molecular cloud. Therefore, the \CCH\Jone/\thCO\Jone{}
ratio is a potential tracer of the presence of massive stars.

\section{Conclusion}
\label{sec:conclusion}

The ORION-B project aims at imaging a statistically significant fraction of
the Orion\,B giant molecular cloud over the 3\mm{} atmospheric window,
starting with approximately 1 square degree towards NGC\,2024, NGC\,2023,
Horsehead, and IC\,434. The mean dust temperature in the mapped region is
26\K, and the probability distribution function of the visual extinction
shows a wide distribution from less than 1 to a few 100 magnitude, with
most of the surface and volume lying below 15 magnitude. These dust
properties suggests that the South-Western edge of Orion\,B is permeated by
far UV fields from massive stars at relatively large spatial scales. Most of
the cloud mass is contained in regions of relatively low \Av{}, implying
that photon-dominated regions are everywhere in the field of view. This is
the reason why CO is over-luminous, resulting in a CO-traced mass that is
about three time as much as the dust traced mass in this region.

Over the 84 to 116\GHz{} frequency range, we easily detected the
fundamental lines of CO isotopologues (from \twCO{} to \CseO), \HCOp, \HCN,
\HCN, and their $^{13}$C isotopologues, \twCN, \CCH, and \NNHp{}, as well
as higher J lines of \twCS{}, \ttSO{}, SiO{}, \cCCCHH, and methanol. The
faintest averaged spectra of the species presented in this article, the
spectra of \HNthC{} \Jone, is $\sim2\,400$ fainter than \twCO{} \Jone{}
line.  Still fainter lines, tentatively detected in the averaged spectra,
are present. They will be discussed in a future paper.

The main CO isotopologues and a few other species clearly display two
velocity components: The main component centered at $\sim10\kms$, where
most of the gas lies, and a more diffuse component centered at $\sim5\kms$
that contains about 10\% of the molecular gas along the line of
sight. Tomography studies of the interstellar medium using visible
absorption against background stars indicates that both components belongs
to Orion\,B: Their relative distance is compatible with the projected
extension of Orion\,B.

A significant fraction of the \HCOp{}, \HCN, and \HNC{} \Jone{} flux
integrated over the observed field of view is coming from translucent (41,
36, and 29\% respectively) and from gas forming the filamentary structure
(typically $40-45\%$). Only 16, 18, and 29\% (for \HCOp, HCN, and HNC,
respectively) of the flux is coming from dense cores ($\Av>15$). The common
assumption that lines of large critical densities ($\sim 10^5\pccm$) can
only be excited by gas of similar density is clearly incorrect. Another
unrelated result is that these lines are also sensitive to the amount of
far UV illumination, even though less sensitive than the fundamental line
of small hydrocarbon chains and \twCN{}.

While the \HCOp{} and \NNHp{} \Jone{} lines have similar critical densities
and similar peak temperatures over the mapped field of view, the
repartition of their flux have a completely different behavior: The \NNHp{}
\Jone{} line is emitted only from regions with $\Av>15$, while the \HCOp{}
\Jone{} line is also emitted from regions of visual extinction as low as
1-2 magnitude. This explains why the surface filling factor of detected
emission highly differs between these two species. This is in part due to
the fact that there exists another regime of radiative transfer (the weak
excitation limit) that produces detectable lines of strongly polar
molecules~\citep[][and reference therein]{liszt16}. The difference of
behavior between \HCOp{} and \NNHp{} can be easily explained by the way
these molecules are produced and destroyed in diffuse and dense gas.

We observe a strong correlation of the line integrated intensities with
\Av, \ie, the line strength increases with the quantity of material along
the line of sight. This correlation will be quantified in~\citet{gratier16}
through a Principal Component Analysis that allows us to find significant
correlations beyond this one with other physical parameters, like the UV
illumination. However, the best tracers of the column density are the
\CeiO{} and \HNC{} \Jone{} lines, followed by the \twCS{} \Jtwo{}
line. This validates the use of these lines to normalize the intensities in
extra-galactic studies. The different species clearly show \Av{} threshold
values above which they start to emit. Dividing the line integrated
intensities by the visual extinction enables us to remove this correlation
and to bring forward chemical differences between the species.

When the visual extinction is not available, making the ratio of the line
integrated intensities also emphasizes different chemistry
behavior. However, proposed dense-gas tracers such as the
HCN\Jone/\HCOp\Jone{} or HCN\Jone/\twCO\Jone{} line ratios should be
superseded by \NNHp\Jone{}/\twCO\Jone{} that more accurately
traces dense gas $\ga 10^4\pccm$. The fact that the CN\Jone{}/HCN\Jone{}
line ratio has a pretty flat spatial distribution in our field of view, may
be because the gas is permeated by far UV fields. However, we find that the
\CCH\Jone/\thCO\Jone{} line ratio is an excellent tracer of the variations
of the far UV illumination in our field of view.

The increasing capabilities of millimeter receivers make it possible to
observe multiple molecular lines in large fields of view, and thus to use
the spatial distribution of low-J line ratios to classify each line of
sight depending on their molecular content.

\begin{acknowledgements}
  We thank CIAS for its hospitality during the two workshops devoted to
  this project. We thank R.~Lallement for useful discussions about the
  distance of the gas in Orion\,B. We thank P.~Andre and N.~Schneider for
  giving us access to the Herschel Gould Belt Survey data, and M.~Lombardi
  for delivering his fit of the spectral energy distribution of the
  Herschel data.  We thank the referee, J.~G.~Mangum, for his careful
  reading of the manuscript and useful comments that improved the
  article. This work was supported by the CNRS program “Physique et Chimie
  du Milieu Interstellaire” (PCMI). JRG thanks MINECO, Spain, for funding
  support under grant AYA2012-32032. This research used data from the
  Herschel Gould Belt survey (HGBS) project
  (\texttt{http://gouldbelt-herschel.cea.fr}). The HGBS is a Herschel Key
  Programme jointly carried out by SPIRE Specialist Astronomy Group 3 (SAG
  3), scientists of several institutes in the PACS Consortium (CEA Saclay,
  INAF-IFSI Rome and INAF-Arcetri, KU Leuven, MPIA Heidelberg), and
  scientists of the Herschel Science Center (HSC).
\end{acknowledgements}

\bibliographystyle{aa} %
\bibliography{orionb} %

\newpage{}

\appendix{} %

\section{Noise properties}
\label{sec:noise}

\TabObservations{} %
\FigTsysVsFreq{} %
\FigNoise{} %

Table~\ref{tab:obs} summarizes the single-dish observations described in
section~\ref{sec:observations}. Figure~\ref{fig:tsys:vs:freq} shows the
average system temperature as a function of the rest frequency over the
observed bandwidth. The mean (solid line) was computed over all the
calibration measurements made during the observation and the shaded
backgrounds display the $\pm3\sigma$ interval that reflects both variations
in the quality of the tuning and of the elevation of the source. The
variations with the frequency come from a combination of hardware and
atmosphere chromaticity effects. The optic and mixer performances vary with
the observed radio-frequency and the cryogenic low noise amplifier
performances vary with the instantaneous intermediate frequency. The
atmospheric transparency mostly varies because of ozone lines present in
the bandpass and because of the 118.75\GHz{} (N=1-1, J=1-0) oxygen line
whose wings strongly increase the \Tsys{} at the upper end of the 3\mm{}
atmospheric window.

Figure~\ref{fig:noise} show the spatial distribution of the noise at an
angular resolution of $31''$ for two lines, each one representing one of
the two tunings. The region around the Horsehead and NGC\,2023 were
observed at the start of the project, before our observing strategy
converged to the one described in Section~\ref{sec:observations}. In
particular, this area was covered at least once along the $\delta x$ axis,
and once along the $\delta y$ axis. This explains why the noise properties
of this region are different from other parts of the map. Except for this
part of the map, the variations in the noise level reflect the variations
of weather or elevation during the observations.

\clearpage{} %

\section{Mean line profiles from different gas regimes}
\label{sec:mean}

In Section~\ref{sec:flux}, we derived the fraction of fluxes that come from
different gas regimes in the lines averaged over the observed field of
view. We here intend to deliver the tools to estimate the typical line
fluxes when varying the number of lines of sight that sample the different
gas regimes. We thus compute the mean line profiles over the same gas
regimes as defined in Sections~\ref{sec:flux:av}
and~\ref{sec:flux:tdust}. Mixing these line profiles in different
proportions would deliver a proxy for the line profile observed in GMCs
that have different ratios of diffuse/dense gas or cold/warm gas.

Figures~\ref{fig:spec:mask:mean:Av} and~\ref{fig:spec:mask:mean:tdust} show
the mean line profiles for the different \Av{} and \Td{} regimes.
Tables~\ref{tab:av:mask:mean} and~\ref{tab:tdust:mask:mean} present, for
each transition, line integrated intensity averaged over the different
\Av{} and \Td{} masked regions. Figure~\ref{fig:line:W:Av}
and~\ref{fig:line:W:Tdust} show these line integrated intensities in
percentage of the integrated intensity averaged over the full field of
view. These later plots indicate the potential intensity gain that can be
expected when resolving the regions that are compact.

As expected, the intensity increases with the visual extinction, \ie{} with
the column density of matter. However, the increase is only a factor 2 for
the \twCO{} \Jone{} line while it reaches a factor of almost 30 for the
\NNHp{} \Jone{} line. All lines except \NNHp{} \Jone{} have a mean
integrated intensity that increases by a factor 2 in the filamentary gas
$(6\le\Av<15)$. The mean spectra at 10\pc{} resolution are closer to the
mean spectra over the translucent gas for all the lines, but for the
\HthCOp{}, \methanol{}, and \NNHp{} low-J lines.

The \CCH, \cCCCHH, \twCN, \twCO, \HCN, and \HCOp{} low-J lines have similar
mean integrated intensities on regions of cold and hot dust, probably
because some of the high visual extinction lines of sight are also
associated with high dust temperature. The mean spectra of hot dust regions
have significantly wider linewidths than the mean spectra of the cold dust
regions, confirming that the gas temperature is significantly higher in at
least the PDR parts along these lines of sight. The \HNC, \thCO, \twCS,
\CseO, \ttSO, \CeiO, \HthCOp, \methanol, and \NNHp{} low-J line mean
integrated intensities are between 2 and 20 times brighter in the cold than
in the hot dust regions. These lines are thus more characteristics of cold
dense gas than the previous category. Finally, the mean line integrated
intensity over the full field of view is closer to the line intensity
integrated over the regions of warm and lukewarm dust regions than over the
cold or hot dust regions.

\TabAvMaskedMean{} %
\FigMaskedSpectraMeanAv{} %
\FigLineAreaAv{} %

\TabTdustMaskedMean{} %
\FigMaskedSpectraMeanTdust{} %
\FigLineAreaTdust{} %


\end{document}